\documentclass[preprint]{aastex}
\usepackage{amsfonts,amsmath,amssymb,amsthm,amsbsy,natbib,amsxtra, graphics}
\usepackage{lineno}
\usepackage[pdftex]{color}

\usepackage{array}
\usepackage{enumerate}
\usepackage{array}
\usepackage{float}
\usepackage{lscape}

\usepackage{listings}
\usepackage{boxedminipage}

\usepackage{tikz, calc}
\usetikzlibrary{shapes.geometric}
\usetikzlibrary{arrows}
\usetikzlibrary{decorations.markings}
\graphicspath{{images/}}

\makeatletter
\newcommand{\customlabel}[2]{%
\protected@write \@auxout {}{\string \newlabel {#1}{{#2}{}}}}
\makeatother

\newcommand{\dif}[3][\partial]{\frac{\mathrm{#1}#2}{\mathrm{#1}#3}}

\renewcommand{\b}{\boldsymbol}

\usepackage{mdwlist}

\newcommand*\patchAmsMathEnvironmentForLineno[1]{%
  \expandafter\let\csname old#1\expandafter\endcsname\csname #1\endcsname
  \expandafter\let\csname oldend#1\expandafter\endcsname\csname end#1\endcsname
  \renewenvironment{#1}%
     {\linenomath\csname old#1\endcsname}%
     {\csname oldend#1\endcsname\endlinenomath}}%
\newcommand*\patchBothAmsMathEnvironmentsForLineno[1]{%
  \patchAmsMathEnvironmentForLineno{#1}%
  \patchAmsMathEnvironmentForLineno{#1*}}%
\AtBeginDocument{%
\patchBothAmsMathEnvironmentsForLineno{equation}%
\patchBothAmsMathEnvironmentsForLineno{align}%
\patchBothAmsMathEnvironmentsForLineno{flalign}%
\patchBothAmsMathEnvironmentsForLineno{alignat}%
\patchBothAmsMathEnvironmentsForLineno{gather}%
\patchBothAmsMathEnvironmentsForLineno{multline}%
}

\title{A general condition for adaptive genetic polymorphism in temporally and spatially heterogeneous environments}	

\author{Hannes Svardal \altaffilmark{2,3,6}, Claus Rueffler \altaffilmark{2,4,7} and Joachim Hermisson \altaffilmark{2,5,8}}
\affil{$^{2}$Mathematics and Biosciences Group, Department of Mathematics, University of Vienna, 1090 Vienna, Austria}
\affil{$^{3}$present address: Gregor Mendel Institute, Austrian Academy of Sciences, 1030 Vienna, Austria}
\affil{$^{4}$Animal Ecology, Department of Ecology and Genetics, Uppsala University, 752 36 Uppsala, Sweden}
\affil{$^{5}$Max F. Perutz Laboratories, 1030 Vienna, Austria}
\altaffiltext{6}{corresponding author: hannes@svardal.at}
\altaffiltext{7}{claus.rueffler@ebc.uu.se}
\altaffiltext{8}{joachim.hermisson@univie.ac.at}

\makeindex

\begin{document}



\begin{abstract}
Both evolution and ecology have long been concerned with the impact of variable environmental conditions on observed levels of genetic diversity within and between species.
We model the evolution of a quantitative trait under selection that fluctuates in space and time, and derive an analytical condition for when these fluctuations promote genetic diversification. As ecological scenario we use a generalized island model  with soft selection within patches in which we incorporate generation overlap.
We allow for arbitrary fluctuations in the environment including spatio-temporal correlations and any functional form of selection on the trait.
Using the concepts of invasion fitness and evolutionary branching, we derive a simple and transparent condition for the adaptive evolution and maintenance of genetic diversity.  This condition relates the strength of selection within patches to expectations and variances in the environmental conditions across space and time. 
Our results unify, clarify, and extend a number of previous results on the evolution and maintenance of genetic variation under fluctuating selection. 
Individual-based simulations show that our results 
are independent of the details of the genetic architecture and on whether reproduction is clonal or sexual. The onset of increased genetic variance is predicted accurately also in small populations in which alleles can go extinct due to environmental stochasticity.

\end{abstract}

\newpage

\section{Introduction}

Explaining observed levels of genetic variation within natural populations is one of the major challenges in the study of evolution. 
One can distinguish adaptive explanations from non-adaptive explanations, with neutral variation and mutation-selection balance being the prime examples for the latter. Amongst adaptive explanations we can distinguish those based on genetic constraints such as over-dominance, where only heterozygote individuals can realize highest fitness, from those that do not rely on such constraints and consequently also apply to haploid species. In the latter case, genetic diversity is an adaptive response to the environment. Our aim is to characterize the conditions that select for adaptive diversity in this sense.

Both temporal and spatial fluctuations are omnipresent in  natural populations. From early on there has been a perception in evolutionary research that such fluctuations should be favorable for the evolution and maintenance of genetic diversity. Modelling efforts to support this claim started in the second half of the 20th century (reviewed in \citealp{Felsenstein:76} and \citealp{Hedrick:76}).
Spatial heterogeneity was soon identified as a potent factor to maintain diversity, even if individuals move freely among patches. The pioneering model here is the Levene model \citep{Levene:53,Gliddon:75}, an island model \citep{Wright:43} with different selective pressures among islands (or patches) and random dispersal of all individuals in each generation. As pointed out by \citet{Dempster:55}, a crucial feature of the Levene model is that density regulation takes place locally within patches.
This ``soft-selection'' regime \citep{Wallace:75,Christiansen:75} can adaptively maintain a protected genetic polymorphism. 
If only a part of the population disperses, this further aids the maintenance of genetic variation as sub-populations can become locally adapted \citep{Deakin:66,Deakin:68,Spichtig:04}.

In the absence of genetic constraints, it was initially thought that purely temporal fluctuations in selection are not sufficient to maintain genetic polymorphism, but would generally favor the genotype with the highest geometric mean fitness \citep{Gillespie:73,Gillespie:74,Felsenstein:76}. It was therefore concluded that temporal fluctuations can only account for a limited amount of the observed genetic variance in diploid populations \citep{Dempster:55} and that there is no tendency to maintain polymorphism in haploid populations \citep{Cook:74}. 
However, two mechanisms were subsequently identified that can maintain
genetic polymorphism under temporal fluctuations, named the ``storage
effect of generation overlap'' \citep{Chesson:81, Chesson:84}  and
the ``effect of relative non-linearity'' \citep{Chesson:94,Szilagyi:10}. 
In both cases, selection is no longer a function of time alone: With the
storage effect, selection acts only on a short-lived stage of the life
cycle (e.g., juveniles), while a long-lived stage (e.g., adults or
persistent dormant stages) is not affected by the fluctuations. With the
non-linearity effect, temporal fluctuations lead to fluctuations in the
population density and polymorphism is maintained by an additional
density-dependent selection component.

Environmental heterogeneities simultaneously occurring in space and time have also been studied. 
After an early attempt by \citet{Levins:62},  it was mainly Gillespie (\citeyear{Gillespie:74}, \citeyear{Gillespie:75}, \citeyear{Gillespie:76b}; but see also \citealp{Hedrick:78}) who treated this topic. He considered fluctuations in an island model, in which the distribution of environmental conditions is identical across all patches, but in which the realized environment at any given point in time may differ among patches. Gillespie's main conclusion was that these transient spatial differences can be sufficient to maintain genetic diversity.

Adaptive maintenance of genetic diversity can also be addressed from an ecological perspective, where species coexistence is a classical research focus (see \citealt{Chesson:00a} for a review). Models and methods are closely related to their genetic counterparts, although this connection is often not made explicit. For clonal inheritance, it is only a matter of semantics whether maintenance of polymorphism within species  (population genetics) or species diversity (ecology) is considered. Conditions for species coexistence in temporally and spatially fluctuating environments have been studied by \citet{Chesson:85,Chesson:00b} and \citet{Comins:85}. Both combine an island model with environmental fluctuations in time and space (similar to Gillespie's model) with Chesson's lottery model introducing generation overlap. They find that in such a scenario both temporal and spatial environmental fluctuations can promote species coexsistence.

Most models described above focus on the maintenance of diversity among two discrete and immutable alleles or types. Thus, stability of the polymorphism is considered from the short-term evolutionary perspective of the dynamics of allele (or phenotype) frequencies. From a long-term evolutionary perspective, one can further ask whether a polymorphism remains stable also in the presence of mutations leading to gradual adaptive changes in the allelic values or phenotypes. In particular, evolutionary stability in this sense should also guarantee that the polymorphism cannot be lost due to the appearance of a single superior (generalist) type. This long-term evolutionary stability has increasingly gained attention with the development of adaptive dynamics and the discovery of evolutionary branching points \citep{Metz:92,Metz:96a,Dieckmann:96,Geritz:98}. Evolutionary branching points are trait values that are attractors of the evolutionary dynamics, but once the population has evolved sufficiently close to such trait values, selection turns disruptive and alternative alleles can invade and coexist. In short, evolutionary branching indicates that the emergence and maintenance of genetic polymorphism is an adaptive process. Several recent studies have used this approach to ask how environmental heterogeneity affects the existence of evolutionary branching points. This has been done for purely spatial heterogeneity \citep{Meszena:97,Geritz:98, Day:00,Nilsson:10a,Nilsson:10b}, under purely temporal variation \citep{Ellner:94,Svardal:11, Abrams:13} and for a combination of the two \citep{Kisdi:02,Parvinen:04,Nurmi:08,Nurmi:11}. Note that the latter studies by Parvinen and coworkers are meta-population models in which  temporal variation is introduced through catastrophes wiping out local populations. The general conclusion from the above studies is that in spatially heterogeneous environments  low migration and large spatial differences favor evolutionary branching, while under purely temporal fluctuations a sufficiently large generation overlap is necessary for branching.

In this article, we follow the recent line of research and ask how environmental heterogeneity affects the scope for the adaptive evolution and maintenance of genetic polymorphism. We consider a modified island model with local population regulation resulting in constant patch occupancies (soft selection), which combines features from the approaches above and extends them in several directions. We follow the evolution of a quantitative trait with a continuum of alleles. The strength and direction of selection within a patch depends on the realized environmental condition. In particular, we allow for an arbitrary distribution of environmental conditions across space and time, including spatial and temporal correlations. The functional dependence of fitness on the trait is also arbitrary. For example, selection can be stabilizing with the optimal trait value depending on the realized local environment, or directional with the direction fluctuating in space and time. We analytically derive a condition for the existence of an evolutionary branching point and investigate the robustness our finding with individual-based simulations.

\section{Model}\label{sec:model}

\subsection{Population structure and life cycle}

Consider the classical island model of population genetics \citep{Wright:43}.  A population occupies $n$ patches that are connected by dispersal. 
We assume that the population dynamics is regulated locally by a limiting resource (e.g., space) so that the adult population size within each patch is at a stable equilibrium and stays constant over time. 
For our analytical treatment, we assume that local populations are sufficiently large so that stochastic effects due to drift can be ignored. This assumption is relaxed in section \ref{sec_robustness}, where we present simulation results for small populations. For large populations, our results are independent of the total population size and it is sufficient to follow relative  population sizes in the following. 

\begin{figure}[h!]
\begin{tikzpicture}[scale=4]
\node at (-1.7,0.3) {\framebox{time-step $t$}};
\node at (-1.,-0.32) {$(1)$};
\node at (0.,-0.35) {$(1)$};
\node at (1.,-0.35) {$(1)$};

\node at (-1.7,0) {reproduction (1)};
\node at (-1.7,-0.13) {adult death (2)};
\node at (-1.7,-0.45) {selection (3)};

\fill[black!20!white,draw=black] (-1,0) -- +(60:0.25cm)  arc (60:120:0.25cm) -- (-1,0);
\node (p1)[shape=circle,draw,minimum size=2cm] at (-1,0) {};
\node at (p1) [below] {$k_1$ adults};
\node at (p1) [above=0.9cm] {\small $(1-\gamma)$};
\node at (p1) [above=0.4cm] {(2)};
\node at (-0.5,0) {...};

\fill[black!20!white,rotate=60,draw=black]  (0,0) --  +(0.25cm,0em)  arc (0:60:0.25cm) -- (0,0);
\node (p2)[shape=circle,draw,minimum size=2cm] at (0,0) {};
\node at (p2) [below] {$k_i$ adults};
\node at (p2) [above=0.9cm] {\small $(1-\gamma)$};
\node at (p2) [above=.4cm] {(2)};

\node at (0.5,0) {...};

\fill[black!20!white,draw=black]  (1,0) --  +(60:0.25cm)  arc (60:120:0.25cm) -- (1,0);
\node (p3)[shape=circle,draw,minimum size=2cm] at (1,0) {};
\node at (p3) [below] {$k_n$ adults};
\node at (p3) [above=0.9cm] {\small $(1-\gamma)$};
\node at (p3) [above=0.4cm] {(2)};


\draw (p1.south) +(-0.05,0) -- +(-0.3,-0.5);
\draw (p1.south) +(0.05,0) -- +(0.3,-0.5);
\draw (p1.south) +(0.0,-0.55) circle [x radius=0.81em,y radius=0.38em] node (o1) {offspring pool};

\draw (p2.south) +(-0.05,0) -- +(-0.3,-0.5);
\draw (p2.south) +(0.05,0) -- +(0.3,-0.5);
\draw (p2.south) +(0.0,-0.55) circle [x radius=0.81em,y radius=0.38em] node (o2) {offspring pool};

\draw (p3.south) +(-0.05,0) -- +(-0.3,-0.5);
\draw (p3.south) +(0.05,0) -- +(0.3,-0.5);
\draw (p3.south) +(0.0,-0.55) circle [x radius=0.81em,y radius=0.38em] node (o3) {offspring pool};

\node at (-1.7,-1.1) {density regulation (4)};
\draw (o1.south) +(-0.31,0.05) -- +(-0.05,-0.53);
\draw (o1.south) +(+0.31,0.05) -- +(+0.05,-0.53);
\node (k1) at (o1.south) [below=5.9em] {$c_1$};
\draw (k1) ellipse (0.2cm and 0.1cm);

\draw (o2.south) +(-0.31,0.05) -- +(-0.05,-0.53);
\draw (o2.south) +(+0.31,0.05) -- +(+0.05,-0.53);
\node (k2) at (o2.south) [below=5.9em] {$c_i$};
\draw (k2) ellipse (0.2cm and 0.1cm);

\draw (o3.south) +(-0.31,0.05) -- +(-0.05,-0.53);
\draw (o3.south) +(+0.31,0.05) -- +(+0.05,-0.53);
\node (k3) at (o3.south) [below=5.9em] {$c_n$};
\draw (k3) ellipse (0.2cm and 0.1cm);

\node at (-1.7,-2.1) {dispersal (5)};
\node (disp) at (0.,-2.1) [shape=ellipse,draw,minimum size=4em] {global migrant pool};

\tikzstyle{arrow}=[decoration={markings,mark=at position 1 with {\arrow[scale=2.5]{>}}},
    postaction={decorate},
    shorten >=0.4pt]
\tikzstyle{backarrow}=[decoration={markings,mark=at position 0.08 with {\arrow[scale=2.5]{<}}},
    postaction={decorate}]

\draw [arrow](k1.south east) +(0.04cm,-0.02cm) -- (disp.north west) node [above,midway] {$m$};
\draw [arrow](k2.south) +(0.06cm,-0.04cm) -- +(0.06,-0.34) node [right,midway] {$m$};
\draw [arrow](k3.south west) +(-0.04cm,-0.02cm) -- (disp.north east) node [above,midway] {$m$};

\node (k12) [shape=ellipse,draw=black,minimum width=4.2em] at (o1.south) [below=17em] {$c_1$};

\node (k22) [shape=ellipse,draw=black,minimum width=4.2em] at (o2.south) [below=17em] {$c_i$};

\node (k32) [shape=ellipse,draw=black,minimum width=4.2em] at (o3.south) [below=17em] {$c_n$};


\begin{scope}[yshift=-1em]

\node at (-1.7,-2.3) {recruitment (6)};
\node at (-1.7,-2.5) {\framebox{time-step $t+1$}};

\fill[black!20!white,draw=black] (-1,-2.7) -- +(60:0.25cm)  arc (60:120:0.25cm) -- (-1,-2.7);
\node (p1p)[shape=circle,draw,minimum size=2cm] at (-1,-2.7) {};
\node at (p1p) [below] {$k_1$ adults};
\node at (-0.5,-2.7) {...};

\fill[black!20!white,draw=black] (0,-2.7) --  +(60:0.25cm)  arc (60:120:0.25cm) -- (0,-2.7);
\node (p2p)[shape=circle,draw,minimum size=2cm] at (0,-2.7) {};
\node at (p2p) [below] {$k_i$ adults};
\node at (0.5,-2.7) {...};

\fill[black!20!white,draw=black]  (1,-2.7) --  +(60:0.25cm)  arc (60:120:0.25cm) -- (1,-2.7);
\node (p3p)[shape=circle,draw,minimum size=2cm] at (1,-2.7) {};
\node at (p3p) [below] {$k_n$ adults};

\end{scope}

\draw [backarrow] (k12.15)  -- (disp.south west);
\node (das1) [xshift=0.1cm] at  (p2p.north) {};
\node (das2) [xshift=0.24cm] at  (disp.south) {};
\draw [backarrow] (k22.north) + (0.06,0.02) -- (das2.north);
\draw [backarrow] (k32.165)  -- (disp.south east);

\draw[arrow]
    (k1.south) +(0cm,-0.04cm) -- (k12.north) node [left,near start] {$1-m$};
\draw[arrow] (k3.south) +(0cm,-0.04cm) -- (k32.north) node [right,near start] {$1-m$};
\draw[dashed,arrow]
  (k2.south) +(-0.06cm,-0.04cm) -- (k22.120)  node [left,very near start] {$1-m$};

\draw[arrow] (k12.south) -- (p1p.north);
\draw[arrow ] (k22.south) -- (p2p.north);
\draw[arrow ] (k32.south) -- (p3p.north);

\node (disp) at (0.,-2.1) [shape=ellipse,draw,minimum size=4em,fill=white] {global migrant pool};

\end{tikzpicture}
\caption{Life cycle: (1) reproduction, (2) adults die with probability $1-\gamma$, (3) viability selection on juveniles, dependent on trait and patch-specific environmental condition $ \theta_{it}$, (4) density regulation within patches resulting in the relative offspring contribution $c_i$, (5) a proportion $1-m$ of the offspring stays in the patch of origin, the rest contributes to a global migrant pool from where offspring are redistributed over all patches, (6) offspring replace the deceased adults and offspring that cannot establish die.}
\label{fig_model}
\end{figure}
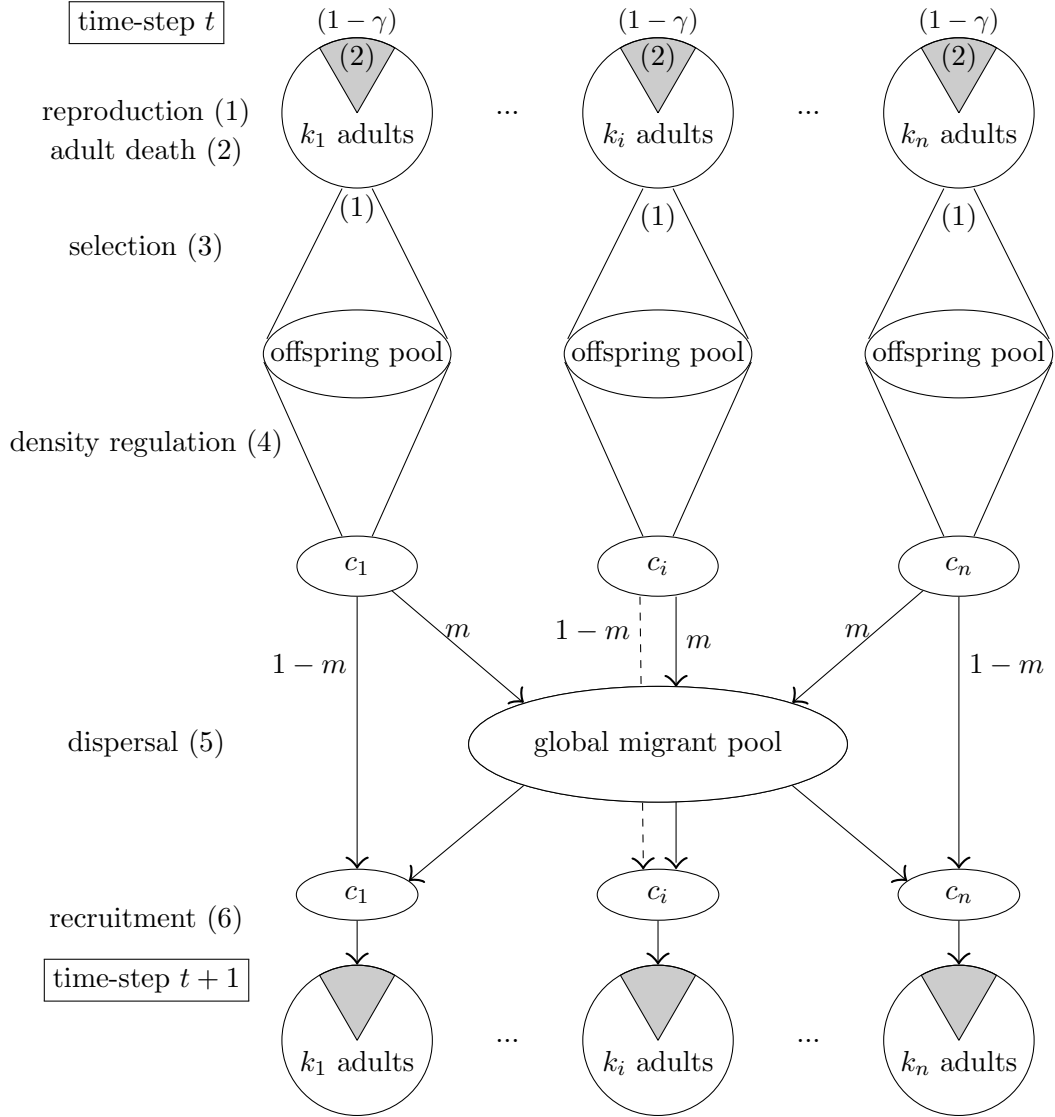

The life cycle is shown in figure \ref{fig_model} and an overview of our notation can be found in table \ref{tbl_notation}. The relative carrying capacity for adults may depend on the patch, with a fraction of $k_i$ adults living in the $i$th patch, i.e., $\sum_{i=1}^n k_i=1$. Adults reproduce within their patch and a fraction $1-\gamma$ dies after reproduction. The remaining fraction $\gamma$ survives to the next reproductive season. Hence, we allow for overlapping generations and the parameter $\gamma$ will be called ``generation overlap'' in the following. The case $\gamma=0$ corresponds to the classical island model with non-overlapping generations and for $\gamma \to 1$ the individuals approach immortality. Note that the long-lived life-stage could also be a dormant stage such as resting eggs or a seed bank \citep{Ellner:94}. 

Juveniles are subject to local viability selection, which depends on their phenotype and the realized environmental condition in the patch. After selection, trait-independent local density regulation further decreases offspring to a patch specific relative juvenile carrying capacity $c_i$, where $\sum_{i=1}^n c_i=1$. A fraction $m$ of the surviving offspring  disperses globally so that the probability of arriving in a certain patch is independent of the patch of origin. This can be modeled as a common dispersal pool for the dispersing offspring of all patches. Note that this includes the possibility that dispersers return to their patch of origin. A fraction $1-m$ of juveniles stays in their native patch.

Offspring individuals are recruited to the adult population of patch $i$ until its adult carrying capacity $k_i$ is reached. The offspring population is assumed large enough so that this is always possible. Hence, a fraction   $(1-\gamma)$ of the adults in each patch derive from the offspring of the previous reproductive season. New recruits are taken with equal probabilities from the non-dispersing offspring in patch $i$ and the immigrants. All non-recruited juveniles die.

\begin{table}[htb] 
\caption{Overview of the notation}
\begin{tabular}{lp{0.8\textwidth}}
\hline
$m$&probability that a juvenile individual disperses out of its native patch\\
$\gamma$& generation overlap; probability that an adult individual survives to the next reproductive season\\
$k_i$, $c_i$&relative adult and juvenile carrying capacity in patch $i$\\
$y$, $x$&trait values of the quantitative trait; under the adaptive dynamics approximation, $y$ is the mutant trait value and $x$ the resident trait value\\
$x^*$&trait value at which directional selection is zero on average (singular point)\\
$\phi_{it}(y)$& frequency of individuals with trait value $y$ in patch  $i$ at time $t$ \\
$\theta_{it}$&environmental condition in patch $i$ at time $t$\\
$r(y,\theta_{it})$& expected number of offspring before density regulation of an individual with trait value $y$ under environmental condition $\theta_{it}$ \\
$\rho(y,\phi_{it},\theta_{it})$& relative reproductive success of an individual with trait value $y$ in patch $i$ at time $t$; under the adaptive dynamics approximation we write $\rho(y,x,\theta_{it})$ for the reproductive succes of a mutant individual with trait value $y$ in a population with resident trait value $x$\\
$s_{it}$&$=\ln({\rho(y,x,\theta_{it})})$; local selection coefficent\\
$\partial s_{it},  \partial^2 s_{it}$& local selection gradient  and local selection curvature evaluated at the singular point; $\frac{\partial s_{it}}{\partial y}|_{y=x=x^*}$ and $\frac{\partial^2 s_{it}}{\partial  y^2}|_{y=x=x^*}$, respectively \\
$l_{ij}(y,\phi_{jt},\theta_{jt})$&expected contribution of individuals with trait value $y$ from patch $j$ to the adult population of patch $i$ at the next time step; under the adaptive dynamics approximation we write  $l_{ij}(y,x,\theta_{jt})$\\
${\sf L}(y,\phi_{1t},..,\phi_{nt},\theta_{1t},..,\theta_{nt})$&population projection matrix with elements $l_{ij}(y,\phi_{jt},\theta_{jt})$; under the adaptive dynamics approximation we write ${\sf L}(y,x,\theta_{1t},...,\theta_{nt})$\\
$w(y,x)$& invasion fitness\\
${\rm E_S}[.]$, ${\rm E_T}[.]$&spatial and temporal averages\\
${\rm Var_S}[.]$, ${\rm Var_T}[.]$& spatial and temporal variances\\
\hline
\end{tabular}
\label{tbl_notation}
\end{table}

\subsection{Phenotypes and environments}

Individuals are characterized by a quantitative trait that can take any value $y$ in the real numbers. 
For concreteness, we assume that the trait is expressed in juveniles and affects juvenile viability, but our results also apply to a trait affecting adult fecundity (see \textit{Discussion}).
We denote by $\phi_{it}$ the density function of individuals with different trait values in patch $i$ at time $t$. 
Hence, $\phi_{it}(y){\rm d}y$ is the frequency of individuals with a trait value between $y$ and $y+{\rm d}y$ and we have $\int_{-\infty}^{\infty}\phi_{it}(y){\rm d}y=1$. 
For simplicity, we refer to $\phi_{it}(y)$ as frequency of individuals with trait value $y$ in the following.
We assume that genotypes
uniquely map to phenotypes. Therefore, selection for phenotypic diversity at the trait level also selects for genetic diversity.

The condition of the environment in patch $i$ at time $t$ will be denoted by $\theta_{it}$. 
In general, $\theta_{it}$, can be a vector containing an arbitrary number of external environmental factors, such as temperature, humidity, or the presence/absence of a pathogen, each taking either continuous or discrete values. For simplicity, we will assume in the following that $\theta_{it}$ is scalar, but note that all our general derivations also holds for vector-valued  $\theta_{it}$.
We refer to the realization of the environmental conditions in all patches at time $t$ as environmental state and denote it by  ($\theta_{1t},..,\theta_{nt}$). The set of possible environmental states, i.e., the sample space, will be denoted by $\Omega$. Our analytical treatment relies on the assumption that the probability for a certain environmental state does not explicitly depend on time.   In other words, we assume  a stationary distribution of environmental states and denote the probability density for environmental state $(\theta_{1},..,\theta_{n})$ by $f(\theta_{1},..,\theta_{n})$. For example, this is the case if the environmental states are determined by an ergodic Markov process.
It follows that our formalism does not capture scenarios of prolonged directional change, such as global warming. 
However, our model allows for any form of spatial and temporal correlations in the environmental conditions.

\subsection{Selection}

We denote by $r(y, \theta_{it})$ the reproductive success, defined as the number of offspring after viability selection, of an individual with trait value $y$ under environmental condition $\theta_{it}$. 
 Our main result holds for any function $r(y,\theta_{it})$. As a concrete example, we consider Gaussian stabilizing selection, where the scalar environmental condition $\theta_{it}$ determines the selective optimum for the trait,
\begin{equation}\label{eq:selection}
r(y,\theta_{it})=r_{\rm max}\exp{\left[-\frac{( \theta_{it}-y)^{2}}{2\sigma^{2}}\right]}.
\end{equation}
Here, $r_{\rm max}$ is the maximal number of offspring. Given that it is the same for all individuals, it cancels out in the following. $\sigma^2$ parametrizes the strength of stabilizing selection. Note that small values of $\sigma^2$  correspond to strong  selection and vice versa.

In a second step, the relative number of offspring in each patch is reduced by local density regulation (due to limitations in space or other resources) to the juvenile carrying capacity $c_i$.
Importantly, the density-dependent reduction in total patch offspring is a trait-independent
 random sampling step so that the expected contribution of an adult individual from patch $i$ to $c_i$ is proportional to its reproductive success relative to the average reproductive success in the patch. We denote an individual's relative reproductive success by
\begin{equation} \label{eq_rel_success}
\rho(y,\phi_{it},\theta_{it})=\frac{r(y, \theta_{it})}{\int_{-\infty}^\infty {\phi_{it}(x) r(x, \theta_{it}){\rm d}x}},
\end{equation}
where the integral in the denominator normalizes the reproductive success by the average reproductive success in the patch. Hence, we assume soft selection \citep{Wallace:75} and $\rho$ depends on the local frequency distribution of all phenotypes, $\phi_{it}$, in patch $i$ at time $t$.

\subsection{Dispersal}

A fraction $m$ of the offspring surviving density regulation disperses over the whole population. We assume that the number of migrants that each patch $i$ contributes to and receives from the global migrant pool is proportional to its juvenile carrying capacity $c_i$.
While being mathematically convenient, this choice is also biologically sensible. It means that the chance that an individual arrives in a certain patch is proportional to the resources this patch provides for offspring.
Finally, from the non-dispersing offspring of patch $i$ and the dispersing offspring that arrive in patch $i$, a random subset  is selected to replace the fraction of $(1-\gamma)$ deceased adults. All non-recruited offspring die.
Note that out model allows the relative adult carrying capacity of a patch ($k_i$) and the relative number of juveniles it sends out and receives ($c_i$) to differ.

\subsection{Dynamics} The frequency of adult individuals with  trait value $y$ in generation $t+1$ is given by 
 \begin{equation} \label{eq_popdyn_restricted}
\phi_{i,t+1}(y)=\sum_{j=1}^n l_{ij}(y,\phi_{jt},\theta_{jt}) \phi_{jt}(y),
\end{equation}
where $l_{ij}$ are the elements of the population projection matrix ${\sf L}(y,\phi_{1t},..,\phi_{nt},\theta_{1t},..,\theta_{nt})$. They are given by 
\begin{subequations} \label{eq_L_main}
\begin{align}
&l_{ij}(y,\phi_{jt},\theta_{jt})=(1-\gamma)mc_j  \rho(y,\phi_{jt},\theta_{jt})  \text{ for } j\neq i \label{lji}\\
&l_{ii}(y,\phi_{jt},\theta_{jt})=\gamma+(1-\gamma)\left(1-m+m c_i\right) \rho(y,\phi_{it},\theta_{it}) \label{lii}
\end{align}
\end{subequations}
(\mbox{appendix \ref{app_model}}).
Each entry $l_{ij}$ gives the expected contribution of individuals with trait value $y$ from patch $j$ to the adult population of patch $i$ at the next time step.
 The term $\gamma$ on the right-hand side of equation \eqref{lii} reflects the possibility that an adult can survive from one time step to the next. All other terms represent the recruitment of offspring to the adult population. By describing the adult population in terms of the relative frequency of individuals with a given trait value, rather than in absolute numbers, equations \eqref{eq_popdyn_restricted} and \eqref{eq_L_main}, and therefore all further derivations, become independent of the adult carrying capacities, $k_i$. Thus, our  results will not depend on the relative patch sizes in the adult population.

In table \ref{tb_special_cases} we list previously analyzed models with ecological settings that are special cases of our model. Furthermore, the table gives the parameter values for which our ecological model becomes equivalent to the  model in the reference.

\begin{landscape}
\begin{table}[htb]
\caption{Previously analyzed models that are special cases of our model}
\begin{tabular}{lp{0.4\textwidth}p{0.45\textwidth}}
\hline
reference &model assumptions in our notation&note\\
\hline
\citet{Levene:53,Geritz:98}&$\gamma=0$, $(\theta_{1},..,\theta_{n})$ fixed over time, $m=1$& Levene model, see appendix \ref{app_gauss_special_cases}\\
\citet{Deakin:66}&$\gamma=0$, $(\theta_{1},..,\theta_{n})$ fixed over time&see appendix, inequality \eqref{eq:br_levene}\\
\citet{Spichtig:04}&$\gamma=0$, $(\theta_{1},..,\theta_{n})$ fixed over time, $n=2$& studies the effect of the number of loci\\
\citet{Gillespie:73}&$\gamma=0$, $n=1$& \\
\citet{Gillespie:74}&$\gamma=0$, $m=1$, $\theta_i$ IID& see appendix, inequality \eqref{eq:br_iid} \\
\citet{Gillespie:75}&$\gamma=0$, $n=2$&\\
\citet{Gillespie:76b}&$\gamma=0$, $n\to\infty$&\\
\citet{Gillespie:76}&$\gamma=0$, $m=1$, $n\to\infty$, $\theta_i$ IID& \\
\citet{Chesson:81};& & \\
\citet{Ellner:94};& &\\
\citet{Svardal:11}&$n=1$&lottery model, see appendix,  inequality \eqref{eq_lottery}\\
\citet{Chesson:85}&$m=1$, no temporal correlations, $n \to\infty$&patchy environment lottery (PEL) model\\
\citet{Comins:85}&$\gamma=0$, $\theta$ identically distributed & \\
\hline
\end{tabular}
\label{tb_special_cases}
\note{
While our model covers and extends most ecological assumptions of these studies, genetic assumptions often differ. IID: identically and independently distributed.}
\end{table}
\end{landscape}

\subsection{Invasion analysis}

For our analytical results we rely on the adaptive dynamics approximation  \citep{Dieckmann:96, Metz:96a, Geritz:98}, which is based on the assumption of rare mutations of small effect. A comprehensive account of the analytical methodology is given in appendix \ref{app_proof}. Under these assumptions, directional evolution is a sequence of substitution events leading toward trait values at which directional selection vanishes. These trait values are attractors of the trait substitution sequence. At such points, selection can be stabilizing or disruptive. In the first case, genetic variation can only be maintained at mutation-selection equilibrium, whereas in the second case genetic variation increases due to adaptive evolution. Trait values of the latter type are known as evolutionary branching points. Importantly, in the vicinity of evolutionary branching points a genetic polymorphism cannot be replaced by a single genotype. In this article, we identify the conditions for the existence of evolutionary branching points. With  computer simulations we show that -- in reasonably large populations -- our results accurately predict the adaptive evolution of genetic polymorphism, even  if the various assumptions of the adaptive dynamics approximation are violated.

We assume that at most one mutation segregates in the population at each point in time and, shortly after a new mutation has occurred, most individuals have the common ``resident'' trait value, $x$, and only few individuals have a very similar ``mutant'' trait value $y$. Then,  the phenotype density in each patch is dominated by the resident phenotype (technically, $\phi_{it}(x') \approx \delta(x'-x)$, where $ \delta(x'-x)$ is the Dirac delta function). Hence, the relative reproductive success of a rare mutant individual with trait value $y$ in patch $i$ at time $t$ simplifies from equation \eqref{eq_rel_success} to 
\begin{equation} \label{eq_rho_ad}
\rho(y,\phi_{it}, \theta_{it})\approx \frac{r(y, \theta_{it})}{r(x, \theta_{it})}=:\rho(y,x, \theta_{it}),
\end{equation}
where in the following we include the resident trait value $x$ as  second argument to $\rho$ and omit the trivial phenotype density function. In other words, mutant individuals are so rare that they only compete with resident individuals during density regulation.
We call the logarithm of $\rho(y,x, \theta_{it})$ ``local selection coefficient'' and denote it by
\begin{equation}\nonumber
s_{it}:=\ln{(\rho(y,x, \theta_{it}))}.
\end{equation}
Furthermore, we refer to its first derivative with respect to the mutant trait value, $\frac{\partial s_{it}}{\partial y}|_{y=x}$, as ``local selection gradient''. 
It measures within-patch  directional selection at a given time step. The second derivative, $\frac{\partial^2 s_{it}}{\partial y^2}|_{y=x}$, called ``local selection curvature'' in the following, measures within-patch quadratic selection  at a given time step. It can be stabilizing ($<0$) or disruptive ($>0$). 

The relative number of mutants in the next generation in the total population is then given by equation  \eqref{eq_popdyn_restricted}, where $\rho(y,\phi_{it}, \theta_{it})$ in equations (\ref{eq_L_main} a,b) is replaced with $\rho(y, x, \theta_{it})$ from equation \eqref{eq_rho_ad}. The ultimate fate of a new mutation is determined by its long term growth rate, which is called invasion fitness. More precisely, invasion fitness is the long term average change in density of individuals with the rare mutant trait value $y$ in a population dominated by trait value $x$.
For a given sequence of environmental states $\{(\theta_{1,0},..,\theta_{n,0}),..,(\theta_{1,T},..,\theta_{n,T})\}$, we call $\lambda_T(y,x)$ the leading eigenvalue of the product of the population projection matrices, $\prod_{t=0}^T {\sf L}(y,x,\theta_{1t},..,\theta_{nt})$. Then, invasion fitness is given by
\begin{equation} \label{eq_inf_fit_lambda}
w(y,x) = \lim_{T\rightarrow \infty} \frac{1}{T} \lambda_T(y,x)
\end{equation} 
\citep{Metz:92, Metz:08c}. Our assumptions on the distribution of the $(\theta_{1t},..,\theta_{nt})$ guaranty convergence of $w(y,x)$ to a fixed value which is independent of the sequence of environmental states (c.f.  appendix \ref{app_proof}).

The  direction of the trait substitution sequence is given by the  ``global'' selection gradient, $S(x)=\dif{w(y,x)}{y}|_{y=x}$. Directional selection vanishes for trait values $x^*$ for which $S(x^*)=0$. Finally, a trait value $x^*$ that is an attractor of the evolutionary dynamics and simultaneously a fitness minimum, i.e.,  $\frac{\partial^{2} w(y,x^*)}{\partial y^2}|_{y=x^{*}}>0$, is an evolutionary branching point (see appendix \ref{app_proof} for details).

\subsection{Simulations} We complement the analytical invasion analysis with two computational approaches. First, we use numerical iterations to calculate the branching condition. This approach implements the assumptions of adaptive dynamics, but mutations change trait values by a discrete amount and the long term success of a mutation is determined from a long but finite sequence of environmental states (cf. online supplementary \ref{app_num_calc}). Second, we employ individual-based computer simulations of the system described by equations \eqref{eq:selection}-\eqref{eq_L_main}. This approach considers finite populations and allows us to study different genetic systems and  mutation rates and effect sizes. We consider sexually reproducing diploid genetics with the trait being determined by up to ten additive loci, each with infinitely many alleles. Mutational effect sizes are drawn from a normal distribution. Alternatively, we assume a large number of loci  (50-500) with two  alleles each. More details on the simulations  and the commented source code can be found in online supplementary \ref{app_sim} and \ref{app_sc_ind}, respectively.

In the simulation study, we assume Gaussian stabilizing selection (cf. equation \ref{eq:selection}) and investigate many different distributions of environmental conditions. For the figures presented in this article, we focus on the case that the selective optima in different patches occur independently of each other and, if not mentioned otherwise, can take values 0 and 1.  Furthermore, all our simulations for the case of identically distributed patches assume that both selective optima occur with equal probability (of 0.5) in all patches, while in the case of  differently distributed patches we assume that selective optimum 1 occurs with probability 2/3 in half of the patches and with probability 1/3 in the other half (we always assume an even number of patches in this case). The additional parameters used in the simulation study are summarized in table \ref{tbl_sim_params}.
 In the simulations, genetic variance is measured as the variance in trait values averaged over 
the last $8\cdot10^4$ time-steps of 10 simulations running for a total of  $10^5$ time-steps each. 

\begin{table}[htb]
\caption{Standard simulation parameters}
\begin{tabular}{llc}
\hline
symbol&explanation&standard value\\
\hline
$N$&population size&6000\\
$c_i$, $k_i$&juvenile and adult carrying capacities&$1/n$\\
$T_{\rm max}$&number of time steps simulated&100000\\
$\mu_{\rm trait}$& trait mutation probability per generation&0.01\\
$\sigma_\mu$&expected mutational effect size; effect is drawn from $N(0,\sigma_\mu^2)$&0.01\\
$ms$&bin size for mutational effects; effects are  rounded to multiples of $ms$&0.01\\
$k$&number of loci&4\\
$rec$&recombination probability&0.01\\
\hline
\end{tabular}
\note{Values of other parameters are given in the respective figure legends.}
\label{tbl_sim_params}
\end{table}

\section{Results}
\customlabel{sec:full_dispersal}{\textit{Results}}

Let $a_i$ and $b_t$ be arbitrary functions of the environmental conditions $\theta_{it}$. 
We define spatial and temporal averages 
as
\begin{equation}\label{eq_es}
{\rm E_S}[a] = \sum_{i=1}^n c_i a_{i}
\end{equation}
\begin{equation} \label{eq_et}
{\rm E_T}[b] = \lim_{T\to\infty} \frac{1}{T} \sum_{t=1}^T b_{t}
\end{equation}
Note that the $a_i$ can explicitly depend on time (such as $a_i$ = $\theta_{it}$), in which case also ${\rm E_S}[a]$ will be a function of time.  Similarly,  $b_t$ can either be patch-specific (such as $b_t = \theta_{it}$) or a function of the environmental conditions across several patches. In particular, we often encounter cases such as $b_t = \sum_i c_i \theta_{it} = {\rm E_S[\theta]}$ in double averages ${\rm E_T}[{\rm E_S}[\theta]] ={\rm E_S}[{\rm E_T}[\theta]]$.
Note that by using the self-averaging property (ergodicity) of the
process that describes the environmental fluctuations, we can write equation \eqref{eq_et} as
\begin{equation} \label{eq_etd}
{\rm E_T}[b] =\int_\Omega b \cdot  f(\theta_1,..,\theta_n) {\rm d} \theta_1..{\rm d}\theta_n.
\end{equation}
The temporal average is thus an expectation with respect to the stationary distribution $f(\theta_1,..,\theta_n)$ of environmental states.  
We define variances and covariances in an analogous way, e.g., ${\rm Var_S}[a]={\rm E_{S}}[a^2]-{{\rm E_{S}}[a]}^2$.

For brevity we introduce the following short-hand notation for the local selection gradient and the local selection curvature  evaluated at a trait value $x^*$ for which the global selection gradient (defined below equation \ref{eq_inf_fit_lambda}) is zero:
\begin{equation}
\partial s_{it}:=\frac{\partial s_{it}}{\partial y}\biggm|_{y=x=x^*} \text{  and  } \partial^2 s_{it}:=\frac{\partial^2 s_{it}}{\partial y^2}\biggm|_{y=x=x^*}.
\end{equation}
In appendix \ref{app_proof}  we prove that trait values $x^*$ can be found by solving
\begin{equation} \label{eq_sp_general}
{\rm E_T}\left[{\rm E_S}\left[{\partial s}\right]\right]=0,
\end{equation} 
For the case of Gaussian selection, there is only one such trait value given by 
\begin{equation}\label{eq:sp}
x^*=\mathrm{E}_{\rm T}[\mathrm{E}_{\rm S}[\theta]]
\end{equation}
(appendix \ref{app_special_cases}).
Equation \eqref{eq_sp_general} says that global directional selection vanishes for trait values for which the spatio-temporal average of the local selection gradients equals zero. From equation \eqref{eq:sp} follows that for Gaussian selection such trait values match the spatio-temporal average selective optimum. Furthermore, such trait values $x^*$ are evolutionary attractors if and only if
\begin{equation} \label{eq_cs_general}
{\rm E_T}\left[{\rm E_S}\left[\partial^2 s\right]\right]<0.
\end{equation}
Thus, $x^*$ is an attractor of the evolutionary dynamics if for this trait value the spatio-temporal average of local selection curvatures is negative, that is, if local selection is on average stabilising. Under Gaussian selection, condition \eqref{eq_cs_general} becomes $-1/\sigma^2<0$ and is always fulfilled.
Note that equations \eqref{eq_sp_general} and \eqref{eq_cs_general} are independent of the dispersal probability, $m$, the generation overlap, $\gamma$, and of temporal and spatial correlations. The order of temporal and spatial averages can be exchanged in these formulas.

The following inequality is our main result. We show in appendix \ref{app_proof} that an evolutionary attractor $x^*$ is an evolutionary branching point if
\begin{equation} \label{eq_branch_general}
\begin{split}
\hspace*{-1.5em} {\rm E_T}\left[{\rm Var_S}\left[{\partial s}\right]\right]+\gamma{\rm Var_T}\left[{\rm E_S}\left[{\partial s}\right]\right]+2\frac{1-m}{m}{\rm Var_S}\left[{\rm E_T}\left[{\partial s}\right]\right]+{\rm \mathcal{C}}\left[{\partial s}\right]>-{\rm E_T}\left[{\rm E_S}\left[{\partial^2 s}\right]\right].
\end{split}
\end{equation}
If this condition is fulfilled, there is selection for genetic polymorphism.
For the case of Gaussian selection, the condition can be rewritten as
\begin{equation} \label{eq_br_gauss_noTC}
\sigma^2_{\text{crit}}:={\rm E_T}\left[{\rm Var_S}\left[\theta\right]\right]+\gamma{\rm Var_T}\left[{\rm E_S}\left[\theta\right]\right]+2\frac{1-m}{m}{\rm Var_S}\left[{\rm E_T}\left[\theta\right]\right]+{\rm \mathcal{C}}[\theta]>\sigma^2,
\end{equation}
where we call the left-hand side $\sigma^2_{\text{crit}}$.
In both cases,
\begin{equation} \label{eq_covt}
{\rm \mathcal{C}}[{a}]:=2(1-\gamma)(1-m)\sum_{\Delta t=1}^\infty(1-(1-\gamma)m)^{\Delta t-1}\left({\rm E_S}\left[{\rm Cov_T}\left[{a_t},{a_{t+\Delta t}}\right]\right] -{\rm Cov_T}\left[{\rm E_S}\left[{a_t}\right], {\rm E_S}\left[{a_{t+\Delta t}}\right]\right]\right),
\end{equation}
with $a=\partial s$ or $\theta$, respectively. ${\rm \mathcal{C}}[{a}]$ summarizes the effect of temporal autocorrelations as discussed below. Note that the covariances ${\rm Cov_T}[.]$ are averages over all times $t$, but we leave the summation index $t$ explicit to express the dependence on the time difference $\Delta t$. 
Conditions \eqref{eq_branch_general} and \eqref{eq_br_gauss_noTC} are readily evaluated for a given pattern of fluctuations in the environmental conditions. We compute and discuss various special cases in appendix \ref{app_gauss_special_cases}, including environmental distributions with continuous and discrete environmental states and with and without temporal and spatial correlations.  In the following, we give an interpretation  for each  term in the branching condition.

\subsection{Influence of the shape of within-patch selection}

Inequalities \eqref{eq_cs_general} and \eqref{eq_branch_general} define upper and lower bounds for the average curvature of local selection. Spatio-temporal average selection must be stabilizing for the trait value $x^*$ to be an evolutionary attractor (inequality \ref{eq_cs_general}).
However, for evolutionary branching to occur the ``diversifying" factors on the left-hand side of condition \eqref{eq_branch_general} have to dominate stabilizing selection.  Otherwise, the trait value $x^*$ is an evolutionary end point (sometimes called ``continuously stable strategy'' or CSS, \citealt{Eshel:83}; inequality \ref{eq_branch_general} with reversed inequality sign).

For the case of Gaussian stabilizing selection, the right-hand side of inequality \eqref{eq_branch_general} is  $1/\sigma^2$, but the terms on the left-hand side scale with $1/\sigma^4$. Hence, stronger stabilizing selection (smaller $\sigma^2$) always promotes branching (inequality \ref{eq_br_gauss_noTC}).
Our formalism can also accommodate stabilizing selection that differs in strength among patches (but still remains constant over time). In this case, the terms in the calculation of the spatial mean and variance in condition \eqref{eq_br_gauss_noTC}  are weighted by the patch-specific factor $1/\sigma_{i}^2$  (see appendix \ref{app_special_cases_varying_selection}). Thus, patches in which selection is strong contribute relatively more to the total variance in selective optima and thus to branching.

\subsection{Influence of spatio-temporal fluctuations in selection} The first two terms on the left-hand side of condition \eqref{eq_branch_general} describe the influence of spatial and temporal fluctuations in selection on evolutionary branching. For Gaussian selection, these fluctuations can be expressed by the fluctuations in the selective optima. 
In the limit of very long-lived adults ($\gamma \to 1$) both terms simply combine to the total variance over time and space and the decomposition into two parts follows the law of total variance. For $\gamma<1$, temporal fluctuations in the spatial average selection only enter proportional to $\gamma$. Hence, such ``global'' temporal fluctuations only contribute to selection for genetic variation if generations overlap. This effect is known as ``storage effect of generation overlap'' \citep{Chesson:81}.
In contrast, expected spatial differences as described by the first term  promote evolutionary branching independent of other factors.

Both terms are also affected by spatial correlations. Note that, as we assume an island model, spatial correlations are an effect across all patches and do not have a particular spatial scale. Expected spatial variation in selection is reduced by positive spatial correlations and increased by negative correlations. 
The opposite is true for temporal variation in the spatial mean selection gradient, because temporal fluctuations in the patches contribute more to global temporal fluctuations under positive and less under negative spatial correlations.
For sufficiently small $\gamma$ the first term dominates the second term so that negative spatial correlations promote branching and positive correlations impede it. 
Specific examples for environmental fluctuations with spatial correlation are analyzed in appendix \ref{app_gauss_special_cases}. Temporal correlations, on the other hand, do not affect the two terms in question.

\subsection{Influence of spatial differentiation}  
The third term on the left-hand side of conditions \eqref{eq_branch_general} and \eqref{eq_br_gauss_noTC}, $2\frac{1-m}{m}{\rm Var_S}\left[{\rm E_T}[.]\right]$,  describes an additional contribution of spatial differences in expected selection that can be interpreted as selection for local adaptation. It is zero if environmental conditions are identically distributed across patches or if all offspring disperse ($m=1$)  and increases with decreasing migration probability. In fact, if ${\rm Var_S}\left[{\rm E_T}[.]\right]\neq0$, then for every $\sigma^2$ exists a critical value of $m$ such that for all migrations probabilies below this critical value condition \eqref{eq_br_gauss_noTC}  is fulfilled.   It is unaffected by temporal or spatial correlations.  Figure \ref{fig_TC}B (solid line) shows how between-patch differences promote evolutionary branching as dispersal decreases.

\subsection{Influence of temporal correlation} Temporal autocorrelations in the local selection gradients only affect the last term, ${\rm \mathcal{C}}[.]$, on the left-hand side of conditions \eqref{eq_branch_general} and \eqref{eq_br_gauss_noTC}. This term is zero in the absence of temporal correlations. In general, positive  temporal autocorrelations within patches promote branching and negative autocorrelations impede it. However, this is not true if these ``local'' autocorrelations are just the consequence of global temporal correlations. This can be seen from definition \eqref{eq_covt}, where the temporal covariance in the spatial mean selection gradient is subtracted from the average temporal covariance within patches. Hence, local but not global temporal autocorrelations promote evolutionary branching. The reason is that local autocorrelations can lead to extended periods over which individual patches consistently experience similar environmental conditions resulting in transient phases of differentiation among patches.

The covariance terms are weighted by the factor $(1-\gamma)(1-m)(1-(1-\gamma)m)^{\Delta t-1}$,
 where $\Delta t$ is the length of the time interval considered. This factor is always smaller than one and can be understood as follows. Similar to the effect of spatial differentiation, the diversifying effect of transient patch differences produced by positive autocorrelations requires that genotypes preferentially stay in the same patch. Hence, the effect of autocorrelations on invasion fitness is strongest for low dispersal and vanishes when all offspring disperse ($m=1$). Generation overlap, $\gamma$, has a two-fold effect on the influence of autocorrelations. Starting from small values, increasing  generation overlap first strengthens the effect of autocorrelations by increasing the probability that genetic material survives in the same patch over the considered time interval. However, as $\gamma$ approaches one, the influence of temporal correlations vanishes  because individuals are more likely to experience a representative sample of the distribution of environmental conditions.
Finally, the just-mentioned effects of dispersal and adult death affect the genotype composition at each time-step. Hence, temporal correlations between distant time points (large $\Delta t$) enter with reduced weight.

\subsection{Influence of patch number and relative patch size} The influence of the number of patches, $n$, can most easily be seen in the case without spatial and temporal correlations, where the branching condition can be simplified. For brevity we only give the condition for the Gaussian case here. The general result and its derivation is given in appendix \ref{app_uncorr_patches}. 
In the absence of correlations, the branching condition can be expressed as 
\begin{equation}\label{eq:br_indep_cov}
\frac{2-m}{m}{\rm Var_S[E_T[}\theta]]+\left(1-\frac{1-\gamma}{n}\right)\mathrm{E_S}\left[\mathrm{Var_T}[\theta]\right]-(1-\gamma)\mathrm{Cov_S}\left[c,\mathrm{Var_T}\left[\theta\right]\right]
>\sigma^2,
\end{equation}
where $\mathrm{Cov_S}[.,.]$ is the spatial covariance. Here, the first term corresponds to permanent patch differences and it draws contributions from both the first and the third term in equation \eqref{eq_br_gauss_noTC}. 
The second term in condition \eqref{eq:br_indep_cov} corresponds to the effect of within-patch fluctuations in environmental conditions and draws contributions from the first and the second term in equation \eqref{eq_br_gauss_noTC}. This term reveals that branching becomes easier with an increasing number of patches. This effect becomes weaker with increasing generation overlap. Conversely, the positive effect of generation overlap vanishes with increasing patch number.
The third term on the left-hand side of  condition \eqref{eq:br_indep_cov} is zero if patches are equally sized or if environmental flutuations are equally strong in all patches.
This term shows that a given amount of temporal fluctuation promotes branching more strongly if it is produced by relatively stronger fluctuation in the smaller patches. Conversely, it contributes less to branching if it is produced by relatively stronger fluctuations in few large patches.
We plot the left-hand side of inequality \eqref{eq:br_indep_cov} ($\sigma^2_{\text{ crit}}$) for the case $c_i=1/n$ as a function of the patch number and for different values of generation overlap  in figure \ref{fig_n_gamma}.

\setlength{\unitlength}{\textwidth}
\begin{figure}[!ht]
\begin{center}
\begin{tabular}{>{\centering\arraybackslash} m{.43\textwidth}>{\centering\arraybackslash} m{.58\textwidth}}
\begin{picture}(0.49,0.33)
\put(0,0){\includegraphics[width=0.45\textwidth]{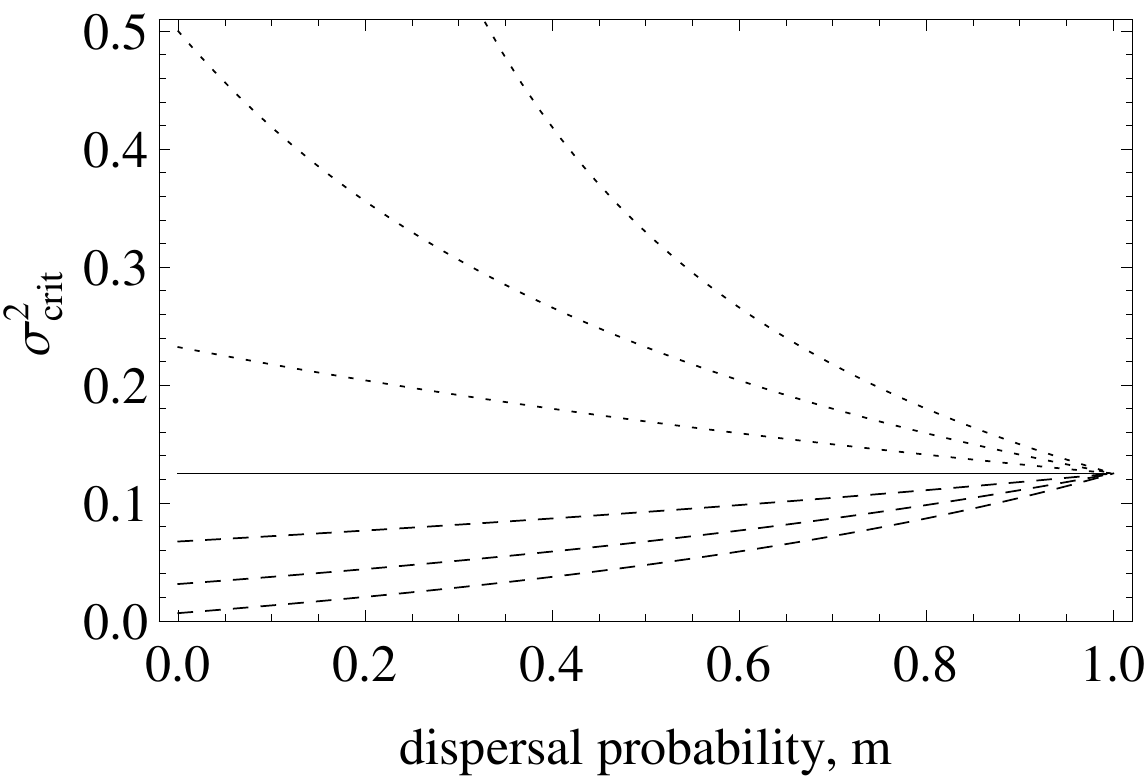}}
\put(0.03,0.32){\textbf A}
\end{picture}
&
\begin{picture}(0.49,0.33)
\put(0,0){\includegraphics[width=0.45\textwidth]{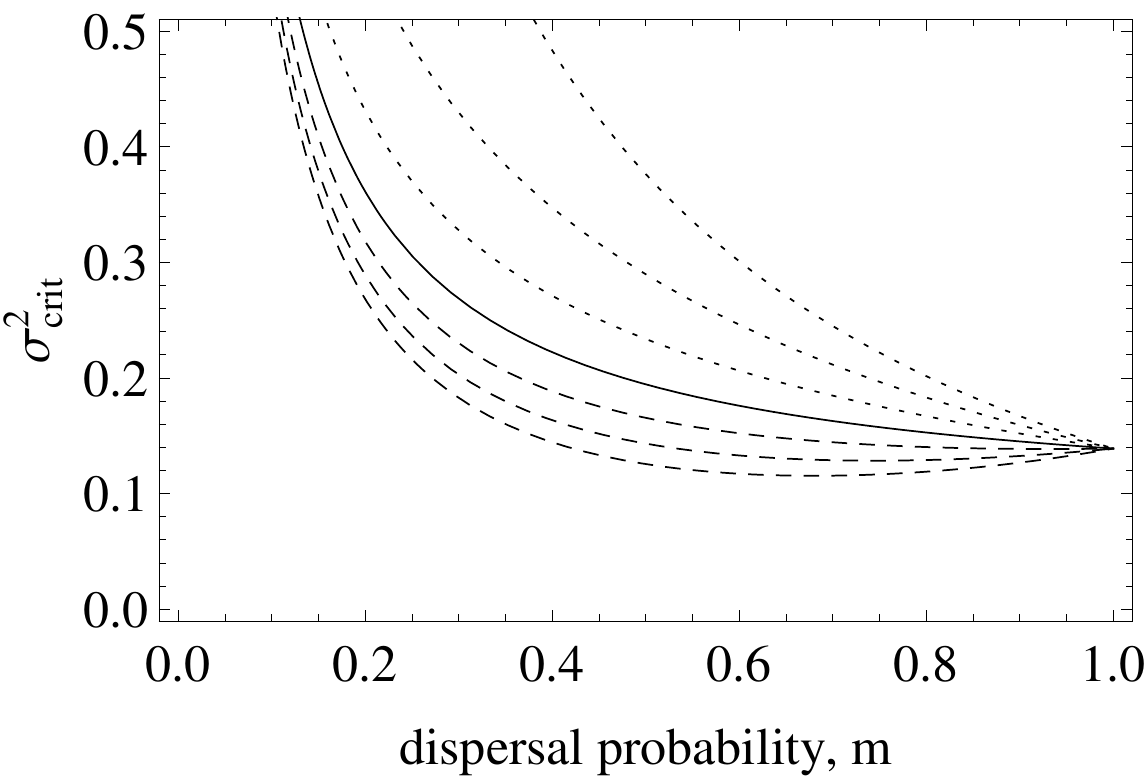}}
\put(0.03,0.32){\textbf B}
\end{picture}
\end{tabular}
\end{center}
\caption{Critical strength of selection below which selection is disruptive as a function of the dispersal probability. Large values of $\sigma_{\rm crit}^2$ indicate a large scope for genetic diversification. 
For these plots, we assume discrete generations ($\gamma=0$) and two patches with selective optima that fluctuate between two possible states. In both panels, solid lines give the case of no temporal correlation, dotted lines the case of positive temporal autocorrelation and dashed lines the case of negative autocorrelation. Autocorrelation increases with the distance of the dotted or dashed lines from the solid line. Details of the parametrization of autocorrelations are given in appendix \ref{app_gauss_special_cases}, where inequalities \eqref{eq_br_NF_TC} and \eqref{eq_br_F_TC} define the curves in panel A and B.
 In panel A, the two alternative selective optima each occur with probability of 1/2 in both patches. Thus, the patches are identically and independently distributed. In this case, the branching condition is independent of $m$ in the absence of temporal correlations. Positive correlations lead to a positive effect of decreased dispersal on the parameter range for branching, while negative correlations have the opposite effect.
In panel B, the patches are not identically distributed, but the first selective optimum occurs in the first patch with probability $1/3$ and in the second patch with probability $2/3$. The opposite is true for the second selective optimum. In this case, decreasing dispersal generally has a positive effect on diversity, unless temporal autocorrelations are strongly negative.
}
\label{fig_TC}
\end{figure}

\begin{figure}[h!]
\begin{center}
\includegraphics[width=0.45\textwidth]{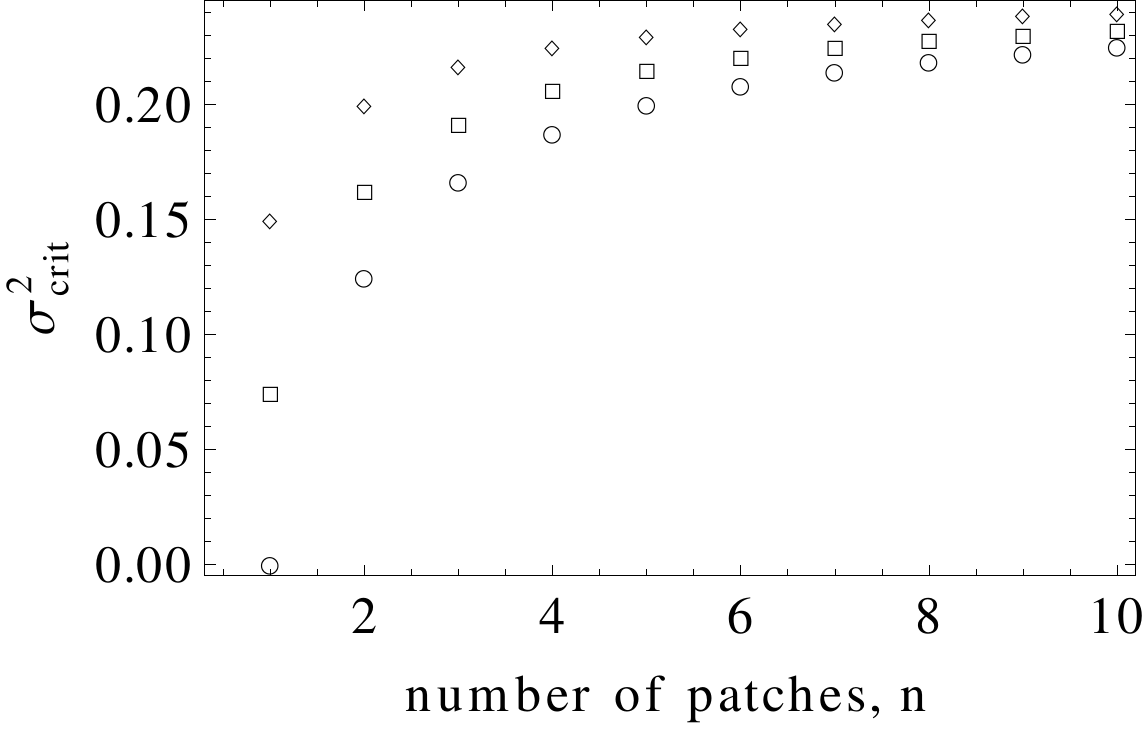}
\end{center}
\caption{Critical strength of selection below which selection is disruptive as a function of the number of patches as given by condition \eqref{eq:br_indep_cov}. Results shown for generation overlap $\gamma=0$ (circles), $\gamma=0.3$ (squares), $\gamma=0.6$ (diamonds), $m=1$, and $c_i=1/n$. 
Increasing the patch-number facilitates diversification (cf.\ second term on the left-hand side of condition \ref{eq:br_indep_cov}).}
\label{fig_n_gamma}
\end{figure}

\setlength{\unitlength}{\textwidth}
\begin{figure}[!ht]
\begin{center}
\begin{tabular}{>{\centering\arraybackslash} m{.43\textwidth}>{\centering\arraybackslash} m{.58\textwidth}}
\begin{picture}(0.49,0.3)
\put(0,0){\includegraphics[width=0.45\textwidth]{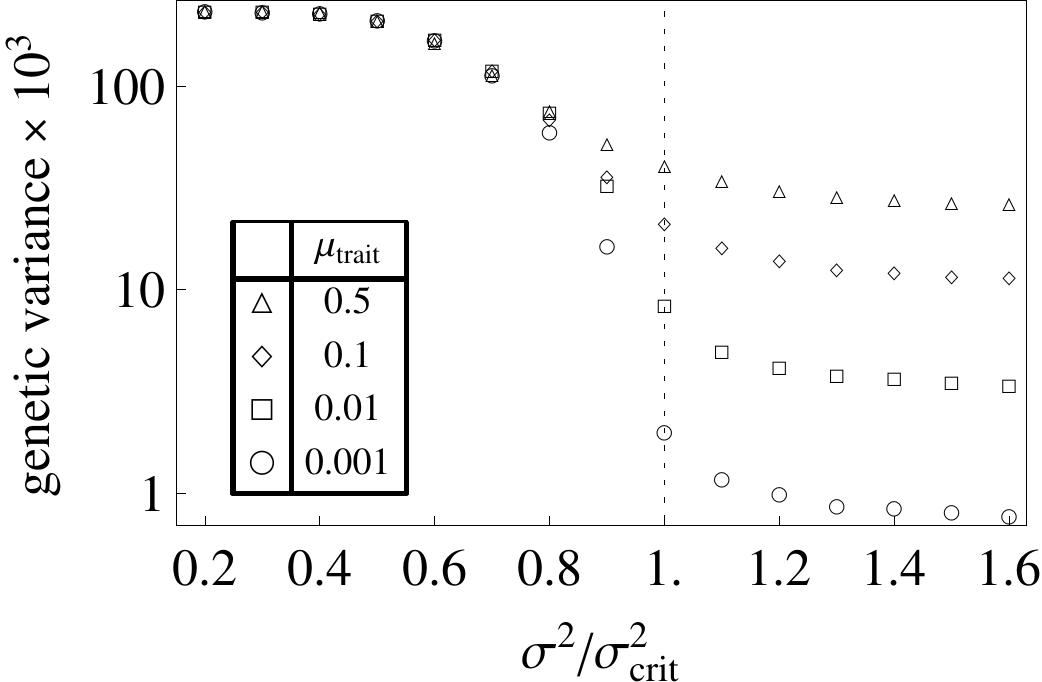}}
\put(0.03,0.3){\textbf A}
\end{picture}
&
\begin{picture}(0.49,0.3)
\put(0,0){\includegraphics[width=0.45\textwidth]{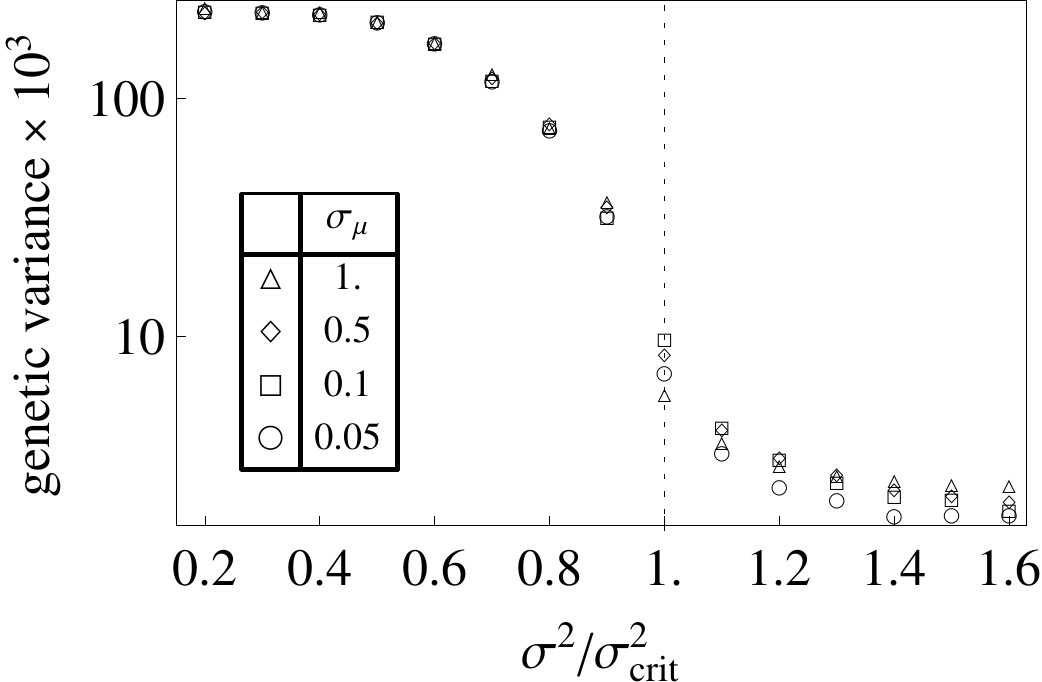}}
\put(0.03,0.3){\textbf B}
\end{picture}
\end{tabular}
\end{center}
\caption{
Long-term average genetic variance (on log-scale) measured from individual based simulations as a function of the strength of Gaussian stabilizing selection for  different (A) mutation rates  and  (B) expected mutational effect sizes. Selective optima fluctuate independently across five patches. 
Spatial differences in expected selection and spatial and temporal correlations are absent. 
The dotted vertical line indicates the analytically expected branching point  $\sigma^2/\sigma^2_{\rm{crit}}=1$ (cf. inequality \ref{eq_br_gauss_noTC}). 
In all cases, the analytical condition for evolutionary branching coincides with the observed incarease in genetic variance in individual based simulations. Note that in panel A the amount of genetic variation is determined by mutation selection balance when the branching condition is not fulfilled (right of the dashed line) but becomes independent of mutation rate when the branching condition is fulfilled (left of the dashed line). Parameters: $\gamma=0.5$, $m=1$;
other parameters as in table \ref{tbl_sim_params}.
}
\label{fig_mutation}
\end{figure}

\begin{figure}[!ht]
\begin{tabular}{m{0.5\textwidth}m{0.5\textwidth}}
\includegraphics[width=0.45\textwidth]{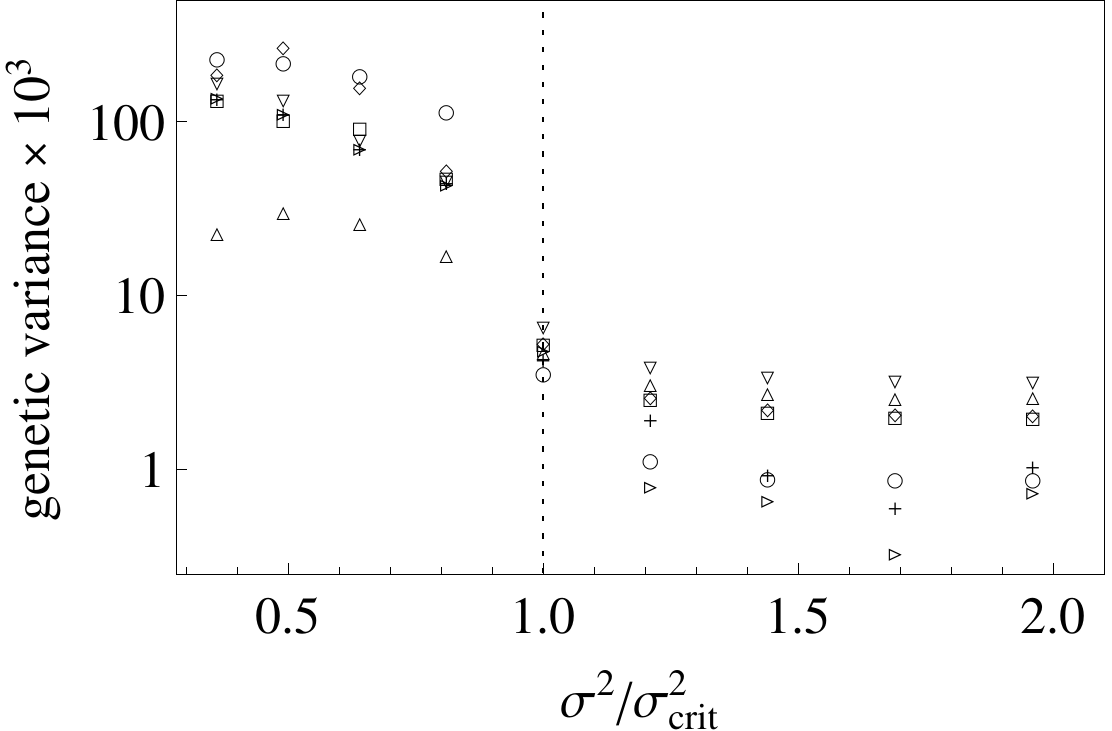}&\includegraphics[width=0.35\textwidth,trim=0cm 2cm 0cm 2cm,clip]{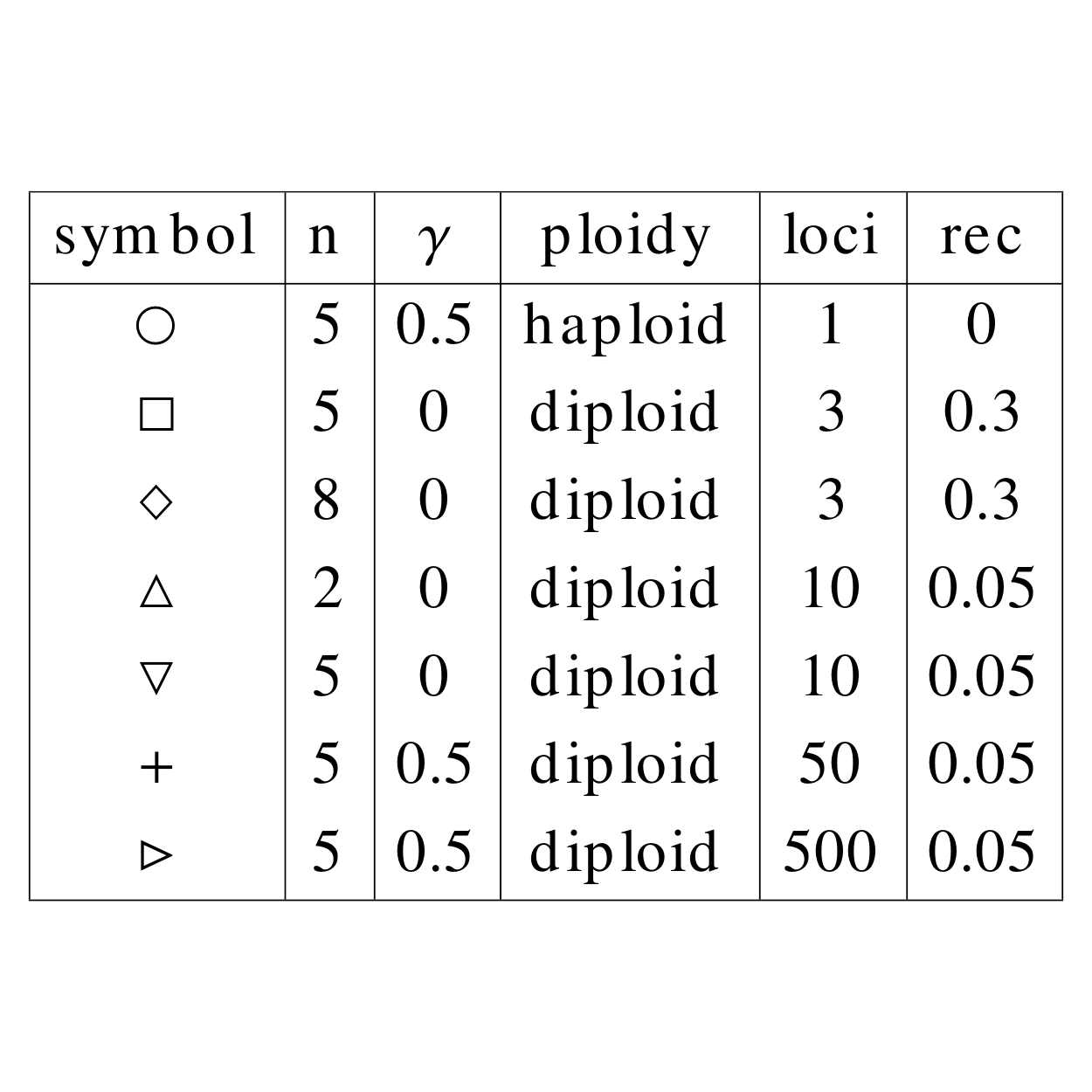}\vspace{0.06\textwidth}\\
\end{tabular}
\caption{Long-term average genetic variance (on log-scale) measured from individual based simulations as a function of the strength of Gaussian stabilizing selection 
 for the case of full dispersal ($m=1$) for various ecological parameters and genetic architectures (see table). The dotted vertical line indicates the analytically expected branching point  $\sigma^2/\sigma^2_{\rm{crit}}=1$ (cf. inequality \ref{eq_br_gauss_noTC}). Genetic variance increases substantially if the branching condition is fulfilled. This holds true for all examined combinations of ecological and genetic scenarios as detailed in the table. Selective optima in the patches are identically and independently distributed. Parameters: $\mu_{\rm trait}=0.01$; other parameters as given to the right of the plot or in table \ref{tbl_sim_params}.}
\label{fig_varana}
\end{figure}

\subsection{Robustness of the results}  \label{sec_robustness}

The analytical results are derived under the assumption of rare mutations of small effect and a very large population size.
Then, if conditions \eqref{eq_cs_general} and  \eqref{eq_branch_general} are fulfilled, there exists a trait value $x^*$  that is an evolutionary attractor at which disruptive selection favors an increase in genetic variance and hence the evolution of adaptive genetic polymorphism.
In order to study the robustness of this result with respect to violations of these assumptions, we perform extensive individual-based simulations for the case of Gaussian stabilizing selection (see  online supplementary \ref{app_sim} for details on the simulation study). 
These simulations show that populations indeed evolve towards the trait value $x^*$ under a wide variety of genetic assumptions and population sizes. 
Furthermore, also our main result -- the condition for adaptive genetic polymorphism  -- is not sensitive to the violation of the adaptive dynamics assumptions. In particular, the branching condition (inequality \ref{eq_branch_general}) accurately predicts an increase in genetic variance due to disruptive selection for all tested mutation rates and mutational effect sizes (figure \ref{fig_mutation}, but see below for the effect of low mutation rates in small populations). 
 Also  sexual reproduction and diploid genetics with several recombining additive loci do not change this picture (figure \ref{fig_varana}). While we assume full dispersal ($m=1$) in this figure, results also hold for other values of $m$ (supplementary figure \ref{fig_restricted_Fixed}).

The only factor that can introduce substantial deviations from our branching condition is population size. For small population sizes genetic variance only starts to increase for stronger selection (smaller $\sigma^2$) than predicted by our condition \eqref{eq_branch_general}. The critical population size below which our results become inaccurate depends on the mutation parameters. In figure \ref{fig_popsize} we used a trait mutation rate of $0.01$ (per locus mutation rate of $\approx 10^{-4}$) and an expected mutational effect size of 0.05. For these values we see deviations from our branching condition only for total population sizes of 100 individuals or less (figure \ref{fig_popsize}A). For smaller mutation rates and effect sizes, deviations appear earlier (i.e., already for larger population sizes). This leads to the counter-intuitive conclusion that the branching condition becomes increasingly robust for small population size as the adaptive dynamics assumptions of rare mutations and small mutational effects are violated.  The reason for this behavior is that in small populations random fluctuations in selection can easily lead to a loss of genetic variation, which is necessary for disruptive selection to act upon. Importantly, this effect is almost entirely due to random fluctuations in the expected size of the mutant sub-population produced by environmental stochasticity and not due to classic genetic drift  (i.e., variance among replicates for a fixed sequence of environments). Indeed, the effect remains unchanged if the random sampling step in recruitment is removed and the number of offspring of each individual is equal its expectation (rounded to integer). Figures \ref{fig_popsize_mP0p001} and \ref{fig_popsize_appendix} show the accuracy of our results for other mutational and ecological parameters.
 In figure \ref{fig_popsize}B we use a constant total population size and test the influence of local population sizes by varying the number of patches. We only get substantial deviations from the analytical prediction if the patches are very small (five or ten individuals per patch). Hence, while a large number of patches promotes evolutionary branching (cf. inequality \ref{eq:br_indep_cov}), extremely small patch sizes are detrimental to adaptive genetic polymorphism.

\setlength{\unitlength}{\textwidth}
\begin{figure}[!ht]
\begin{center}
\begin{tabular}{>{\centering\arraybackslash} m{.43\textwidth}>{\centering\arraybackslash} m{.58\textwidth}}
\begin{picture}(0.49,0.3)
\put(0,0){\includegraphics[width=0.45\textwidth]{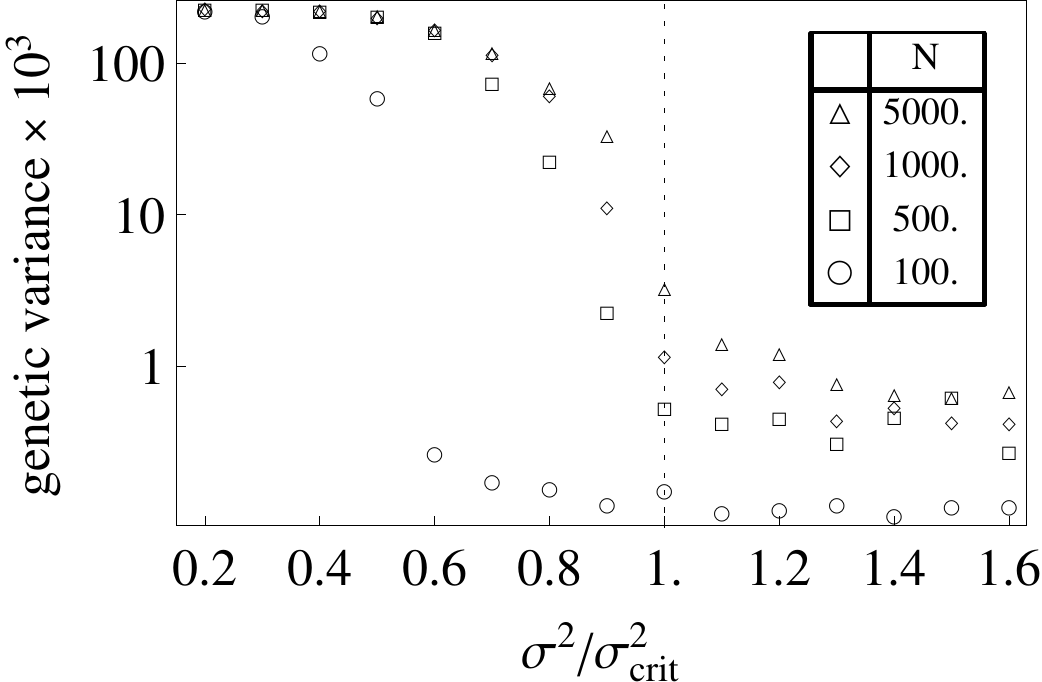}}
\put(0.03,0.3){\textbf A}
\end{picture}
&
\begin{picture}(0.49,0.3)
\put(0,0){\includegraphics[width=0.45\textwidth]{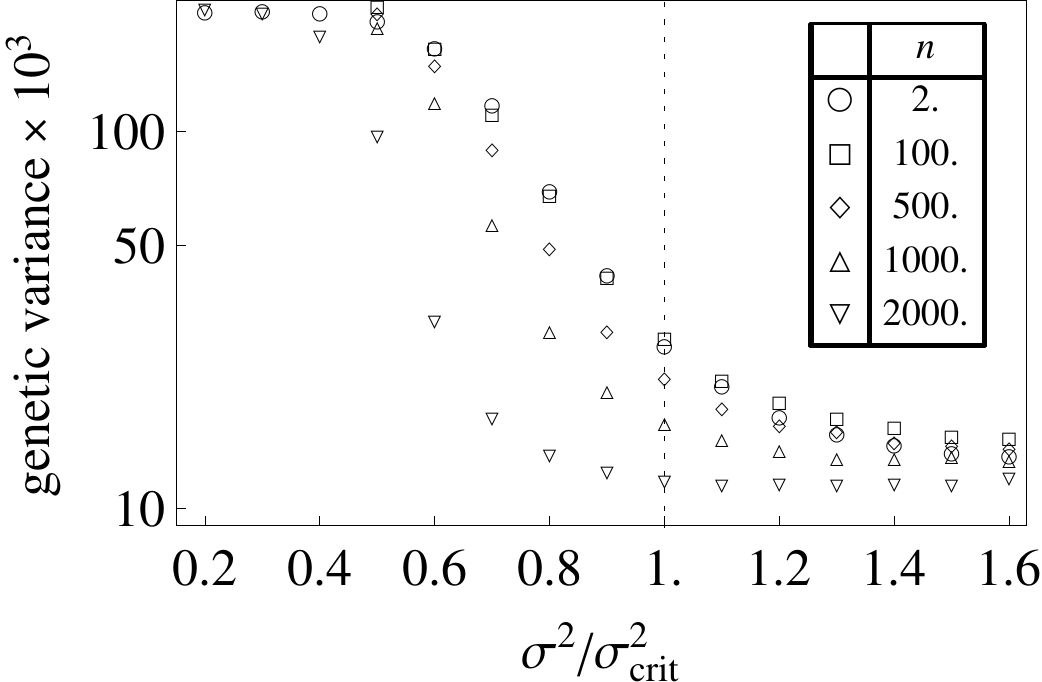}}
\put(0.03,0.3){\textbf B}
\end{picture}
\end{tabular}
\end{center}
\caption{ 
Long-term average genetic variance in a population (on log-scale) from individual-based simulations as a function of the strength of Gaussian stabilizing selection for identically and independently  distributed selective optima across patches  (cf. figure \ref{fig_restricted_Fixed}A). In panel A the total population size ($N$) is varied for a constant number of $n=5$ patches. In populations of 100 individuals or less, the increase in genetic variance only happens for smaller values of $\sigma^2$ (stronger selection) than predicted by the branching condition.
In panel (B) the number of patches (and hence the local population size) is varied for a constant total population size of $N=10000$. It shows that only for very small patch size ($n>1000$, less than 10 individuals per patch) the increase in genetic variance is substantially delayed as compared to the analytical predictions. Parameter values: $\gamma=0.5$, $m=1$, $\sigma_{\mu}=0.05$; other parameters as in table \ref{tbl_sim_params}.}
\label{fig_popsize}
\end{figure}

\subsection{Genetic structure of the polymorphism}

In the previous section we have shown that, in reasonably large populations, increased levels of genetic variation evolve 
if the branching condition is fulfilled.
Here we examine the structure of the resulting genetic polymorphism.

In sufficiently large populations of clonally reproducing organisms and in the absence of environmental stochasticity, populations split at a branching point into two discrete sub-populations that evolve away from each other in phenotype space while negative frequency dependence protects them from extinction. 
If there are more than two possible selective optima and if these are sufficiently distant from each other relative to the strength of selection, then the sub-populations can split again until a polymorphic equilibrium is reached at which each type experiences stabilizing selection. Such a scenario is  described in more detail in \citet[figure 6]{Svardal:11}.  

However, in populations of biologically realistic size the genetic polymorphism can look very different, mainly depending on two factors: (1) the strength of demographic fluctuations in the sub-populations characterized by different trait values, due to environmental stochasticity and drift, and (2) the genetic architecture of the trait. 

Environmental stochastiticy can be understood as the deviation of the growth rate of a mutant sub-population over a finite time interval from the invasion fitness (equation \ref{eq_inf_fit_lambda}). Such excursions of the short-term growth rate from its long-term average are caused by the random fluctuation in the environmental conditions. In our model, the strength of environmental stochasticity is increased  by (i) a large temporal variance in environmental conditions within each patch, (ii) a small number of patches, (iii) positive spatial and temporal correlations, (iv) a small spatial variance in expected environmental conditions, (v) small generation overlap \citep[for the effect of generation overlap, see][figure 7]{Svardal:11}, and (vi) strong stabilizing selection (small $\sigma^2$).

\begin{figure}[!ht]
\begin{center}
\includegraphics[width=0.8\textwidth]{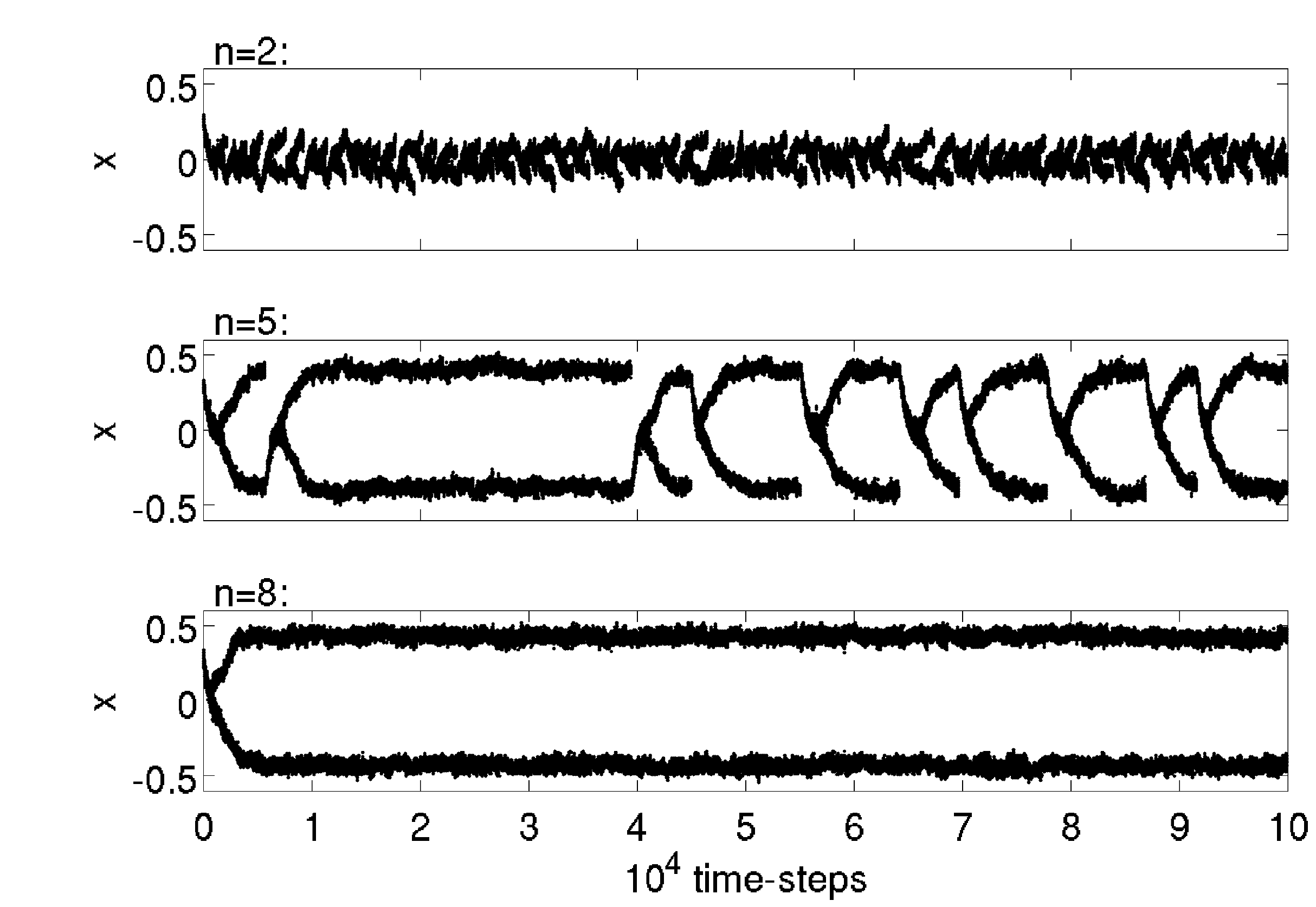}
\end{center}
\caption{Population trait values over time from individual-based simulations for three different patch numbers at a constant total population size of $N=4000$. Selective optima are identically and independently  distributed across patches and we assume a single haploid locus. Note that the two alternative selective optima are at $-0.5$ and $0.5$ instead of $0$ and $1$.  For a small number of patches the genetic dimorphism is frequently lost, and with increasing $n$ the dimorphism becomes increasingly stable. Parameter values: $\mu_{\rm trait}=0.001$, $\gamma=0$, $m=1$, $\sigma^2=0.09$; other parameters as given in table \ref{tbl_sim_params}.}
\label{fig_genhist_haploid_n}
\end{figure}

Figure \ref{fig_genhist_haploid_n} shows the effect of environmental stochasticity  by varying the number of patches. For eight patches (figure \ref{fig_genhist_haploid_n}, bottom panel),  environmental stochasticity is sufficiently weak  so that we observe the above-mentioned stable polymorphism. For the case of five patches (figure \ref{fig_genhist_haploid_n}, middle panel), environmental stochasticity is stronger and
a series of environmental states  unfavorable to one sub-population can lead to its extinction. The remaining sub-population then evolves back to the branching point where it subsequently splits again to restore the genetic polymorphism. With only two patches (figure \ref{fig_genhist_haploid_n}, top panel), environmental stochasticity is so strong that extinction of sub-populations happens very frequently and no clear split of the population into discrete branches is visible (figure \ref{fig_genhist_haploid_n}, top panel). However, even in such a regime there is a detectable increase in genetic variance when the branching condition is fulfilled (except for the cases of very small population size mentioned above). Note that the precise number of patches, for which each of these regimes can be observed, depends on the other factors discussed above.
 
Interestingly, population size -- while being important for the general accuracy of our branching condition -- only plays a minor role for the stability of the genetic polymorphism. Its influence on the stability of polymorphism seems to be logarithmic. 
Finally, while strong selection increases the effect of environmental stochasticity and therefore favors the extinction of subpopulations it is necessary that selection is sufficiently strong for polymorphism to emerge in the first place. This two-fold effect can be seen in figures \ref{fig_varana}  (tip-up triangles)  and \ref{fig_restricted_Fixed}, where for some parameter combinations genetic variance decreases for very small $\sigma^2$.

\begin{figure}[!ht]
\begin{center}
\includegraphics[width=0.6\textwidth]{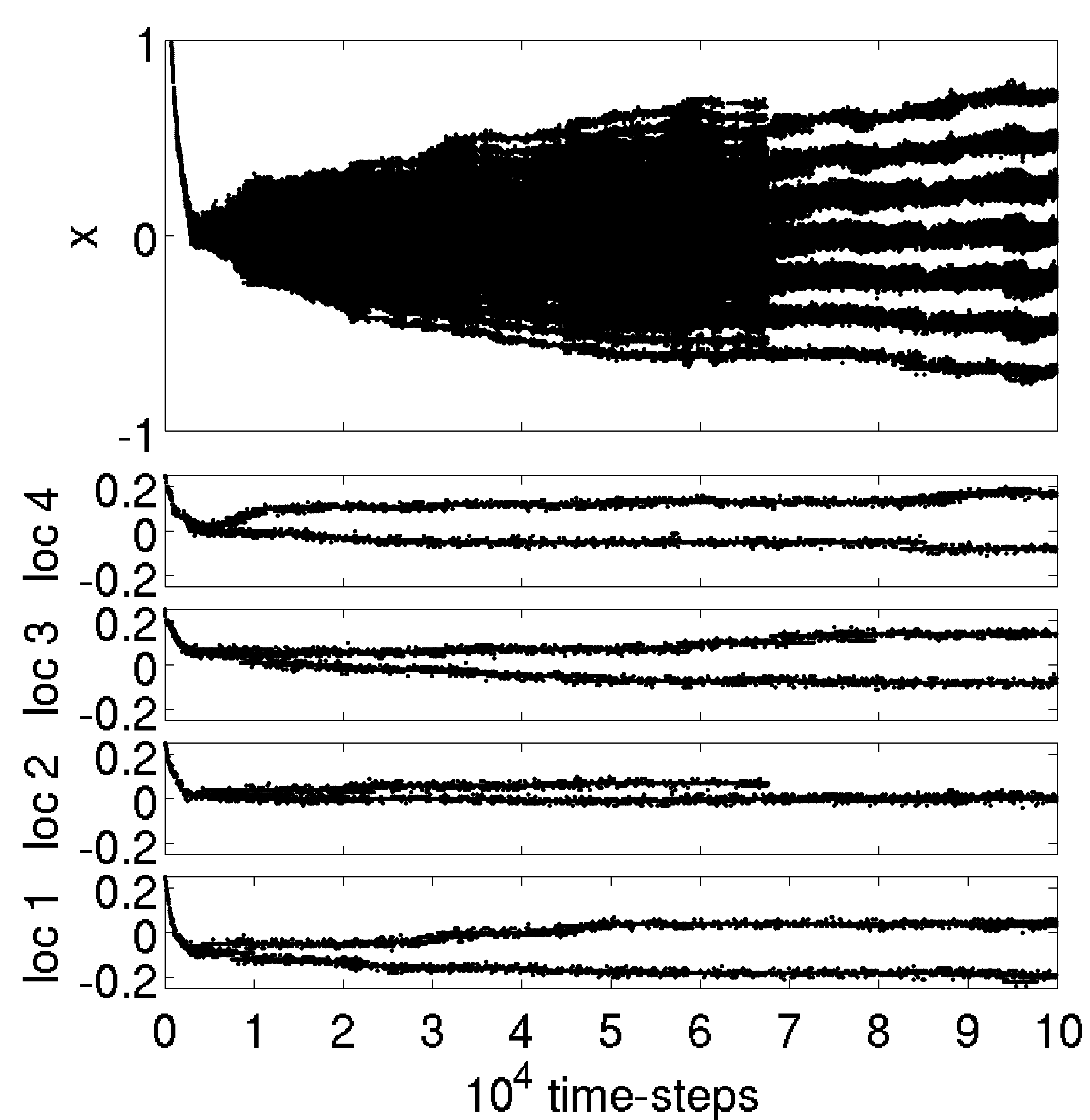}
\end{center}
\caption{Population trait values (top panel) and allelic values at four diploid additive loci (lower panels) over time  from individual based simulations. The two alternative selective optima are at $\pm 1$. Branching initially happens at all four loci. At the phenotypic level this results in a cloud of different phenotypes. After the polymorphism at locus 2 goes extinct around time-step 70000, discrete phenotypic clusters become visible. Parameter values: $N=4000$, $n=4$, $m=1$, $\gamma=0.1$,  $rec=0.3$, $\mu_{\rm trait}=4\times10^{-4}$; other parameters as in table \ref{tbl_sim_params}.}
\label{fig_genhist_rec}
\end{figure}

The second factor that influences the structure of the genetic polymorphism is the genetic basis of the trait. While clonal populations can split into discrete phenotypic clusters that each evolve towards a peak in the fitness landscape, this is not possible for sexual populations because mating between individuals from two different clusters results in intermediate phenotypes \citep[][]{Dieckmann:99,Kisdi:99}. 
In figure \ref{fig_varana} we show by means of individual-based simulations that also sexually reproducing diploid populations evolve increased levels of genetic variance for a wide range of ecological and genetic assumptions. This is in particular true for various multi-locus cases. The structure of the genetic polymorphism depends on the ploidy, the number of loci, and the recombination rate. The patterns observed range from one intermediate  phenotype (heterozygote) between two diverging branches for a single diploid locus, to a cloud of phenotypes for many recombining loci. 
Figure \ref{fig_genhist_rec} shows an example of a simulation run for a diploid population in which the trait is determined by four additive and strongly recombining loci. Disruptive selection initially leads to evolutionary branching at all loci resulting in a  cloud of phenotypes. Only after the dimorphism at one locus is lost, discrete phenotypic clusters become visible. The long-term expectation is that the genetic polymorphism becomes concentrated at a single locus, while all other loci become monomorphic \citep{Kopp:06,VanDoorn:06,Yeaman:11}, 
because this minimizes the amount of maladapted intermediate phenotypes.

Another effect of sexual reproduction and a multi-locus genetic architecture is that it
counteracts the extinction of alleles in the polymorphism due to environmental stochasticity because extreme types that have disappeared during adverse conditions can be restored from intermediate types by recombination.

\section{Discussion}

When does temporally and spatially fluctuating selection favor the evolution and maintenance of genetic diversity in a population? We investigate this question for a generalized island model \citep{Wright:43} that we combine with the simple age structure of the lottery model \citep{Chesson:81} so that generations can overlap. Our model is characterized by the following features.
The population decomposes into patches of locally competing individuals from which a certain fraction of juveniles disperses each generation over the whole population. The population is thus subdivided, but there is no isolation by distance. Environmental conditions can fluctuate over time, either independently in different patches or with arbitrary spatial and temporal correlations. The environment exerts selection on a quantitative trait. Selection acts only on a single, short-lived life stage (e.g. juveniles), 
 while a long lived stage survives from one reproductive season to the next with a constant probability, independent of trait and locality. The total reproductive output of each patch is constant (soft selection).

Our ecological model includes several standard models of ecology and population genetics as special cases (see table \ref{tb_special_cases}). 
In contrast, our genetic assumptions differ from most previous models listed in table \ref{tb_special_cases}. While the latter usually focus on two alleles or two species, we consider a quantitative trait and we ask under which conditions selection on the trait turns disruptive, so that genetic polymorphism is selected for. 
Technically, we do this by using the adaptive dynamics approximation \citep{Metz:96a,Geritz:98} to identify conditions for the existence of evolutionary branching points. These are trait values that are on the one hand attractors of the evolutionary dynamics, but at which, once the majority of the population is sufficiently close to them, selection turns disruptive.
Evolutionary branching points are the signature of negative frequency-dependent selection due to intraspecific competition. We obtain a very general condition for evolutionary branching and thus for the adaptive evolution and maintenance of genetic diversity in temporally and spatially heterogeneous environments (equations \ref{eq_sp_general}-\ref{eq_br_gauss_noTC}). This is our main result.

\subsection{Factors promoting adaptive diversification}
We can distinguish four main factors that promote adaptive genetic polymorphism, which each capture an aspect of the temporal and spatial heterogeneity. A detailed quantitative analysis is provided in the results section and the supplements. A more intuitive interpretation of the results can be given in terms of ecological niches.  

First, ecological niches are created by spatial differences in selection. On the one hand, such differences can arise from spatial differences in the expected environmental conditions ("permanent niches"). On the other hand, temporal fluctuations within patches can lead to spatial differences in selection even in the absence of permanent differences ("fluctuating niches").
Permanent niches enter our condition for genetic diversification via two terms. The first contribution (first term in inequality  \ref{eq_br_gauss_noTC})  is independent of dispersal. Note, that this is the only factor creating ecological niches in the classical Levene model with no fluctuations within patches and full migration \citep{Levene:53}; our results reproduce previous findings for the Levene model by \citet{Geritz:98} (appendix \ref{app_gauss_levene_model}). 
The second contribution to permanent niches  (third term in inequality  \ref{eq_br_gauss_noTC}),  only exists under restricted dispersal ($m<1$) and becomes increasingly important as dispersal decreases by allowing for local adaptation \citep{Deakin:66,Deakin:68,Gillespie:75,Snyder:03}. 
In spatially heterogeneous environments, rare types can accumulate in patches with favorable environmental conditions leading to a positive fitness-density covariance for rare types and thus favoring coexistence (cf. \citealp{Chesson:00b}).
 In the absence of spatial correlations, both contributions to permanent niches can be combined into a single term (first term in condition \ref{eq:br_indep_cov}).
Fluctuating niches are also captured by the first term in condition \eqref{eq_br_gauss_noTC} and, hence, their importance is  independent of the strength of dispersal as long as it is positive. Generally, fluctuating niches are less efficient in promoting adaptive differentiation than permanent niches, but their contribution increases with an increasing number of patches and with increasing negative spatial correlations across patches, while positive correlations have the opposite effect.
The effect of independent temporal fluctuations on genetic polymorphism was already observed in population genetic models by Gillespie (\citeyear{Gillespie:74,Gillespie:75,Gillespie:76b}; \citealp{Gillespie:76}) and ecological models by \citet{Chesson:85} and \citet{Comins:85}. 
\citet{Chesson:85} found local temporal fluctuations as effective in promoting polymorphism as permanent spatial niches. However, from our results it is clear that Chesson's finding only holds under full dispersal and for sufficiently many patches (see equation \ref{eq:br_indep_cov}).

Second, if generations overlap, temporal fluctuations can create ecological niches even in the absence of spatial variation (second term in equation \ref{eq_br_gauss_noTC}). This phenomenon, known as ``storage effect of generation overlap'' \citep{Chesson:84} or ``time-dispersal'' \citep{Comins:85}, can be understood as the appearance of ecological niches in the same patch over the life time of an individual (which increases with increasing generation overlap). The storage-effect was first described in the lottery model \citep{Chesson:81}. We find the branching condition for this model as derived by \citet[equation 9]{Ellner:94} to be a special case of ours (appendix \ref{app_gauss_lottery}).

Third, under limited dispersal, positive temporal correlations in environmental conditions
stabilize ecological niches and therefore increase the scope for genetic polymorphism while negative temporal correlations reduce it (fourth term in equation \ref{eq_br_gauss_noTC}). Intuitively, if dispersal is restricted, and if there is a higher than average chance that the local environment stays similar (positive temporal correlation), rare types can accumulate in patches where they have higher growth rates.  In contrast, if individuals stay in their patch of origin but the local environment is likely to change, specialized phenotypes are selected against. To our knowledge, this effect has not been described in the literature on the maintenance of polymorphism. However, it is known that positive temporal autocorrelations in the growth rates promote species persistence in a metapopulation \citep[e.g.][]{Schreiber:10}. This is a closely related phenomenon since adaptive diversification depends on the persistence of rare mutant invaders.

Fourth, ecological niches created by the above-mentioned factors only promote genetic diversity if ``specialist'' genotypes that evolve adaptation to these niches have an advantage over ``generalist'' genotypes that equally exploit all niches. In other words, for genetic diversification to be favored, there have to be sufficiently strong trade-offs for the performance in different niches. In our model, stronger trade-offs are reflected by increasing average within-patch stabilizing selection. Hence, increasing the strength of stabilizing local selection promotes population-level disruptive selection for genetic polymorphism.

\subsection{Generality of the results}Our analytical conditions are derived under the assumption of large populations and rare mutations of small effect. However, we show with individual-based simulations that the branching condition reliably predicts the emergence and maintenance of genetic variance
for a wide range of genetic and ecological details (figures \ref{fig_mutation}, \ref{fig_varana} and \ref{fig_restricted_Fixed}). Sexual reproduction, if anything, further facilitates the maintenance of increased levels of genetic diversity due to its capacity to restore lost phenotypes by recombination.

We only see substantial deviations from our analytical result in small populations with low mutation rates  (i.e., low new mutational input). The reason is that in such a situation demographic fluctuations quickly remove the genetic variation introduced by mutation. Hence, disruptive selection looses its target. 
Such an effect has been described for the action of demographic stochasticity  \citep[``sampling drift'', cf.][]{Ajar:03,Claessen:07,Wakano:13}. However, in our case it is rather caused by fluctuations in the expected number of mutants due to environmental stochasticity \citep[for a similar result in a different model, see][]{Johansson:06}. This can be seen, because the effect is unchanged if random sampling is removed during density regulation, whereas it vanishes in the absence of temporal fluctuations.
For biologically reasonable mutation rates, such deviations only become important for total population sizes of below 100-1000 individuals (figures \ref{fig_popsize} and \ref{fig_popsize_mP0p001}).  
Furthermore, very small local population sizes (less than 10 individuals per patch) lead to similar deviations from our branching condition.  Stable coexistence  in a protected polymorphism is based on the selective advantage of rare genotypes.  However, if patches are very small, already a small number of copies of a genotype corresponds to an appreciable frequency and hence the potential for negative frequency-dependent selection is weakened. In the extreme case of only two individuals per patch, there is no negative frequency dependence and  evolutionary branching is impossible (see online figure \ref{fig_popsize_mP0p001}). The effect of small patch sizes has previously been described in competition models \citep{Day:01, Ajar:03}.

Individual-based simulations further reveal that the genetic structure of the evolving polymorphism depends on two factors.
First, strong environmental stochasticity can lead to the extinction of genotypes in a polymorphism  and therefore to a repeated pattern of branching and extinction events \citep[cf.][]{ Claessen:07, Johansson:10}. 
 Second, the polymorphism pattern depends on the genetic architecture  of the trait. Under clonal reproduction we readily obtain discrete phenotype clusters. For sexual reproduction, in contrast, diverging alleles at different loci recombine and a connected cloud  
of phenotypes can emerge. In the long run, however, several mechanisms can evolve in sexual populations under disruptive selection  that prevent the production of intermediate phenotypes \citep{Rueffler:06b}. These include a single major effect locus \citep{Kopp:06,VanDoorn:06}, dominance modifiers \citep{Dooren:99,Peischl:10}, and assortative mating \citep{Dieckmann:99,Pennings:08}.
Finally, we note that, while our results are formulated in terms of genetic diversity within a species, they 
can also inform us about the outcome of an immigration event where an already reproductively isolated species is added to a species that has evolved sufficiently close to a singular point. If the singular point is a branching point and if the immigrating species is sufficiently similar to the resident species, then the two species are able to coexist and will subsequently undergo character displacement.

Previous studies on polymorphism in heterogeneous environments can broadly be classified in two categories: Studies following the ``classic'' approach of deriving conditions for the stable coexistence of a given set of alleles or species in a protected polymorphism and studies that investigate evolutionary branching in a quantiative trait.
As discussed above, our findings are in qualitative agreement with classic results on coexistence in variable envrionments \citep[e.g.][]{Levene:53,Dempster:55,Deakin:66,Deakin:68,Christiansen:74,Gillespie:73,Gillespie:74,Gillespie:75,Karlin:82, Chesson:81,Chesson:85,Comins:85}, but some differences arise due to the difference in genetic assumptions.
In particular, our condition can be both more stringent or less restrictive than the condition for stable coexistence of two fixed alleles. On the one hand, our condition is more restrictive since it requires long-term evolutionary stability in the sense that the polymorphism cannot be replaced by a generalist with intermediate phenotype \citep[cf.][]{Kisdi:99,Spichtig:04}.
Indeed, as soon as any temporal or spatial heterogeneity in selection exists (left-hand side of inequality \ref{eq_branch_general} non-zero) coexistence of suitable pairs of alleles is possible (appendix \ref{app_prot_poly}; \citealp{Kisdi:99}). However, such polymorphism will only persist on an evolutionary time scale if intermediate phenotypes are inferior (conditions \ref{eq_cs_general} and \ref{eq_branch_general} fulfilled).
On the other hand, our branching condition can be less restrictive, because it measures disruptive selection on the quantitative trait only locally, i.e., for phenotypically similar genotypes. This local condition does not necessarily imply coexistence of alternative phenotypes that are far away in trait space.
This leads to the main discrepancy between our results and earlier studies: If the patches are identically distributed, our local condition for genetic polymorphism is independent of the dispersal probability. In contrast, \citet{Gillespie:75} and \citet{Comins:85} found for the same scenario that the condition for protected polymorphism of discrete alleles becomes more restrictive with decreasing dispersal. Numerical calculations (appendix \ref{app_num_calc}) and individual based simulations (figure \ref{fig_restricted_Fixed}B) show how these results can be reconciled: The local condition for the existence of disruptive selection is independent of the dispersal probability, but the maximal phenotypic distance between genotypes that can coexist in a polymorphism decreases with decreasing dispersal. Hence, while the onset of selection for genetic polymorphism is independent of the dispersal probability, the amount of genetic variance that can evolve after branching decreases with reduced dispersal (figure \ref{fig_restricted_Fixed}B).

Previous models investigating evolutionary branching in subdivided populations either assume two patches connected by migration \citep[e.g.][]{Meszena:97, Day:00, Kisdi:02} or many patches connected through a common dispersal pool \citep[e.g.][]{Parvinen:04,Nurmi:08}. None of these models considers the combination of spatially and temporally heterogeneous environments and overlapping generations. Although these models also assume within-patch density regulation, the population size is not fixed but can change as a result of the evolutionary dynamics. The prize for relaxing the assumption of fixed within-patch densities is that the condition for branching can generally not be derived analytically. Nevertheless, qualitatively these models are in agreement with our branching condition in the sense that limited migration and pronounced patch differences favor the existence of evolutionary branching points. However, while in our model under Gaussian selection the generalist strategy is always an attractor of the evolutionary dynamics, 
in the references cited above 
it becomes evolutionarily repelling for very large patch differences in combination with low migration rate.  A cursory analysis shows that this qualitative difference is indeed due to the trait dependence of local population densities.

 The life cycle in our model can be found in a wide range of organisms. In this article, we equated the long-lived life stage with adult individuals. Obvious examples are perennial plants, corals, some fungi or vertebrates. Alternatively, the long-lived stage can be a resting stage \citep{Chesson:84,Ellner:94} such as a seed bank in plants, resting eggs in crustaceans, fungal spores or bacterial endospores. In such cases, the parameter $1-\gamma$ gives the rate of recruitment from the resting stage into the reproducing population.
Furthermore, while we assumed that selection acts upon juvenile viability (before density regulation), our results are equally valid if selection acts upon an adult trait contributing to fertility. We briefly discuss the case of selection on recruitment probability in appendix \ref{app_sel_recruitment}. 
An important aspect of our result for empirical applications is that the condition for adaptive diversification does not depend on any details in the distribution of environmental conditions beyond its mean values and (co-)variances. Given a sufficiently long time series of measured data, these quantities can be estimated in natural systems.

\subsection{Limitations and Extensions} The model presented in this article has several limitations. First, we assume that dispersal is global, that is, that there is no isolation by distance. \citet{Comins:85} studied the case that dispersal is limited to six neighboring patches and found that local dispersal has little effect on the overall conclusions. If patches are equally sized and their environmental conditions are identically distributed, then our results equally apply to a stepping stone dispersal model (results not shown). Furthermore,  if selection only varies spatially, it has been shown by \citet{Debarre:10} that isolation by distance has no qualitative effect on branching compared to an island model. 
Second, we assume that offspring are regulated locally within the patches (soft selection) to a constant density  before dispersal. This assumption is most obviously fulfilled in systems with competition for space. However, in any biological system there will be traits that do not (or only slightly) influence the carrying capacity and these traits can still be under selection. Models with local density regulation but trait dependent equilibrium population densities have been discussed in the previous paragraph. In the case of global density regulation, genetic polymorphism is generally not adaptively maintained \citep{Dempster:55}. In nature, density regulation is often due to a mixture of local and global factors. \citet{Debarre:11} showed that, as long as there is some local density regulation component, there is scope for the evolution of genetic polymorphism, but the condition will be more stringent.
Third, we assume that the fraction of juvenile immigrants into a patch is proportional to the juvenile carrying capacity of that patch  (conservative migration). This assumption is necessary for analytical tractability and has the consequence that our results are independent of the adult carrying capacities. Hence, fluctuations in adult population sizes do not change our conclusions. 
 Fourth, in our model there is no frequency-dependent competition within patches. Frequency dependence comes from the fluctuations in selection over time and across patches  in combination with soft selection. Resource competition with frequency dependence within patches has been analyzed in the context of spatially structured populations by \citet{Day:00,Day:01}, \citet{Ajar:03} and \citet{Nilsson:10a}, and for the future it would be interesting to combine both effects in a single study. 
Fifth, we assume the dispersal probability, $m$, to be a parameter rather than an evolvable trait. Several models that treat the joint evolution of habitat specialization and dispersal propensity  suggest that polymorphic coexistence is further facilitated by the evolution of different dispersal strategies \citep{Kisdi:02, Nurmi:11} and even more so under the evolution of habitat choice \citep{Ravigne:09}.
Finally, in our study, mutations and finite population size are only treated in simulations. There is a number of theoretical studies that use diffusion approximations to derive allele frequencies under mutation-selection-drift balance in similar ecological scenarios \citep{Wright:48,Kimura:54,Kimura:55,Gillespie:73b,Karlin:74,Taylor:08}. Especially \citet{Taylor:08} gives a comprehensive account of the balance of these forces under fluctuating selection.

In this study, we focused on the evolution of genetic variation in response to negative frequency-dependent disruptive selection. In principle, however, any mechanism increasing phenotypic variation is favored under this condition \citep[reviewed in][]{Rueffler:06b}.
 Adaptive genotype-environment interaction (phenotypic plasticity, \citealp{West-Eberhard:03,Via:87,Gillespie:89}), sexual dimorphism \citep{Bolnick:03,VanDoorn:04} and random phenotype determination \citep{Leimar:05,Svardal:11} are well studied alternatives. In cases where more than one mechanism exists that can increase phenotypic variation, it is interesting to ask which of these is more likely to evolve.
The odds for the evolution of genetic polymorphism versus the evolution of randomly determined alternative phenotypes have been compared by \citet{Leimar:05} and \citet{Svardal:11}. \citet{Svardal:11} show that under temporal fluctuations in a single panmictic population ($n=1$, lottery model) genetic polymorphism and random phenotype determination are equally favored only in the limit of infinite generation overlap ($\gamma\to1$). For $\gamma<1$,  random phenotype determination is the favored evolutionary response to disruptive selection, if it can evolve without constraints, because it also serves as a ``bet-hedging'' strategy under temporally fluctuating selection. For the current model, we speculate the following. For disruptive selection due to spatial differences in the expected environmental conditions (``permanent niches'') and full dispersal, genetic polymorphism and random phenotype determination are equally favored. However, as dispersal decreases, genetic polymorphism gains an advantage over random phenotype determination because the genotype serves as cue for future environmental conditions resulting in local adaptation \citep{Leimar:08}. This effect is amplified by positive temporal correlations. On the other hand, under temporally fluctuating selection, and more so with negative temporal correlations, random phenotype determination is favored, because then a benefit due to bet-hedging arises and the genotype becomes a misleading cue for future environments. A comparison of the advantage of genetic phenotype determination relative to adaptive phenotypic plasticity in a spatially heterogeneous environment was performed by \citet{Leimar:06}. These authors show that -- in line with previous research -- the benefit of plasticity strongly depends on the reliability of environmental cues. Furthermore, under restricted migration and strong local selection genetic polymorphism can be favored more strongly because the genotype can be an even better predictor of the coming selective environment. Applying this rationale to our model, we predict that negative temporal autocorrelations in the environmental conditions disfavor genetic polymorphism relative to plasticity because the genotype loses its value as a predictor for future environmental conditions.

Varying selective pressures and their implications for genetic diversity within species and for species diversity have always been major topics in ecology and evolution. Empirical examples for genetic polymorphism under heterogeneous selection include melanism in moth \citep{Cook:03} and mice \citep{Nachman:03,Vignieri:10}, pesticide resistance in insects \citep{McKenzie:96} and rats \citep{Pelz:05}, and pathogen resistance in animals (e.g. MHC polymorphism, \citealp{Spurgin:10}) and plants (e.g. R-gene polymorphism, \citealp{Bergelson:01}). For a review see \citet{Hedrick:06}.
Furthermore, empirical studies show that genetic variance is positively correlated with environmental heterogeneity both in natural \citep{Nevo:78} and lab conditions \citep{Mackay:81,Kassen:02} and that adaptive traits show substantial genetic variation \citep{Houle:92}. 
\citet{Byers:05} argued that no comprehensive models exist that give the potential of heterogeneous environments to maintain genetic variation in traits of adaptive significance. Our present study should be seen as a step in the lasting endeavor to fill this gap.

\section{Acknowledgements}

We thank Peter Chesson, Troy Day, Hans Metz, Reinhard B\"urger and the reviewers for comments that helped improving the manuscript. The individual based computer simulations presented have been achieved using the Vienna Scientific Cluster (VSC). The authors gratefully acknowledge funding from the Vienna Science and Technology Fund (WWTF) through grants MA06-01 and MA07-015.

\newpage

\begin{appendix}

\numberwithin{table}{section}
\renewcommand*\thetable{\Alph{section}\arabic{table}}

\section{Appendix: Ecological model}\label{app_model}

\subsection{Population  projection matrix}

Each generation, a fraction $(1-\gamma)$ of the adults in each patch is replaced by juveniles. Hence, we have
\begin{equation}\label{eq_phi_tp1}
\phi_{i,t+1}(y)=\gamma\phi_{it}(y)+(1-\gamma)\phi_{it}^o(y),
\end{equation}
where $\phi_{it}^o(y)$ is the frequency of individuals with trait value $y$ among offspring individuals in patch $i$ at time $t$. To calculate $\phi_{it}^o(y)$, we make two assumptions. First, local and dispersing juveniles have the same chance for recruitment into the adult population and, second, dispersing juveniles arrive in patch $i$ with a probability given by the relative juvenile carrying capacity of this patch, $c_i$. Then, we have
\begin{equation} \label{eq_phi_off}
\phi_{it}^o(y)=\frac{(1-m)c_i\phi_{it}(y)\rho(y,\phi_{it},\theta_{it})+m c_i\sum_{j}c_j\phi_{jt}(y)\rho(y,\phi_{jt},\theta_{jt})}{(1-m)c_i\int_{-\infty}^\infty\phi_{it}(x)\rho(x,\phi_{it},\theta_{it}){\rm d}x+m c_i\sum_{j}c_j\int_{-\infty}^\infty\phi_{jt}(x)\rho(x,\phi_{jt},\theta_{jt}){\rm d}x},
\end{equation}
where the numerator describes the frequency of offspring with phenotype $y$ and the denominator normalizes with all offspring that compete for establishment in patch $i$. The terms proportional to $(1-m)$ correspond to non-dispersing individuals, the terms proportional to $m$ correspond to individuals that disperse into patch $i$. Noting that $\int_{-\infty}^\infty\phi_{.t}(x)\rho(x,\phi_{.t},\theta_{.t}){\rm d}x=1$ and that $\sum_j c_j=1$, the denominator of \eqref{eq_phi_off} simplifies to $c_i$. This means that the relative proportions of local offspring and immigrants among the newly recruited individuals in each patch are $1-m$ and $m$, respectively, and
we obtain
\begin{equation} \label{eq_phi_off2}
\phi_{it}^o(y)=(1-m)\phi_{it}(y)\rho(y,\phi_{it},\theta_{it})+m \sum_{j}c_j\phi_{jt}(y)\rho(y,\phi_{jt},\theta_{jt}).
\end{equation}
Plugging this into equation \eqref{eq_phi_tp1} we obtain the elements of the population projection matrix given in equation \eqref{eq_L_main}.

\subsection{Selection on recruitment probability} \label{app_sel_recruitment}
Here we treat the case that phenotype-dependent selection happens on the recruitment probability into the adult population. In particular, all adult individuals initially produce an equal amount of offspring, $r(y,\theta_{it})=r$, but  offspring have trait-dependent recruitment performances $b(y,\theta_{it})$  into the adult population.  
For simplicity we assume for this part that  all offspring disperse ($m=1$).
Then, equation \eqref{eq_phi_tp1} still holds, but $\phi_{it}^o(y)$ is now given by
\begin{equation} \label{eq_recruitment_sel}
\phi_{it}^o(y)=\sum_j c_j \phi_{jt}(y) \frac{b(y, \theta_{jt})}{\int_{-\infty}^\infty{{\rm E_S}[ \phi_{t}(x)]b(x, \theta_{jt})}{\rm d}x} 
\end{equation}
where ${\rm E_S}[ \phi_{t}(x)]=\sum_k c_k  \phi_{kt}(x)$ is the frequency of individuals with trait value $x$  in the dispersal pool.  Note that $l_{ij}$ takes the same form as in equation \eqref{lji}, with $\rho$ replaced by the fraction in equation \eqref{eq_recruitment_sel}.  Hence, under full dispersal, selection on juvenile survival and recruitment are similar, except that in the second case there is a global mix of competitors. For $m<1$ the denominator in equation \eqref{eq_recruitment_sel} becomes more complicated and depends on $m$.

\section{Appendix: Analytical methods --  evolutionary invasion analysis}
\label{app_proof}

Following the adaptive dynamics approach \citep{Dieckmann:96, Metz:96a, Geritz:98}, we assume that mutations change a trait value $x$ by a small amount to $y={x + \delta x}$. Mutations are rare enough so that the
previous mutation has either gone to fixation or disappeared from the population before a new mutation arises.
For each new mutant it is then determined whether it can invade
 and replace the resident type and thus become the new resident itself.
The fundamental tool to predict this dynamics is the invasion fitness $w(y, x)$, which is defined as the long-term average per capita exponential growth rate of an infinitesimally small mutant sub-population with trait value $y$ in the resident population with trait value $x$ \citep{Metz:92, Metz:08c}.

In sufficiently large populations, a mutant has a positive probability to invade if $w(y,x) > 0$ and is doomed to extinction if $w(y, x) < 0$. Furthermore, if mutations have a sufficiently small phenotypic effect, successful invaders will go to fixation and replace the resident (\citealp{Geritz:05}). 
The evolutionary dynamics is then given by a series of mutation-substitution events. The direction of the evolutionary dynamics is predicted by the selection gradient,
\begin{equation}\label{eq_grad}
S(x)=\dif{w(y,x)}{y}\bigg|_{y=x}.
\end{equation}
Points $x^{*}$ where directional selective forces vanish, i.e., where $S(x^{*})= 0$, are called evolutionarily singular points \citep{Metz:96a, Geritz:98}. Singular points can be classified according to whether they are an attractor of the evolutionary dynamics and whether they are invadable.

A singular point $x^*$ is an attractor of the evolutionary dynamics if
\begin{equation}\label{eq:jacobian_main}
\frac{\partial S(x)}{\partial x}\bigg|_{x=x^{*}}<0
\end{equation}
and repelling when the inequality is reversed \citep{Eshel:83, Abrams:93, Metz:96a, Geritz:98}. In appendix \ref{app_special_cases} we show that in our model in the case of Gaussian selection a unique singular point exists that is always an attractor.

Since by definition the selection gradient at a singular point equals zero, the fitness landscape locally around a singular point is described by the second partial derivative of invasion fitness with respect to the mutant trait. The singular point is a local fitness minimum if 
\begin{equation}\label{eq:Hessian}
\frac{\partial^{2} w(y,x^*)}{\partial y^2}\bigg|_{y=x^{*}}>0
\end{equation}
and a local maximum if the inequality is reversed.
If the singular point is a local maximum of the fitness
landscape, the corresponding singular trait value cannot be invaded by any nearby mutant.
If such a trait value is also an attractor of the evolutionary dynamics, it is a final stop of evolution. If, however, the singular trait value is a minimum of the fitness landscape,
it can be invaded by nearby mutants. Singular points that are both an attractor of the evolutionary dynamics and invadable are known as evolutionary branching points \citep{Metz:96a,Geritz:98}.
In conclusion, in the neighborhood of an evolutionary branching point, monomorphic populations experience directional selection towards this point and once the evolutionary dynamics is sufficiently close selection turns disruptive. Then, invading mutants are able to coexist with the resident at the evolutionary branching point -- a dimorphism has evolved. Selection on this dimorphism continues to be divergent and thus leads to increased genetic variance (see figure \ref{fig_pip} for more information).  Importantly, such a stable pair of residents cannot be replaced by any mutant with intermediate phenotype.

In our model, invasion fitness (equation \ref{eq_inf_fit_lambda}) can also be written as 
\begin{equation} \label{eq_inv_fit_2}
w(y,x)=\lim_{T\to \infty}\frac{1}{T}\ln\left(\b u \prod\limits^T_{t=1}{\sf L}(y,x,\theta_{1t},..,\theta_{nt}) \b v\right)
\end{equation}
\citep{Tuljapurkar:90}, 
where  ${\sf L}(y,x,\theta_{1t},..,\theta_{nt})$ is defined in equation \eqref{eq_L_main}. If we exclude the trivial case $m=0$, invasion fitness as given by equation \eqref{eq_inv_fit_2} is independent of the entries in the row vector $\b u$ and the column vector $\b v$ as long as they are non-negative. Hence, below we can make convenient choices for these vectors. Note that in equation \eqref{eq_inv_fit_2} population projection matrices are multiplied from the left with increasing $t$.

\begin{table}[htb]
\caption{Shorthand notations used in the appendices.}
\begin{tabular}{lccc}
\hline
population projection matrix&${\sf L}(y,x,\theta_{1t},..,\theta_{nt})$&$\rightarrow$&${\sf L}_t$\\
relative reproductive success&$\rho(y,x,\theta_{it})$&$\rightarrow$&$\rho_{it}$\\
first derivative&$\frac{\partial }{\partial y}a$&$\rightarrow$&$\partial a$\\
second derivatives&$\frac{\partial^2 }{\partial y^2}a$&$\rightarrow$&$\partial^2a$ \\
& $\frac{\partial^2 }{\partial x \partial y}a$&$\rightarrow$&$\partial_x \partial_ya$ \\
evaluation at singular trait value&$a|_{y=x=x^*}$&$\rightarrow$&$a^*$\\
\hline
\end{tabular}
\note{Here, $a$ can be any function of the mutant and resident trait value. \\See  table \ref{tbl_notation} for general notation.}
\label{tbl_notation_short}
\end{table}

\subsection{Singular points} 

To keep the following derivations concise, we introduce the shorthand notations given in table \ref{tbl_notation_short}. From the condition for a singular point \eqref{eq_grad} we obtain with equation \eqref{eq_inv_fit_2}
\begin{equation} \label{eq_sp_00}
0=\frac{\partial w(y,x^*)} {\partial y}\biggm|_{y=x^*} =\lim_{T\to \infty}\frac{1}{T}\left(\frac{1}{\xi_T}\sum_{\tau=1}^{T}\b u\prod^{T}_{t_2=\tau+1} {\sf L}_{t_2}\partial {\sf L_{\tau}}\prod^{\tau-1}_{t_1=1} {\sf L}_{t_1}\b v\right)^*,
\end{equation}
where
\begin{equation} \label{eq_xi}
\xi_T=\b u \prod\limits^{T}_{t=1}{\sf L}_{t} \b v.
\end{equation} 
Evaluating at $y=x=x^*$, the matrix ${\sf L}_t^*$ has the entries
\begin{subequations}  \label{ l_star}
\begin{align} 
l_{ij}^* &=  (1-\gamma)m c_j \text{ for }j\neq i \label{lji_star}\\
l_{ii}^* &=  1-(1-\gamma)m\left(1-c_i\right)\label{lii_star}.
\end{align}
\end{subequations}
Hence, ${\sf L}_t^*$ is independent of $x^*$ and $t$ and
equation \eqref{eq_sp_00} simplifies to
\begin{equation}\label{eq_sp02}
0=\lim_{T\to \infty}\frac{1}{T}\frac{1}{\xi^*_T}\sum_{\tau=1}^{T}\b u({\sf L}^*)^{T-\tau} (\partial{\sf L}_{\tau})^* ({\sf L}^*)^{\tau-1}\b v.
\end{equation}
Note that $\sf L^*$ is a stochastic matrix, i.e., the entries in each row sum up to one. 
We choose $\b u$ and $\b v$ to be the leading left and right eigenvectors of $\sf L^*$ to the eigenvalue one. The elements of these eigenvectors are $u_i=c_i$ and $v_i=1$, respectively, and we obtain $\xi_T^*=\b u {\sf L^*}\b v =1$. Equation \eqref{eq_sp02} simplifies to
\begin{equation} \label{eq_sp_before_average}
\begin{split}
0=\lim_{T\to \infty}\frac{1}{T}\sum^{T}_{\tau=1}\b u (\partial{\sf L}_{\tau})^* \b v
\end{split}
\end{equation}
Note that we can write
\begin{equation} \label{eq_dLStar}
\b u (\partial{\sf L}_{\tau})^*\b v=\sum_{i=1}^n\sum_{j=1}^n  u_i  (\partial {l_{ij}}_{\tau})^*v_j=(1-\gamma)\sum_{i=1}^n c_i(\partial\rho_{i\tau})^*.
\end{equation}
With the definitions of spatial and temporal expectation given in equation \eqref{eq_es} and \eqref{eq_et},  equation \eqref{eq_sp_00} becomes
\begin{equation}
{\rm E_T}\left[{\rm E_S}\left[(\partial\rho)^*\right]\right]=0.
\end{equation}
Expressing this equation in terms of the local selection gradient $s_{it}=\ln\rho_{it}$ gives equation \eqref{eq_sp_general}.

\subsection{Invadability}\label{app_invadability}
A singular point $x^*$ is invadable if inequality \eqref{eq:Hessian} is fulfilled. 
Analogously to above we obtain
\begin{equation}
\frac{\partial^2 w(y,x^*)}{\partial y^2}\biggm|_{y=x^*}=\lim_{T\to \infty}\frac{1}{T}\left(\partial\left(\frac{1}{\xi_T}\vphantom{\sum^{T}_{\tau=1}\b u\prod^{T}_{t_2=\tau+1} {\sf L_{t_2}}\partial {\sf L}_{\tau}\prod^{\tau-1}_{t_1=1} {\sf L_{t_1}}\b v}\right.\right.\!
  \underbrace{\sum^{T}_{\tau=1}\b u\prod^{T}_{t_2=\tau+1} {\sf L_{t_2}}\partial {\sf L}_{\tau}\prod^{\tau-1}_{t_1=1} {\sf L_{t_1}}\b v}_{=:\zeta_{y,T}}\!%
  \left.\left.\vphantom{\sum^{T}_{\tau=1}\b u\prod^{T}_{t_2=\tau+1} {\sf L_{t_1}}\partial {\sf L}_{\tau}\prod^{\tau-1}_{t_2=1} {\sf L_{t_2}}\b v}\right)\right)^*,
\end{equation}
where $\xi_T$ is given by equation \eqref{eq_xi}. This derivative can be rewritten as
\begin{equation}\label{eq_2nd_derr0}
\begin{split}
\frac{\partial^2 w(y,x^*)}{\partial y^2}\biggm|_{y=x^*}=&\lim_{T\to \infty}\frac{1}{T}\left(-\frac{\zeta_{y,T}^2}{\xi_T^2}+
\frac{1}{\xi_T}\sum^{T}_{\tau=1}\b u\prod^{T}_{t_2=\tau+1} {\sf L}_{t_2}\partial^2 {\sf L}_{\tau}\prod^{\tau-1}_{t_1=1} {\sf L}_{t_1}\b v\right.\\
&+\left.\frac{1}{\xi_T}\sum^{T}_{\tau_1=1}\sum^{T}_{\tau_2=\tau_1+1}\b u\prod^{T}_{t_3=\tau_2+1}{\sf L}_{t_3}\partial {\sf L}_{\tau_2}\prod^{\tau_2-1}_{t_2=\tau_1+1}{\sf L}_{t_2}\partial {\sf L}_{\tau_1}\prod^{\tau_1-1}_{t_1=1}{\sf L}_{t_1}\b v\right.\\
&+\left.\frac{1}{\xi_T}\sum^{T}_{\tau_1=1}\sum^{\tau_1-1}_{\tau_2=1}\b u\prod^{T}_{t_3=\tau_1+1}{\sf L}_{t_3}\partial {\sf L}_{\tau_1}\prod^{\tau_1-1}_{t_2=\tau_2+1}{\sf L}_{t_2}\partial {\sf L}_{\tau_2}\prod^{\tau_2-1}_{t_1=1}{\sf L}_{t_1}\b v\right)^*.
\end{split}
\end{equation}
Analogously to the derivation of the singular point, we  choose $\b u$ and $\b v$ to have elements
 $u_i=c_i$ and $v_i=1$. Then $\xi^* = \b u {\sf L^*}\b v = 1$, and noting that the last two terms in equation \eqref{eq_2nd_derr0} are equal to each other, we get  for the right-hand side
\begin{equation}\label{eq_hessian_before_average}
\begin{split}
\lim_{T\to \infty}\frac{1}{T}\left[-\left(\sum^{T}_{\tau=1}\b u (\partial {\sf L}_{\tau})^*\b v\right)^2
+\sum^{T}_{\tau=1}\b u (\partial^2 {\sf L}_{\tau})^*\b v+
2\sum^{T}_{\tau_1=1}\sum^{\tau_1-1}_{\tau_2=1}\b u(\partial {\sf L}_{\tau_1})^*{\sf L^*}^{\tau_1-1-\tau_2}(\partial {\sf L}_{\tau_2})^*\b v\right].
\end{split}
\end{equation}
We use equation \eqref{eq_dLStar} and the analogue for the second derivative with respect to $y$ to obtain
\begin{equation}\label{eq_hessian_3}
\begin{split}
\frac{\partial^2 w(y,x^*)}{\partial y^2}\biggm|_{y=x^*}=&\lim_{T\to \infty}\frac{1}{T}\left[-\left(\sum^{T}_{\tau=1}(1-\gamma)\sum_{i=1}^n c_i  (\partial\rho_{i\tau})^*\right)^2
+\sum^{T}_{\tau=1}(1-\gamma)\sum_{i=1}^n c_i(\partial^2\rho_{i\tau})^*\right.\\
&+\left.2\sum^{T}_{\tau_1=1}\sum^{\tau_1-1}_{\tau_2=1}\b u(\partial {\sf L}_{\tau_2})^*{\sf L^*}^{\tau_1-1-\tau_2}(\partial {\sf L}_{\tau_1})^*\b v\right].
\end{split}
\end{equation}
Calculating the squared term and identifying temporal and spatial averages as defined in equation \eqref{eq_es} and \eqref{eq_et}, we get
\begin{equation}\label{eq_sequence_average}
\begin{split}
\frac{\partial^2 w(y,x^*)}{\partial y^2}\biggm|_{y=x^*}=&-(1-\gamma)^2\left({\rm E_T}\left[{\rm E_S}\left[(\partial\rho)^*\right]^2\right]+\lim_{T\to \infty}\frac{1}{T}\sum_{\tau_1=1}^{T}\sum_{\substack{\tau_2=1\\\tau_2\neq\tau_1}}^{T}{\rm E_S}\left[(\partial\rho_{\tau_1})^*\right]{\rm E_S}\left[(\partial\rho_{\tau_2})^*\right]\right)\\
&+(1-\gamma){\rm E_T}\left[{\rm E_S}\left[(\partial^2\rho)^*\right]\right]+2\lim_{T\to \infty}\frac{1}{T}\sum_{\tau_1=1}^T\sum_{\tau_2=1}^{\tau_1-1}\b u(\partial {\sf L}_{\tau_2})^*{\sf L^*}^{\tau_1-1-\tau_2}(\partial {\sf L}_{\tau_1})^*\b v.
\end{split}
\end{equation}
Noting that we can write
\begin{equation}
{\sf L}^*=
\mathbb{I}+m(1-\gamma)( \b v\b u-\mathbb{I}),
\end{equation}
where $\mathbb{I}$ is the identity matrix and $ \b v \b u $ is the tensor product, i.e., a matrix with elements $[\b v \b u]_{ij}=v_iu_j=c_j$, we can calculate 
\begin{equation} \label{eq_Ltau}
{{\sf L}^*}^\tau=(1-m(1-\gamma))^\tau(\mathbb{I}- \b v\b u)+ \b v\b u.
\end{equation}
Here, we have used the binomial theorem and that $(\b v\b u )^k= \b v\b u$ for $k>0$. Furthermore,
\begin{subequations}\label{eq_Lpartial}
\begin{align}
\b u(\partial {\sf L}_{\tau_2})^*&=\b u\left({\sf L^*}-\gamma \mathbb{I}\right)\mathbb{I}\left[(\partial\rho_{\tau_2})^*\right]=(1-\gamma) \b u \mathbb{I}\left[(\partial\rho_{\tau_2})^*\right]& \\
(\partial {\sf L}_{\tau_1})^*\b v&=\left({\sf L^*}-\gamma \mathbb{I}\right)\mathbb{I}\left[(\partial\rho_{\tau_1})^*\right]\b v, 
\end{align}
\end{subequations}
where $\mathbb{I}\left[a\right]$ is a diagonal matrix with $a_i$ at position $i$.
Using equations \eqref{eq_Ltau} and \eqref{eq_Lpartial}, the last term in equation \eqref{eq_sequence_average} calculates to
\begin{equation}\label{eq_17lastterm}
\begin{split}
&2\lim_{T\to \infty}\frac{1}{T}\sum_{\tau_1=1}^T\sum_{\tau_2=1}^{\tau_1-1}(1-\gamma) \b u \mathbb{I}\left[{(\partial\rho_{\tau_2})}^*\right]{\sf L^*}^{\tau_1-1-\tau_2}\left({\sf L^*}-\gamma \mathbb{I}\right)\mathbb{I}\left[(\partial\rho_{\tau_1})^*\right]\b v=\\
&2(1-\gamma)^2 \left(\lim_{T\to \infty}\frac{1}{T}\sum_{\tau_1=1}^T\sum_{\tau_2=1}^{\tau_1-1}\left(\b u \mathbb{I}\left[(\partial\rho_{\tau_1})^*\right]\b v\right)\left(\b u \mathbb{I}\left[(\partial\rho_{\tau_2})^*\right]\b v\right)+\right.\\
&\left.(1-m)\lim_{T\to \infty}\frac{1}{T}\sum_{\tau_1=1}^T\sum_{\tau_2=1}^{\tau_1-1}(1-m(1-\gamma))^{\tau_1-1-\tau_2}\big(\b u \mathbb{I}\left[(\partial\rho_{\tau_1})^*(\partial\rho_{\tau_2})^*\right]\b v-\left(\b u \mathbb{I}\left[(\partial\rho_{\tau_2})^*\right]\b v\right)\left(\b u \mathbb{I}\left[(\partial\rho_{\tau_1})^*\right]\b v\right)\big)\right).\\
\end{split}
\end{equation}
Using that $\b u \mathbb{I}[a]\b v={\rm E_S}[a]$, the first term in this expression becomes
\begin{equation} \label{eq_canceling_term}
2(1-\gamma)^2\sum_{\tau_1=1}^T\sum_{\tau_2=1}^{\tau_1-1}{\rm E_S}\left[(\partial\rho_{\tau_1})^*\right]{\rm E_S}\left[([\partial\rho_{\tau_2})^*\right]=(1-\gamma)^ 2\sum_{\tau_1=1}^T\sum_{\substack{\tau_2=1\\\tau_2\neq\tau_1}}^{T}{\rm E_S}\left[(\partial\rho_{\tau_1})^*\right]{\rm E_S}\left[(\partial\rho_{\tau_2})^*\right],
\end{equation}
where we have used that the $(\partial\rho_{i\tau})^*$ have a stationary distribution. This term precisely cancels with the second term in equation \eqref{eq_sequence_average}.
The last term in expression \eqref{eq_17lastterm} can be identified as a spatial covariance,
\begin{equation} \label{eq_CovS}
\begin{split}
2(1-\gamma)^2(1-m)\lim_{T\to \infty}\frac{1}{T}\sum_{\tau_1=1}^T\sum_{\tau_2=1}^{\tau_1-1}(1-m(1-\gamma))^{\tau_1-1-\tau_2}{\rm Cov_S}[(\partial\rho_{\tau_2})^*,(\partial\rho_{\tau_1})^*].\\
\end{split}
\end{equation}
This term contains the spatial covariance between $(\partial\rho_{i})^*$ at two points in time, summed over all
possible pairs of time points. 
Under our time-invariance assumption of the environmental process, covariance terms 
depend only on time differences 
$\tau=\tau_1 - \tau_2$.
The outer sum can then be written as a weighting factor to the terms of the inner sum and we obtain
\begin{equation}
2(1-\gamma)^2(1-m)\lim_{T\to \infty}\sum_{\tau=1}^{T-1}(1-\frac{\tau}{T})(1-m(1-\gamma))^{\tau-1}{\rm E_T}[{\rm Cov_S}[(\partial\rho_{ t})^*,(\partial\rho_{t+\tau})^*]],
\end{equation}
In the limit $T\to\infty$ this becomes
\begin{equation} \label{eq_covariance_1}
\begin{split}
2(1-\gamma)^2(1-m)\sum_{\tau=1}^\infty(1-m(1-\gamma))^{\tau-1}{\rm E_T}[{\rm Cov_S}[(\partial\rho_{ t})^*,(\partial\rho_{t+\tau})^*]]=2(1-\gamma)\frac{1-m}{m}{\rm Var_S}\left[{\rm E_T}\left[(\partial\rho)^*\right]\right]+\\
2(1-\gamma)^2(1-m)\sum_{\tau=1}^\infty(1-m(1-\gamma))^{\tau-1}\left({\rm E_S}\left[{\rm Cov_T}\left[(\partial\rho_{ t})^*,(\partial\rho_{ t+\tau})^*\right]\right] -{\rm Cov_T}\left[{\rm E_S}\left[(\partial\rho_{ t})^*\right], {\rm E_S}\left[(\partial\rho_{ t+\tau})^*\right]\right]\right),
\end{split}
\end{equation} 
where the index $t$ is kept for clarity where necessary.
For the right-hand side of equation \eqref{eq_covariance_1} we used the definition of the covariance, ${\rm Cov}[a,b]={\rm E}[a b]-{\rm E}[a]{\rm E}[b]$, and the relations ${\rm E_T}[{\rm E_S}[a]]={\rm E_S}[{\rm E_T}[a]]$ and ${\rm E_T}[(\partial\rho_{it})^*]={\rm E_T}[(\partial\rho_{it+\tau})^*]$. The transformation applied here reflects the fact that spatial covariance between the $(\partial\rho_{i.})^*$ at different points in time can be produced by two factors. First, differences in the probability distribution of environmental conditions between patches produce spatial covariance. This contribution is independent of $\tau$. Second, spatial covariance is produced by temporal correlations in environmental conditions. 

Going back to the the second derivative of invasion fitness, equation \eqref{eq_sequence_average}, we can reformulate the first and third term using
\begin{equation} \label{eq_vartes}
{\rm E_T}\left[{\rm E_S}\left[(\partial\rho)^*\right]^2\right]={\rm E_T}\left[{\rm E_S}\left[(\partial\rho)^*\right]\right]^2+{\rm Var_T}\left[{\rm E_S}\left[(\partial\rho)^*\right]\right],
\end{equation}
and
\begin{equation} \label{eq_var_relation}
{\rm E_S}\left[(\partial\rho)^*\right]^2={\rm E_S}\left[{(\partial\rho)^*}^2\right]-{\rm Var_S}\left[(\partial\rho)^*\right],
\end{equation}
where it follows from equation \eqref{eq_sp_general} that the first term on the right-hand side of equation \eqref{eq_vartes} is zero.
With equations \eqref{eq_canceling_term}-\eqref{eq_var_relation}, we can write equation \eqref{eq_sequence_average} as
\begin{equation} \label{eq_hessian_general}
\begin{split}
&\frac{\partial^2 w(y,x^*)}{\partial y^2}\biggm|_{y=x^*}=(1-\gamma)\left({\rm E_T}\left[{\rm E_S}\left[(\partial^2\rho)^*-{(\partial\rho)^*}^2\right]\right]+{\rm E_T}\left[{\rm Var_S}\left[(\partial\rho)^*\right]\right]+\gamma{\rm Var_T}\left[{\rm E_S}\left[(\partial\rho)^*\right]\right]\right.\\
&\left.+
2(1-\gamma)(1-m)\sum_{\tau=1}^\infty(1-m(1-\gamma))^{\tau-1}{\rm E_T}\left[{\rm Cov_S}\left[(\partial\rho_{ t})^*,(\partial\rho_{t+\tau})^*\right]\right]\right)=\\
&(1-\gamma)\left({\rm E_T}\left[{\rm E_S}\left[(\partial^2\rho)^*-\left((\partial\rho)^*\right)^2\right]\right]+{\rm E_T}\left[{\rm Var_S}\left[(\partial\rho)^*\right]\right]+\gamma{\rm Var_T}\left[{\rm E_S}\left[(\partial\rho)^*\right]\right]+2\frac{1-m}{m}{\rm Var_S}\left[{\rm E_T}\left[(\partial\rho)^*\right]\right]+\right.\\
&\left.2(1-\gamma)(1-m)\sum_{\tau=1}^\infty(1-m(1-\gamma))^{\tau-1}\left({\rm E_S}\left[{\rm Cov_T}\left[(\partial\rho_{ t})^*,(\partial\rho_{ t+\tau})^*\right]\right] -{\rm Cov_T}\left[{\rm E_S}\left[(\partial\rho_{ t})^*\right], {\rm E_S}\left[(\partial\rho_{ t+\tau})^*\right]\right]\right)\right),
\end{split}
\end{equation}
where the second version is longer, but easier to interpret (see results section and discussion).
With $s_{it}:=\ln(\rho_{it})$ and noting that $(\partial s_{it})^*=(\partial \rho_{it})^*$ and $(\partial^2 s_{it})^*=(\partial^2 \rho_{it})^*-\left((\partial \rho_{it})^*\right)^2$, we can write condition \eqref{eq:Hessian}  as  given in equation \eqref{eq_branch_general}. 

\subsection{Gaussian selection}\label{app_gaussian_selection}

For Gaussian stabilizing selection towards a selective optimum $\theta_{it}$, the functions $r(x,\theta_{it})$ in  $\rho_{it}= \exp(s_{it}) =\frac{r(y,\theta_{it})}{r(x,\theta_{it})}$ are given by equation \eqref{eq:selection} and
we obtain
\begin{equation} \label{eq_rhop}
(\partial_y\rho_{it})^*=(\partial_y s_{it})^*=\frac{\theta_{it}-x^*}{\sigma^2},
\end{equation}
\begin{equation} \label{eq_rhopp}
(\partial^2_y\rho_{it})^*=\frac{(\theta_{it}-x^*)^2}{\sigma^4}-\frac{1}{\sigma^2},
\end{equation}
\begin{equation} \label{eq_rhopp}
(\partial^2_ys_{it})^*=-\frac{1}{\sigma^2}.
\end{equation}
With this, conditions \eqref{eq:sp} and \eqref{eq_br_gauss_noTC} easily follow from conditions  \eqref{eq_sp_general} and \eqref{eq_branch_general}, respectively.

\subsection{Convergence stability} 

In this section, we use subscripts $y$ and $x$ to distinguish derivatives with respect to mutant and resident, respectively. A singular point is an attractor of the evolutioary dynamics if inequality \eqref{eq:jacobian_main} is fulfilled. This inequality can be written as
\begin{equation} \label{eq_def_cs_app}
\frac{\partial^{2} w(y,x)}{\partial x \partial y}\bigg|_{y=x=x^{*}}+\frac{\partial^{2} w(y,x^*)}{\partial y^2}\bigg|_{y=x^{*}}<0
\end{equation}
\citep{Geritz:98},
where the double derivative with respect to $y$ is given by equation \eqref{eq_hessian_general}. The mixed derivative equals
\begin{equation}
\begin{split}
\frac{\partial^2 w(y,x)}{\partial x \partial y }\biggm|_{y=x=x^*}=&\lim_{T\to \infty}\frac{1}{T}\left(\partial_x\left(\frac{1}{\xi_T}\vphantom{\sum^{T}_{\tau=1}\b u\prod^{T}_{t_1=\tau+1} {\sf L_{t_1}}\partial_y {\sf L}_{\tau}\prod^{\tau-1}_{t_2=1} {\sf L_{t_2}}\b v}\right.\right.\!
  \underbrace{\sum^{T}_{\tau=1}\b u\prod^{T}_{t_1=\tau+1} {\sf L_{t_1}}\partial_y {\sf L}_{\tau}\prod^{\tau-1}_{t_2=1} {\sf L_{t_2}}\b v}_{=:\zeta_{y,T}}\!%
  \left.\left.\vphantom{\sum^{T}_{\tau=1}\b u\prod^{T}_{t_1=\tau+1} {\sf L_{t_1}}\partial_y {\sf L}_{\tau}\prod^{\tau-1}_{t_2=1} {\sf L_{t_2}}\b v}\right)\right)^*\\
=&\lim_{T\to \infty}\frac{1}{T}\left(-\frac{\zeta_{y,T}\zeta_{x,T}}{\xi_T^2}
+\frac{1}{\xi_T}\sum^{T}_{\tau=1}\b u\prod^{T}_{t_1=\tau+1} {\sf L}_{t_1}\partial_x\partial_y {\sf L}_{\tau}\prod^{\tau-1}_{t_2=1} {\sf L}_{t_1}\b v\right.\\
&+\left.\frac{1}{\xi_T}\sum^{T}_{\tau_1=1}\sum^{T}_{\tau_2=\tau_1+1}\b u\prod^{T}_{t_1=\tau_2+1}{\sf L}_{t_1}\partial_x {\sf L}_{\tau_2}\prod^{\tau_2-1}_{t_2=\tau_1+1}{\sf L}_{t_2}\partial_y {\sf L}_{\tau_1}\prod^{\tau_1-1}_{t_3=1}{\sf L}_{t_3}\b v\right.\\
&+\left.\frac{1}{\xi_T}\sum^{T}_{\tau_1=1}\sum^{\tau_1-1}_{\tau_2=1}\b u\prod^{T}_{t_1=\tau_1+1}{\sf L}_{t_1}\partial_y {\sf L}_{\tau_1}\prod^{\tau_1-1}_{t_2=\tau_2+1}{\sf L}_{t_2}\partial_x {\sf L}_{\tau_2}\prod^{\tau_2-1}_{t_3=1}{\sf L}_{t_3}\b v\right)^*.
\end{split}
\end{equation}
Here, $\zeta_{x,T}$ is the same as $\zeta_{y,T}$ except that the derivative of ${\sf L}_t$ is with respect to the resident trait value $x$. Using analogous simplifications as in the calculations above, we obtain 
\begin{equation}
\begin{split}
\frac{\partial^2 w(y,x)}{\partial x \partial y}\biggm|_{y=x=x^*}=&\lim_{T\to \infty}\frac{1}{T}\left[-\left(\sum^{T}_{\tau=1}(1-\gamma)\sum_{i=1}^n(\partial_y \rho_{i\tau})^*\right)\left(\sum^{T}_{\tau=1}(1-\gamma)\sum_{i=1}^n(\partial_x \rho_{i\tau})^*\right)
+\right.\\
&\sum^{T}_{\tau=1}(1-\gamma)\sum_{i=1}^n(\partial_x\partial_y \rho_{i\tau})^*+\sum^{T}_{\tau_1=1}\sum^{T}_{\tau_2=\tau_1+1}\b u(\partial_x {\sf L}_{\tau_1})^*{\sf L^*}^{\tau_2-1-\tau_1}(\partial_y {\sf L}_{\tau_2})^*\b v+\\
&\left.\sum^{T}_{\tau_1=1}\sum^{\tau_1-1}_{\tau_2=1}\b u(\partial_y {\sf L}_{\tau_1})^*{\sf L^*}^{\tau_2-1-\tau_1}(\partial_x {\sf L}_{\tau_2})^*\b v\right].
\end{split}
\end{equation}
Using that 
\begin{equation}
(\partial_x {\sf L}_{\tau})^*=-(\partial_y {\sf L}_{\tau})^*\text{,  }
(\partial_x \rho_{i\tau})^*=- (\partial_y \rho_{i\tau})^*
\text{  and  }
(\partial_x\partial_y \rho_{i\tau})^*=-(\partial^2_y \rho_{i\tau})^*,
\end{equation}
we obtain
\begin{equation}\label{eq_q}
\begin{split}
\frac{\partial^2 w(y,x)}{\partial y \partial x}\biggm|_{y=x=x^*}=&\lim_{T\to \infty}\frac{1}{T}\left[\left(\sum^{T}_{\tau=1}\left((1-\gamma)\sum_{i=1}^n(\partial_y \rho_{i\tau})^*\right)\right)^2-\sum^{T}_{\tau=1}(1-\gamma)\sum_{i=1}^n{(\partial^2_y \rho_{i\tau})^*}^2\right.\\
&-\left.2\sum^{T}_{\tau_1=1}\sum^{T}_{\tau_2=\tau_1+1}\b u(\partial_y {\sf L}_{\tau_1})^*{\sf L^*}^{\tau_2-1-\tau_1}(\partial_y {\sf L}_{\tau_2})^*\b v\right].
\end{split}
\end{equation}
Combining equations \eqref{eq_hessian_3} and \eqref{eq_q}, condition \eqref{eq_def_cs_app} becomes
\begin{equation}
(1-\gamma)\lim_{T\to \infty}\frac{1}{T}\left[\sum^{T}_{\tau=1}\sum_{i=1}^n (\partial^2_y\rho_{i\tau})^*-{(\partial_y\rho_{i\tau})^*}^2\right]={\rm E_T}\left[{\rm E_S}\left[ (\partial^2_y\rho)^*\right]-{\rm E_S}\left[{(\partial_y\rho)^*}^2\right]\right]<0.
\end{equation}
Expressing this inequality in terms of $s$ results in equation \eqref{eq_cs_general}.

\renewcommand{\thetable}{C\arabic{table}}
\renewcommand{\thefigure}{C\arabic{figure}}
\setcounter{table}{0}
\setcounter{figure}{0}
\section{Appendix: Special cases and extensions}\label{app_special_cases}

\subsection{Uncorrelated patches}  \label{app_uncorr_patches}
Here, we derive equation \eqref{eq:br_indep_cov} from the main text.
If we assume that there are no correlations in environmental conditions across patches, the probability density function can be written as
\begin{equation} \nonumber
f(\theta_{1},..,\theta_{n})=\prod\limits_{i=1}^n f_i( \theta_i),
\end{equation}
where $f_i(\theta_i)$ is the probability distribution of environmental condition $\theta_i$ in patch $i$. In the following, we write $f_i:= f_i(\theta_i)$. 
Then the first term on the left-hand side of condition \eqref{eq_branch_general} can be written as
\begin{equation} \label{eq_average_poisson}
\mathrm{E}_{\rm T}\left[\mathrm{Var}_{\rm S}\left[(\partial_y s)^*\right]\right]=\int\limits_{-\infty}^{\infty}...\int\limits_{-\infty}^{\infty}{\left(\prod\limits_{i=1}^n f_i\right)\left[\sum\limits_{i=1}^{n}{c_i{(\partial_y s_i)^*}^2}-\sum\limits_{i=1}^{n}\sum\limits_{j=1}^n{c_i c_j(\partial_y s_i)^*  (\partial_y s_j)^*}
\right]\mathrm{d} \theta_1...\mathrm{d} \theta_n}.
\end{equation}
Since  the environmental conditions in the patches are independent, the $i$th term in the first sum is a constant with respect to all integration variables except  $\theta_i$, and the $(i,j)$th term in the second  sum is a constant with respect to all integration variables except $\theta_i$ and $\theta_j$. Noting that $\int f_k {\rm d}\theta_k=1$, we can simplify by exchanging the distribution averaging and the averaging over the patches. For the last term in equation \eqref{eq_average_poisson} we have to distinguish between the indices $i=j$, where the optimum is the same, and $i\neq j$, for which patches are independent. With this we get
\begin{equation}
\begin{split}
\mathrm{E}_{\rm T}\left[\mathrm{Var}_{\rm S}\left[{(\partial_y s)^*}\right]\right]=&\sum\limits_{i=1}^{n}c_i\int\limits_{-\infty}^{\infty}{f_i {(\partial_y s_i)^*}^2\mathrm{d}\theta_i}-\sum\limits_{i=1}^{n}c_i^2\int\limits_{-\infty}^{\infty}{f_i {(\partial_y s_i)^*}^2\mathrm{d} \theta_i}\\
&-\sum\limits_{i=1}^n c_i\sum\limits_{\substack{j=1\\j\neq i}}^n c_j\int\limits_{-\infty}^{\infty}\int\limits_{-\infty}^{\infty}{f_i f_j {(\partial_y s_i)^*}{(\partial_y s_j)^*} \mathrm{d}\theta_i\mathrm{d}\theta_j}.
\end{split}
\end{equation}
Here, the first term comes from the first term in brackets on the right-hand side of equation \eqref{eq_average_poisson}, the middle term is the part of the last term from equation \eqref{eq_average_poisson} where $i=j$ and the last term is the one where $i\neq j$. 
With the definition of the temporal average (equation \ref{eq_etd}), this becomes
\begin{equation}\label{eq_indep_rhs1_interm}
\mathrm{E}_{\rm T}\left[\mathrm{Var}_{\rm S}\left[{(\partial_y s)^*}\right]\right]=\sum\limits_{i=1}^{n}c_i\mathrm{E}_{\rm T}\left[{(\partial_y s_i)^*}^2\right]-\sum\limits_{i=1}^{n}c_i^2\mathrm{E}_{\rm T}\left[{(\partial_y s_i)^*}^2\right]-\sum\limits_{i=1}^n c_i \mathrm{E}_{\rm T}\left[{(\partial_y s_i)^*}\right] \sum\limits_{\substack{j=1\\j\neq i}}^n c_j \mathrm{E}_{\rm T}\left[{(\partial_y s_j)^*}\right].
\end{equation}
Using that $\mathrm{E}_{\rm T}[b^2]=\mathrm{Var}_{\rm T}[b]+{\mathrm{E}_{\rm T}[b]}^2$ for any $b$ and with the definition of the spatial average (equation \ref{eq_es}), we obtain after rearranging
\begin{equation} \label{eq_br_rhs1}
\mathrm{E}_{\rm T}\left[\mathrm{Var}_{\rm S}\left[{(\partial_y s)^*}\right]\right]=\mathrm{E}_{\rm S}\left[\left(1-c\right)\mathrm{Var}_{\rm T}\left[{(\partial_y s)^*}\right]\right]+\mathrm{Var}_{\rm S}\left[\mathrm{E}_{\rm T}\left[{(\partial_y s)^*}\right]\right],
\end{equation}
where we omit the patch index of $c_i$ inside the spatial average. 
We can apply steps analogous to \eqref{eq_average_poisson}-\eqref{eq_br_rhs1} to reformulate the second term on the left-hand side of inequality \eqref{eq_branch_general} and get
\begin{equation} \label{eq_rhs_indep_C2}
\mathrm{Var}_{\rm T}\left[\mathrm{E}_{\rm S}\left[{(\partial_y s)^*}\right]\right]=\sum\limits_{i=1}^{n}{c_i^2\mathrm{E}_{\rm T}[{(\partial_y s_i)^*}^2]}+\sum\limits_{i=1}^{n}c_i\mathrm{E}_{\rm T}[{(\partial_y s_i)^*}]  \sum\limits_{\substack{j=1\\j\neq i}}^{n}{c_j \mathrm{E}_{\rm T}[{(\partial_y s_j)^*}]- \mathrm{E}_{\rm S}[\mathrm{E}_{\rm T}[{(\partial_y s)^*}]]}^2,
\end{equation}
where the middle term on the right-hand side is the same as the last term in \eqref{eq_indep_rhs1_interm} and can be rearranged as above. With the definition of the spatial average we get
\begin{equation} \label{eq_br_rhs2}
\mathrm{Var}_{\rm T}[\mathrm{E}_{\rm S}[{(\partial_y s)^*}]]=\mathrm{E}_{\rm S}\left[{c\mathrm{Var}_{\rm T}[{(\partial_y s)^*}]}\right].
\end{equation}
In the absence of temporal correlations, the term ${\rm \mathcal{C}}\left[({\partial_y s})^*\right]$ in condition  \eqref{eq_branch_general} equals zero. Hence, for uncorrelated patches the condition for invadability, inequality \eqref{eq_branch_general}, can be rewritten as
\begin{equation} \label{eq_br_indep_general}
\begin{split}
\mathrm{Var}_{\rm S}\left[\mathrm{E}_{\rm T}\left[(\partial_y s)^*\right]\right]+\mathrm{E}_{\rm S}\left[\left(1-c\right)\mathrm{Var}_{\rm T}\left[(\partial_y s)^*\right]\right]&\\
+\gamma\mathrm{E}_{\rm S}\left[{c\mathrm{Var}_{\rm T}\left[(\partial_y s)^*\right]}\right]+2\frac{1-m}{m}{\rm Var_S}\left[{\rm E_T}\left[(\partial_y s)^*\right]\right]&>-{\rm E_T}\left[{\rm E_S}\left[{(\partial^2_y s)^*}\right]\right].
\end{split}
\end{equation}
This can be further simplified by using the definition of the covariance, ${\rm Cov}[a,b]={\rm E}[a b]-{\rm E}[a]{\rm E}[b]$, and that ${\rm E_S}[c]=1/n$. We obtain
\begin{equation}\label{eq_br_indep_cov_gen}
\frac{2-m}{m}{\rm Var_S}\left[{\rm E_T}\left[(\partial_y s)^*\right]\right]+\left(1-\frac{1-\gamma}{n}\right)\mathrm{E_S}\left[\mathrm{Var}_T\left[(\partial_y s)^*\right]\right]-(1-\gamma)\mathrm{Cov_S}\left[c,\mathrm{Var_T}\left[(\partial_y s)^*\right]\right]>-{\rm E_T}\left[{\rm E_S}\left(\partial^2_y s)^*\right]\right],
\end{equation}
which for the Gaussian case becomes condition \eqref{eq:br_indep_cov}.
It can be informative to consider the origin of the the terms in  condition \eqref{eq_br_indep_cov_gen} with respect to the original branching condition, inequality \eqref{eq_branch_general}.
For the case of Gaussian selection and equally sized patches ($c_i=1/n$), we have
\begin{equation}\label{eq:br_indep}
\overbrace{\mathrm{Var_S}[\mathrm{E_T}[\theta]]+\left(1-\frac{1}{n}\right)\mathrm{E_S}\left[\mathrm{Var_T}[\theta]\right]}^{\mathrm{E}_{\rm T}[\mathrm{Var}_{\rm S}[ \theta]]=}+\overbrace{\frac{\gamma}{n}\mathrm{E_S}\left[\mathrm{Var_T}[ \theta]\right]}^{\gamma\mathrm{Var}_{\rm T}[\mathrm{E}_{\rm S}[\theta]]=}+2\frac{1-m}{m}{\rm Var_S[E_T[}\theta]]>\sigma^2,
\end{equation}
where the correspondence to terms in condition  \eqref{eq_br_gauss_noTC} is given  above the braces. 
We see that the expected spatial variation results from two sources. The first source are differences in the expected environments among patches and the second source are temporal fluctuations within patches. These fluctuations lead to differences among patch optima for a given season.
 The second factor becomes increasingly important as the number of patches increases. The third term on the left-hand side of condition \eqref{eq:br_indep} shows that temporal fluctuations within the patches contribute less to  ``global'' temporal fluctuations as the patch number, $n$, increases. The reason is that local fluctuations average out over space.

\subsection{Different strength of selection in the patches}\label{app_special_cases_varying_selection}
Here we consider the case of Gaussian selection and relax the assumption that the strength of selection is identical for all patches. We denote by $1/\sigma_{i}$ the strength of selection in patch $i$. Then, inserting equation \eqref{eq_rhop} into equation \eqref{eq_sp_general} gives
\begin{equation}
0=-x^*\sum\limits_{i=1}^{n}{c_i\frac{1}{\sigma_i^2}}+\int\limits_{\Omega}{f(\theta_1,..,\theta_n)\sum\limits_{i=1}^{n}{c_i\frac{\theta_i}{\sigma_{i}^2}}\mathrm{d}\theta_1..\mathrm{d}\theta_n},
\end{equation}
which, using the definition of the  spatial average and solving for $x^*$, gives 
\begin{equation} \label{eq_sp_sigma}
x^*=\frac{{\rm E_T}\left[{\rm E_S}\left[\frac{ \theta}{\sigma^2}\right]\right]}{\rm{E_S}\left[\frac{1}{\sigma^2}\right]}.
\end{equation}
For the branching condition we start with equation \eqref{eq_branch_general} and use the results from appendices \ref{app_invadability} and \ref{app_gaussian_selection}, except that now the $\sigma_{i}$ cannot be factored out from the spatial averages. In the absence of temporal correlations, we get
\begin{equation}
\begin{split}
&\sum\limits_{i=1}^{n}{c_i\frac{1}{\sigma_{i}^2}}<\int\limits_{\Omega}{f(\theta_1,..,\theta_n)\left[\sum\limits_{i=1}^{n}{c_i\left(\frac{\theta_i-x^*}{\sigma_{i}^2}\right)^2}-\left(\sum\limits_{i=1}^{n}{c_i\frac{ \theta_i-x^*}{\sigma_{i}^2}}\right)^2\right]\mathrm{d}\theta_1..\mathrm{d}\theta_n}\\
&+\gamma\int\limits_{\Omega}{f(\theta_1,..,\theta_n)\left(\sum\limits_{i=1}^{n}{c_i\frac{\theta_i-x^*}{\sigma_{i}^2}}\right)^2\mathrm{d}\theta_1..\mathrm{d}\theta_n}+
\sum_{i=1}^nc_i\left(\int_\Omega\frac{\theta_i-x^*}{\sigma_{i}^2}\mathrm{d}\theta_1..\mathrm{d}\theta_n\right)^2-
\left(\sum_{i=1}^nc_i\int_\Omega\frac{\theta_i-x^*}{\sigma_{i}^2}\mathrm{d}\theta_1..\mathrm{d}\theta_n\right)^2.
\end{split}
\end{equation}
With the definition of the spatial variance and using equation \eqref{eq_sp_sigma} we obtain the branching condition, 
\begin{equation} \label{eq_br_sigma}
{\rm E_T}\left[{\rm Var_S}\left[\frac{\theta-x^*}{\sigma^2}\right]\right]+\gamma\mathrm{Var}_{\rm T}\left[\mathrm{E}_{\rm S}\left[\frac{\theta}{\sigma^2}\right]\right]+2\frac{1-m}{m}{\rm Var_S}\left[{\rm E_T}\left[\frac{\theta}{\sigma^2}\right]\right]>{\rm E_S}\left[\frac{1}{\sigma^2}\right].
\end{equation}
In this case, the terms in the calculation of the spatial mean and variance are weighted by $1/\sigma_{i}^2$. Thus, patches in which selection is strong contribute relatively more to the total variance in selective optima and thus to branching.

\subsection{Special cases for Gaussian selection} \label{app_gauss_special_cases}

From equation \eqref{eq_br_gauss_noTC} (or equations \ref{eq:br_indep} or \ref{eq:br_indep_cov} for independently distributed patches), $\sigma^2_{\text{crit}}$ is readily computed for any combination of optima distributions. Empirically, the relevant expectations and variances can be estimated from sufficiently long time-series measurements.
In the following we compute the branching condition for several relevant examples, starting from simple to more complex cases. For classical models we recover the known conditions. 

\subsubsection{No spatial differences} \label{app_gauss_lottery}
 If the selective optimum is always equal across all patches, then the population effectively consists of a single patch. From equation \eqref{eq_br_gauss_noTC} the branching condition becomes
\begin{equation} \label{eq_lottery}
\gamma\mathrm{Var}_{\rm T}[\theta]>\sigma^2.
\end{equation}
For this special case we retrieve the lottery model of species coexistence \citep{Chesson:81}, for which it is known that temporal fluctuations in selection can lead to the evolution and maintenance of genetic variance if there is generation overlap \citep{Seger:87}. Our branching condition is consistent with earlier work (\citealp{Ellner:94}, equation 9; \citealp{Svardal:11}, equation 7).

\subsubsection{No temporal fluctuations} \label{app_gauss_levene_model}
 If selective optima in the patches are fixed, then the time-averages in condition \eqref{eq:sp} and condition \eqref{eq:br_indep_cov} disappear and the branching condition becomes 
\begin{equation} \label{eq:br_levene}
\frac{2-m}{m}\ {\rm Var_S}[ \theta]=\frac{2-m}{m}\sum\limits_{i=1}^{n}\sum\limits_{\substack{j=1\\j\neq i}}^{n}{c_ic_j(\theta_i- \theta_j)^2}>\sigma^2.
\end{equation}
For the special case of the Levene model ($m = 1$), this equals \citeauthor{Geritz:98}'s (\citeyear{Geritz:98}) equation B8.
 The right-hand side of equation \eqref{eq:br_levene} measures the sum of the squared pairwise differences between patches weighted by their relative output. Thus, for given difference between patches, the evolution and maintenance of genetic variation becomes more difficult if the patch sizes become more different. This is in accordance with results by  \citet{Gillespie:74}.

\subsubsection{Independently and identically distributed optima with two possible states} 
If the occurrence of environmental conditions follows an identical distribution in all patches, then the first term on the right-hand side of equation \eqref{eq:br_indep} equals zero. 
Here, we focus on the case that all patches are of identical size and can take the selective optima $\theta_A$ and $\theta_B$, with probabilities $p$ and $1-p$, respectively. We assume that spatial correlations are absent. 
Using equation \eqref{eq:br_indep}, we find for the branching condition
\begin{equation} \label{eq:br_iid}
\left(1-\frac{1-\gamma}{n}\right)(\theta_A-\theta_B)^2p(1-p)>\sigma^2.
\end{equation}
In the limit of infinitely many patches we retrieve the condition for the Levene model (equation \eqref{eq:br_levene} with $m=1$ and 2 patches of relative sizes $p$ and $1-p$). This is because the realized frequency of patch types in a given year approaches the expected frequency. For two patches and $\gamma=0$, the right-hand side of equation \eqref{eq:br_iid} is reduced by a factor 1/2 relative to the Levene model.

\subsubsection{Known distributions of selective optima} If the distribution of the selective optima is known, then this information can be used to obtain an explicit expression for the left-hand side of equation \eqref{eq:br_indep} or \eqref{eq:br_indep_cov}. 
For instance, if the selective optima in the patches are determined by many random processes of small effect, then they are approximately Gaussian distributed. Then $\rm{E}_{T}[ \theta]$ and $\rm{Var}_{T}[\theta]$ are given by the mean and variance of the Gaussian distributions in each patch, $\mu_{\theta i}$ and  $\sigma_{\theta i}^2$, respectively, and the branching condition for the case of equal juvenile carrying capacities is
\begin{equation} \label{eq:br_normal}
\frac{1}{n}\sum\limits_i\left(\mu_{\theta i}-\bar{\mu}_{\theta}\right)^2+\frac{n-1+\gamma}{n^2}\sum\limits_i\sigma_{\theta i}^2>\sigma^2,
\end{equation}
where $\bar{\mu}_{\theta}=\frac{1}{n}\sum{\mu_{\theta i}}$.

Another classic distribution is given when the optimum for the trait under consideration is determined by how often a certain event occurs during the selective period. This would for instance be the case if the optimum is determined by the number of predation events or the number of days with frost. If we assume that these events are independent and that they occur with a constant rate, then they are Poisson distributed. Also other environmental quantities such as the amount of rain per area are approximately Poisson distributed.
In such cases, the branching condition is given by equation \eqref{eq:br_normal} where $\mu_{ \theta i}$ and $\sigma_{ \theta i}^2$ are substituted by the rate parameter of the Poisson distribution.

\subsubsection{Spatial correlations}
It is straightforward to consider different patterns of spatial correlations in our framework. Here we give two simple examples for two equally sized patches with identically distributed environmental conditions. Furthermore, we assume that temporal autocorrelations are absent (${\rm \mathcal{C}}[\theta]=0$ in condition \ref{eq_br_gauss_noTC}). With this assumption, the third term on the left-hand side of condition \eqref{eq_br_gauss_noTC} is independent of spatial correlations. Hence, we can assume without loss of generality that $m=1$.

First, assume that the selective optima in the two patches are multivariate normal distributed with identical mean $\mu=0$ and variance $\sigma_e^2$, and that there are correlations between the patches  of strength $\omega_{\rm S}$. Then, the probability density function of the environmental state vector is given by
\begin{equation}
    f(\theta_1,\theta_2) =
      \frac{1}{2 \pi  \sigma_{e}^2 \sqrt{1-\omega_{\rm S}^2}}
      \exp\left(
        -\frac{1}{2\sigma_{e}^2(1-\omega_{\rm S}^2)}\left(
          \theta_1^2 +
          \theta_2^2 -
          2\omega_{\rm S}\theta_1\theta_2
        \right)
      \right).
\end{equation}
With this, the branching condition can be calculated from equation \eqref{eq_br_gauss_noTC} as
\begin{equation} \label{eq_br_multivariate}
\frac{\sigma_e^2}{2}(1-\omega_{\rm S}+\gamma(1+\omega_{\rm S}))>\sigma^2.
\end{equation}
We interpret this condition below.

Second, assume that the optima in the two patches are identically Bernoulli distributed. Specifically, we assume that in both patches the two selective optima $\theta_A$ and $\theta_B$ occur with probability $1/2$.
We measure spatial correlation with a parameter $\omega_{\rm S} \in [-1,1]$. A value $\omega_{\rm S}=0$ indicates the absence of correlation. Positive values of $\omega_{\rm S}$ linearly increase the probability that the two patches are equal in any generation and negative values linearly increase the probability that the patches are different.
For $\omega_{\rm S}=\pm1$ the optima in the patches are always equal or different, respectively. 
The branching condition calculates to
\begin{equation}\label{eq:br_sc}
\frac{1}{8}\left[(1-\omega_{\rm S})+\gamma(1+\omega_{\rm S})\right]( \theta_A- \theta_B)^2>\sigma^2.
\end{equation}
From both equation \eqref{eq_br_multivariate} and \eqref{eq:br_sc}, we see that spatial correlations have opposite effects on two parts of $\sigma^2_{\rm crit}$. First, positive (negative) correlations decrease (increase) spatial variance by making the patches less (more) likely to be different. Second, positive (negative) correlations increase (decrease) the amount of global temporal fluctuations by making it more (less) likely that patches change in concert. The former effect always out-weights the latter so that positive (negative) correlations generally 
hinder (promote) the evolution of genetic diversity. However, the effect of correlations decreases with increasing generation overlap. 
Furthermore, for completely negatively correlated selective optima between the patches, we retrieve the case of the Levene model with two equally sized patches, for no correlation we obtain the case of two patches with identically and independently distributed selective optima from above, and for completely positively correlated patches we obtain the condition from the lottery model.

\subsubsection{Temporal correlation}

Temporal correlations only affect the term ${\rm \mathcal{C}}\left[{\partial s}\right]$ in condition \eqref{eq_branch_general}, which is given in equation \eqref{eq_covt}. While this term appears rather complicated, it is readily computed for any special case. For illustration, we consider two scenarios.

First, assume  two patches with identically and independently Bernoulli distributed selective optima, specifically, in both patches the two same selective optima 0 and 1 occur with probability $1/2$.
Temporal correlations are measured with a parameter $\omega_{\rm T} \in [-1,1]$. We assume that with probability $1-|\omega_{\rm T}|$ a selective optimum is determined according to the Bernoulli distribution, whereas with a probability $|\omega_{\rm T}|$ the selective optimum in a patch is the identical to (different from) the previous time step if $\omega_{\rm T}>0$ ($\omega_{\rm T}<0$).
With these assumptions, it is easy to see that ${\rm E_T}\left[{\rm Var_S}\left[\theta\right]\right]={\rm Var_T}\left[{\rm E_S}\left[\theta\right]\right]=1/8$,
${\rm E_S}\left[{\rm Cov_T}\left[\theta_{t},\theta_{t+\tau}\right]\right]={\omega_{\rm T}}^\tau/4$,
and
${\rm Cov_T}\left[{\rm E_S}\left[\theta_{it}\right], {\rm E_S}\left[\theta_{it+\tau}\right]\right]={\omega_{\rm T}}^\tau/8$.
The branching condition, equation \eqref{eq_branch_general}, then equals
\begin{equation} \label{eq_br_NF_TC}
\frac{1+\gamma}{8}+\frac{\omega_{\rm T}(1-\gamma)(1-m)}{4(1-\omega_{\rm T}(1-m(1-\gamma)))}>\sigma^2.
\end{equation}
This relation is plotted in figure \ref{fig_TC}A for parameter values $\omega_{\rm T}=0$ (solid line), $\omega_{\rm T}=0.3,0.6,0.9$ (dotted lines, bottom to top) and $\omega_{\rm T}=-0.3,-0.6,-0.9$ (dashed lines, top to bottom).

Second, we consider a scenario in which the probability distributions differ between patches. In particular, we still assume that there are two Bernoulli distributed patches with possible optima $0$ and $1$, but now we take the probability of occurrence of optimum 1 to be 1/3 and 2/3 in patch one and two, respectively. Modeling temporal autocorrelation in the same way as above, we can compute the terms in equation \eqref{eq_br_gauss_noTC}, which yields ${\rm E_T}\left[{\rm Var_S}\left[\theta\right]\right]=5/36$, 
${\rm Var_T}\left[{\rm E_S}\left[\theta\right]\right]=4/9$, ${\rm Var_S}\left[{\rm E_T}\left[\theta\right]\right]=1/36$,
${\rm E_S}\left[{\rm Cov_T}\left[\theta_{t},\theta_{t+\tau}\right]\right]=2{\omega_{\rm T}}^\tau/9$,
and
${\rm Cov_T}\left[{\rm E_S}\left[\theta_{t}\right], {\rm E_S}\left[\theta_{t+\tau}\right]\right]={\omega_{\rm T}}^\tau/9$.
Hence, the branching condition becomes
\begin{equation} \label{eq_br_F_TC}
\frac{5}{36}+\frac{4 \gamma }{9}+\frac{1-m}{18 m}+\frac{2\omega_{\rm T}(1-\gamma)(1-m)}{9(1-\omega_{\rm T}(1-m(1-\gamma)))}>\sigma^2.
\end{equation}
This relation is plotted in figure \ref{fig_TC}B  for parameter values $\omega_{\rm T}=0$ (solid line), $\omega_{\rm T}=0.3,0.6,0.9$ (dotted lines, bottom to top) and $\omega_{\rm T}=-0.3,-0.6,-0.9$ (dashed lines, top to bottom).

\subsection{Protected polymorphism} \label{app_prot_poly}

In this appendix, we show that, as soon as the left-hand side of our branching condition (equation \ref{eq_branch_general})  is positive, pairs of trait values can be found that are able to coexist in a protected dimorphism. However, at least locally around the singular point $x^*$, if inequality \eqref{eq_branch_general} is not fulfilled, such a dimorphism can be invaded and replaced by the singular strategy. We consider the case of Gaussian selection.

\begin{figure}[H]
\begin{center}{\setlength{\unitlength}{1cm}
	\begin{picture}(8,8.)
	\put(0,4.8){\includegraphics[width=3.5cm,keepaspectratio=true,clip=true]{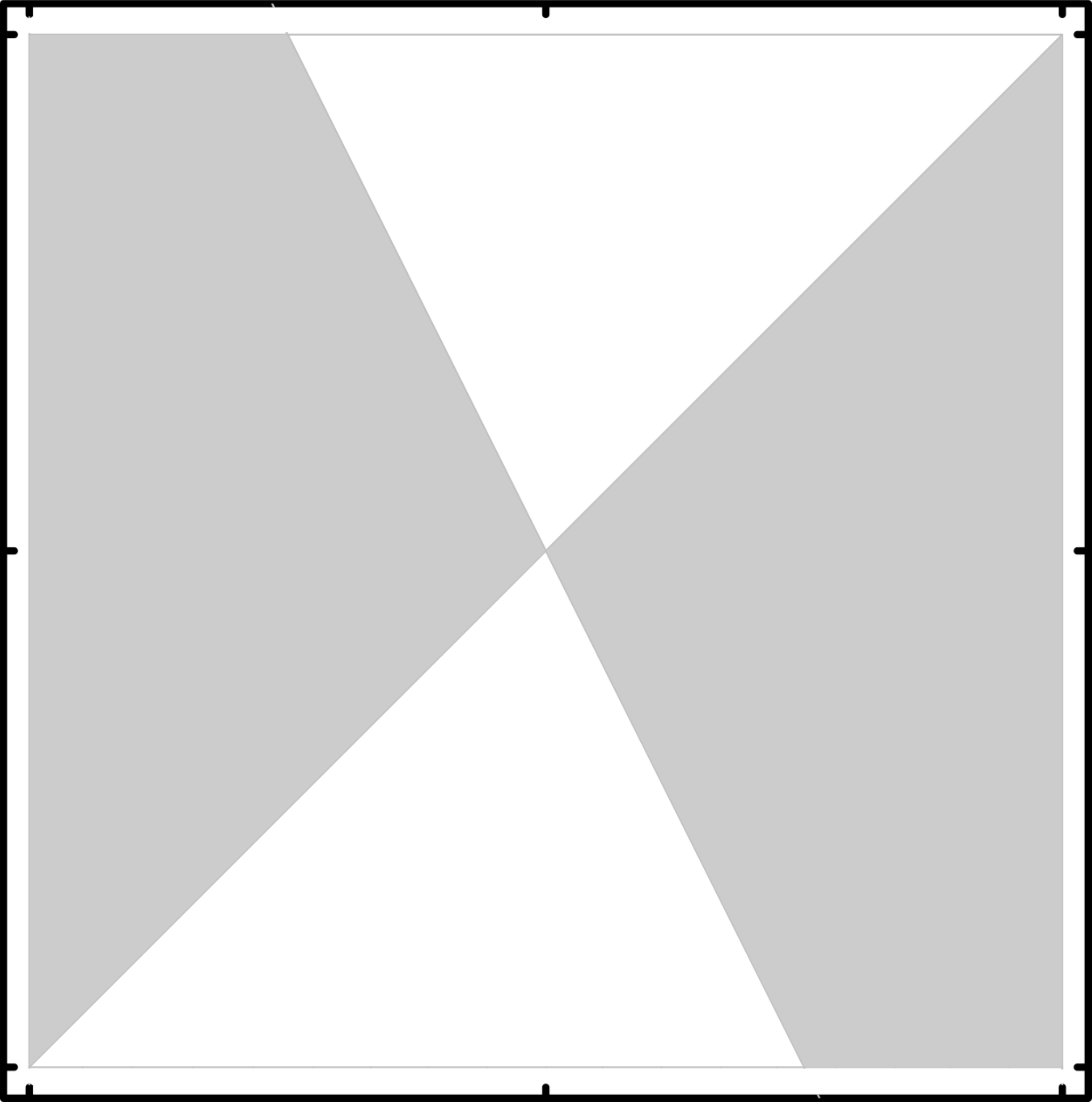}}
	\put(4,4.8){\includegraphics[width=3.5cm,keepaspectratio=true, clip=true]{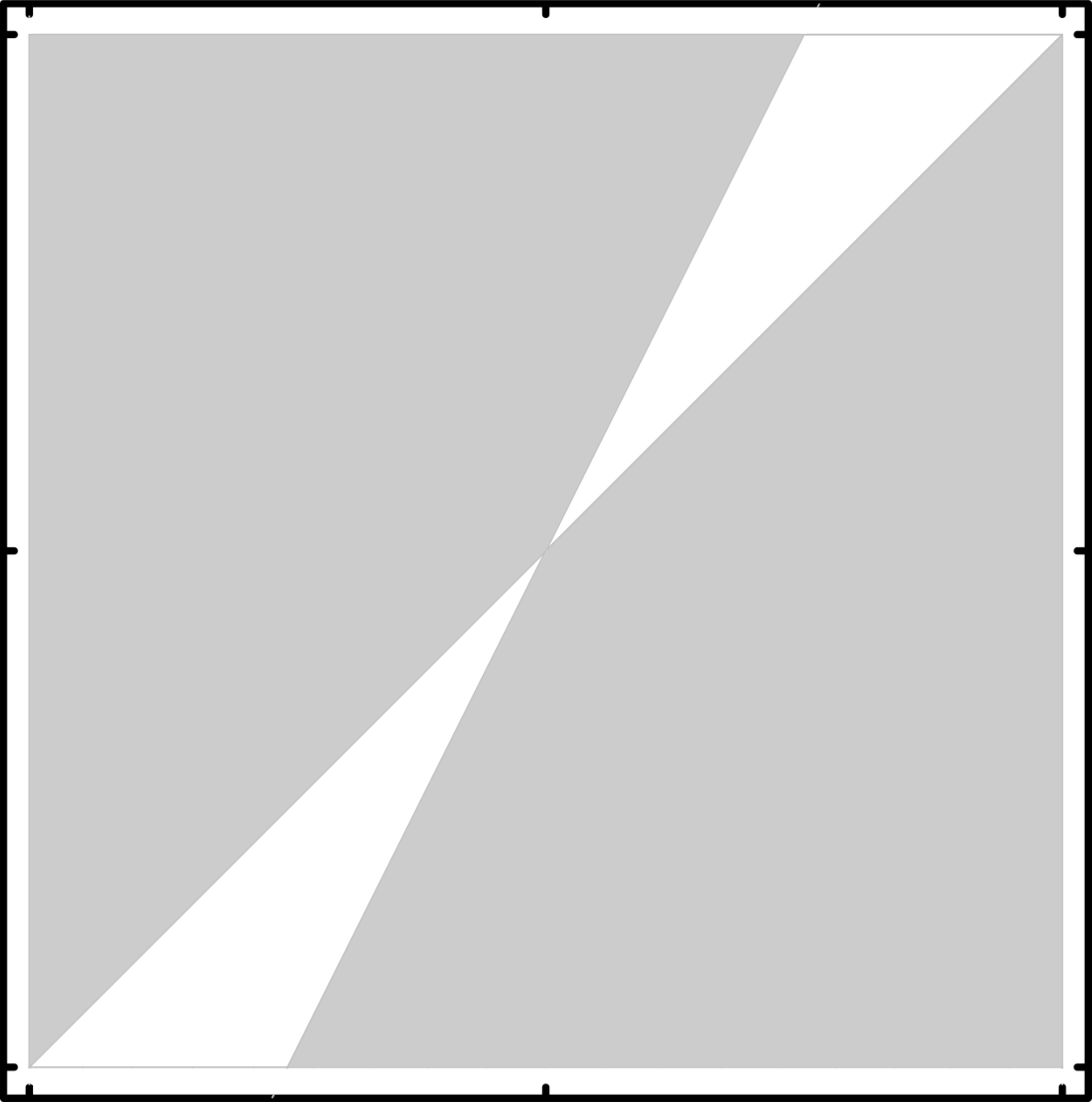}}
	\put(0,.3){\includegraphics[width=3.5cm,keepaspectratio=true,clip=true]{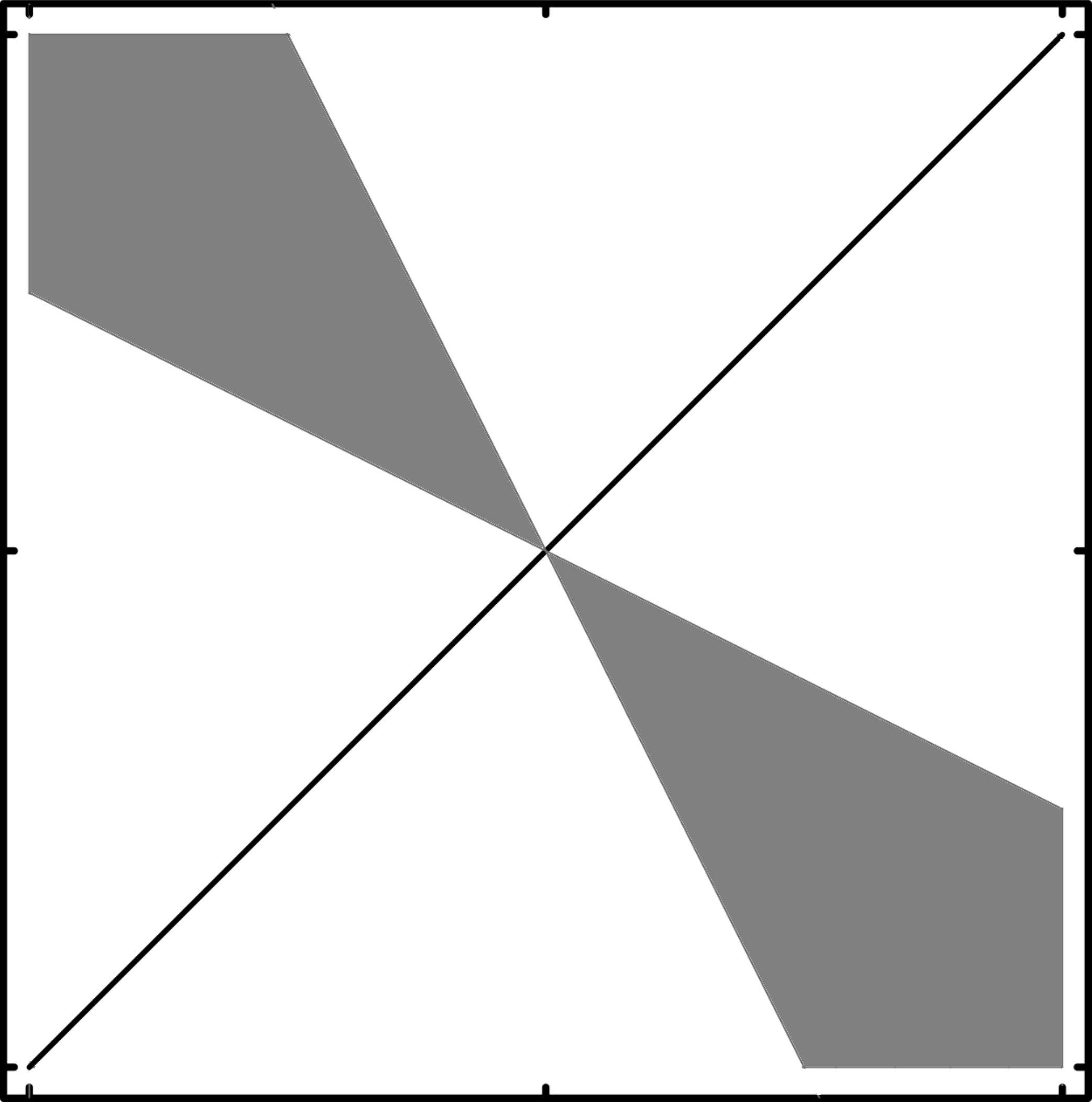}}
	\put(4,.3){\includegraphics[width=3.5cm,keepaspectratio=true, clip=true]{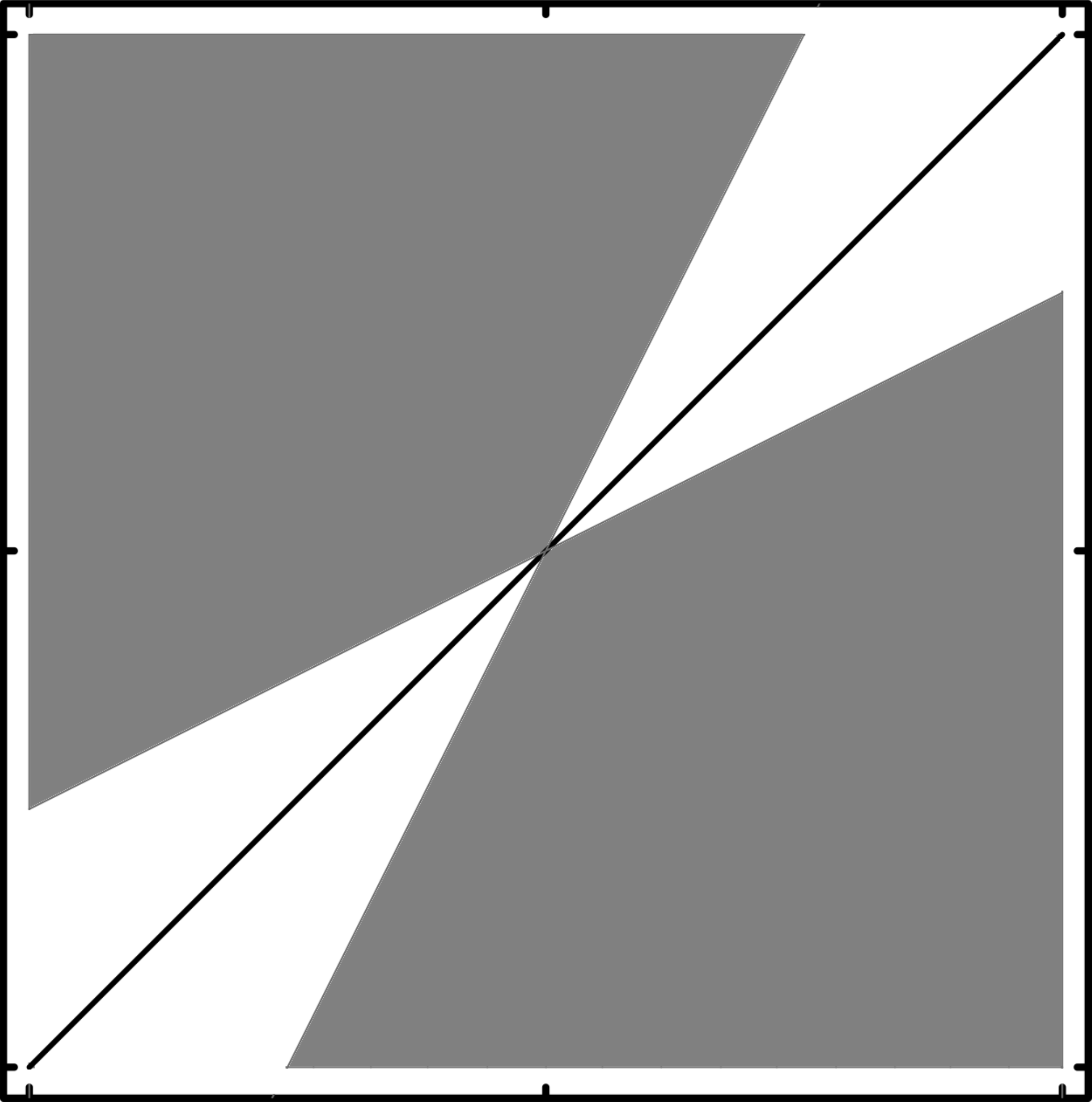}}
	\put(1.6,4.4){\Large $x^{*}$}
	\put(5.6,4.4){\Large $x^{*}$}
	\put(3,4.4){\Large $x$}
	\put(7,4.4){\Large $x$}
	\put(-.5,7.8){\Large $y$}
	\put(2.9,-.1){\Large $x_{1}$}
	\put(6.9,-.1){\Large $x_{1}$}
	\put(-.7,3.3){\Large $x_{2}$}
	\put(0.2,7.8){\bf A}
	\put(4.2,7.8){\bf B}
	\put(0.2,3.3){\bf C}
	\put(4.2,3.3){\bf D}
	\thicklines
	\put(1,1.3){\vector(1,1){.1}}
	\put(5,1.3){\vector(1,1){.1}}
	\put(2.5,2.8){\vector(-1,-1){.1}}
	\put(6.5,2.8){\vector(-1,-1){.1}}
	\thinlines
	\thicklines
	\put(1,2.8){\vector(1,0){.3}}
	\put(1,2.8){\vector(0,-1){.3}}
	\put(2.5,1.3){\vector(-1,0){.3}}
	\put(2.5,1.3){\vector(0,1){.3}}
	\put(5,2.8){\vector(-1,0){.3}}
	\put(5,2.8){\vector(0,1){.3}}
	\put(6.5,1.3){\vector(1,0){.3}}
	\put(6.5,1.3){\vector(0,-1){.3}}
	\thinlines
	\put(.7,6.7){\Large $+$}
	\put(2.5,6.25){\Large $+$}
	\put(1.7,7.5){\Large $-$}
	\put(1.45,5.5){\Large $-$}
	\put(4.7,6.7){\Large $+$}
	\put(6.5,6.25){\Large $+$}
	\put(6.5,7.7){\Large $-$}
	\put(4.7,5.3){\Large $-$}
	\end{picture}}
\end{center}
\caption{(A, B) Representative pairwise invadability plots (PIPs) for the two types of singular points allowing for protected dimorphism that are possible in our model. PIPs are contour plots of the invasion fitness function $w(y,x)$ with a single contour line at height zero. The abscissa gives the resident trait value $x$ and the ordinate gives the mutant trait value $y$. Combinations of resident and mutant trait values located in the gray region, labelled with $+$, indicate that for these types $w(y,x)>0$ while combinations of resident and mutant trait values located in the white region, labelled with $-$, indicate that for these two types $w(y,x)<0$. Panel A shows a singular point that is an attractor of the evolutionary dynamics and uninvadable by nearby mutants and thus an evolutionary endpoint. Panel B shows a singular point that is an evolutionary branching point. (C,D) Coexistence plots for two types in the neighborhood of the singular point $x^*$ corresponding to the PIPs directly above. These plots can be derived by plotting the contour plots for $w(y,x)$ and $w(x,y)$ on top of each other. 
The abscissa gives the trait value $x_1$ of one type and the ordinate gives the trait value $x_2$ of a second type. Combinations of types where $w(x_1,x_2)>0$ and $w(x_2,x_1)>0$, i.e., where the two types can coexist in a protected dimorphism, are shown gray. Arrows indicate the direction of gradual evolutionary change. The arrows on the diagonal indicate that for $x_1=x_2$, i.e., for monomorphic evolution, the singular point is an attractor in both cases.  For more information about the use of PIPs to derive the direction of monomorphic and dimorphic evolution see \citet{Metz:96a}, \citet{Geritz:98} and \citet{Diekmann:04}.}
\label{fig_pip}
\end{figure}

\citet{Metz:96a} and \citet{Geritz:98} showed that singular points in one-dimensional trait spaces can generically be classified into eight different configurations. These configurations can be visualized in terms of pairwise invasibility plots (PIPs, figure \ref{fig_pip}A,B). In our model, four out of these eight configurations are impossible. To see this, note that in our model the singular point is always an attractor of the evolutionary dynamics (cf. appendix \ref{app_special_cases}). In terms of PIPs, this means that to the left of the singular point we have a plus-region above the diagonal and a minus-region below the diagonal while to the right of the singular point this pattern is reversed. This rules out four of the possible eight configurations. From the remaining four configurations only two allow for protected dimorphism
(figure \ref{fig_pip}A,B). For these PIPs, one can show that the set of pairs of types that are able to coexist locally around a singular point is non-empty (figure \ref{fig_pip}C,D). Mathematically speaking, this is the case if and only if
\begin{equation} \label{eq_prot_poly}
\frac{\partial^{2} w(y,x^*)}{\partial y^2}\bigg|_{y=x^{*}}>-\frac{\partial^{2} w(x^*,x)}{\partial x^2}\bigg|_{x=x^{*}}
\end{equation}
\citep{Metz:96a,Geritz:98}. The left-hand side in this inequality is given by equation \eqref{eq_hessian_general} and the right-hand side takes an analogous form where all derivatives with respect to $y$ are replaced by derivatives with respect to $x$. The derivatives $(\partial_y\rho_{it})^*$ appearing in equation \eqref{eq_hessian_general} are given by equations \eqref{eq_rhop} and \eqref{eq_rhopp}. Analogously, we obtain
\begin{equation}
(\partial_x\rho_{it})^*=-\frac{\theta_{it}-x^*}{\sigma^2}
\end{equation}
and
\begin{equation}
(\partial^2_x\rho_{it})^*=\frac{(\theta_{it}-x^*)^2}{\sigma^4}+\frac{1}{\sigma^2}.
\end{equation}
Inserting these expressions into inequality \eqref{eq_prot_poly} we obtain
\begin{equation}
2\sigma^2_{\rm crit}>0,
\end{equation}
where $\sigma^2_{\rm crit}$ is the left-hand side of condition \eqref{eq_br_gauss_noTC}. Thus, in our model, if the the left-hand side of equation (14) is positive,, then pairs of types that are located symmetrically around the singular point can surely coexist in a protected dimorphism. We can see from figure \ref{fig_pip} that coexistence is not restricted to symmetric pairs but is possible, depending on parameter values, also for pairs that are slightly asymmetric (figure \ref{fig_pip}C) or even for pairs that are located on the same side of the singular point (figure \ref{fig_pip}D). Note, that the case of $\sigma^2_{\rm crit}=0$ corresponds to the degenerate configuration where the non-diagonal zero-contour line in the PIP has a slope of $-45^{\circ}$. In this case, the set of pairs of types that can coexist locally around a singular point is empty and types that are located exactly symmetrically around the singular point are selectively neutral with respect to each other.

How do the two possible configurations of singular points differ from each other? The PIP in figure \ref{fig_pip}A shows a singular point that is uninvadable (condition \ref{eq_br_gauss_noTC} not fulfilled) while the 
PIP in figure \ref{fig_pip}B shows a singular point that is invadable by nearby mutants and thus shows an evolutionary branching point (condition \ref{eq_br_gauss_noTC} fulfilled). From the classification of singular points it is also known that locally around a singular point the direction of the coevolutionary dynamics of two coexisting types points towards the singular strategy if the singular point is uninvadable (figure \ref{fig_pip}C). That is, selection is convergent in this case, while selection is divergent if the singular point is a branching point (figure \ref{fig_pip}D). Thus, in the first case a polymorphism is protected only on the ecological timescale but not on the evolutionary timescale where the polymorphism is expected to be eventually replaced by a single phenotype adopting the singular strategy. If, however, the singular point is a branching point, then a protected dimorphism can emerge from a monomorphism. Furthermore, the types present in the dimorphism are expected to subsequently evolve away from each other leading to increased genetic variance.

\setlength{\unitlength}{\textwidth}


\renewcommand{\thetable}{D\arabic{table}}
\renewcommand{\thefigure}{D\arabic{figure}}
\setcounter{table}{0}
\setcounter{figure}{0}
\setcounter{section}{3}
\section{Online Supplementary: Additional methods and results}
\label{online_app}

\subsection{Numerical calculation of $\sigma^2_{ \rm{crit}}$} \label{app_num_calc}

To check the accuracy of our analytical formulas, we also
calculate $\sigma^2_{ \rm{crit}}$ numerically. For this, we compute invasion fitness of a mutant close to the resident located at the singular point for different values of $\sigma^2$ and check below which value of $\sigma^2$ invasion fitness is positive and
the fitness landscape thus has a local minimum at the singular point. Invasion fitness is calculated using the procedure given in \citet{Metz:08c} and, for given parameters, $\sigma^2_{ \rm{crit}}$ is numerically approached using a method of nested intervals. The commented Mathematica \citep{Mathematica:11} source code is given in  online appendix \ref{app_sc_num}.

\begin{figure}[H]
\begin{center}
\begin{tabular}{>{\centering\arraybackslash} m{.43\textwidth}>{\centering\arraybackslash} m{.58\textwidth}}
\begin{picture}(0.4,0.32)
\put(0,0.01){\includegraphics[width=0.43\textwidth]{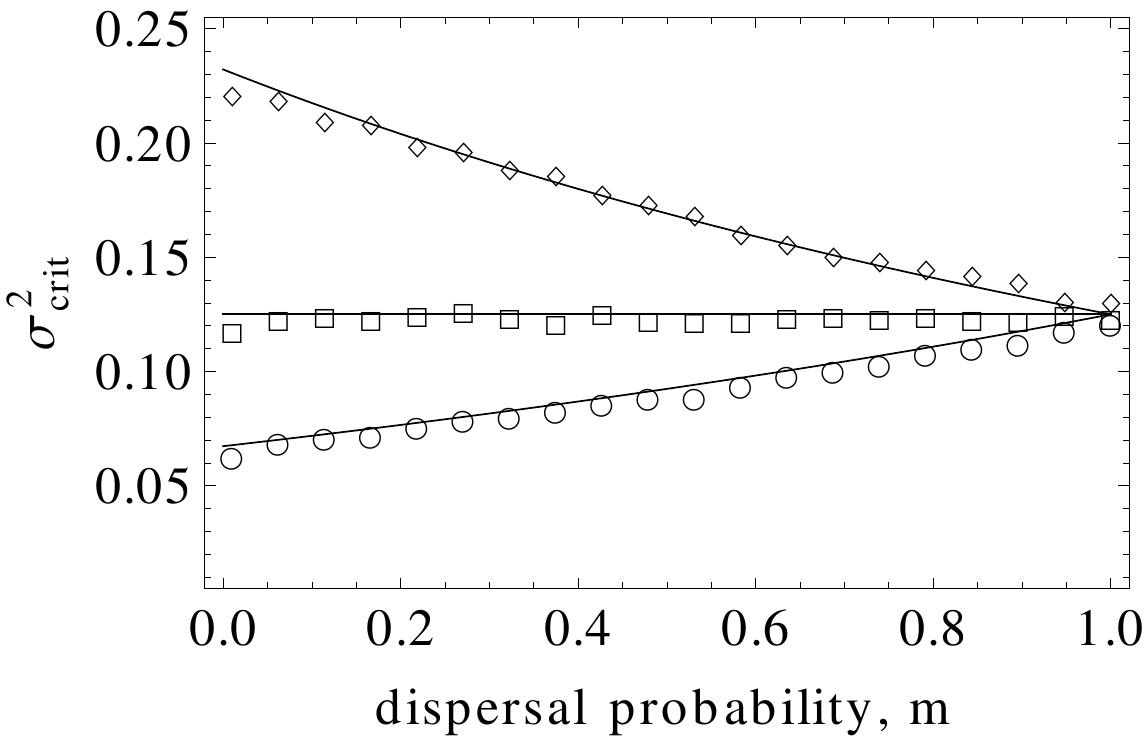}}
\put(0.03,0.31){\textbf A}
\end{picture}
&
\begin{picture}(0.6,0.32)
\put(0,0){\includegraphics[width=0.62\textwidth]{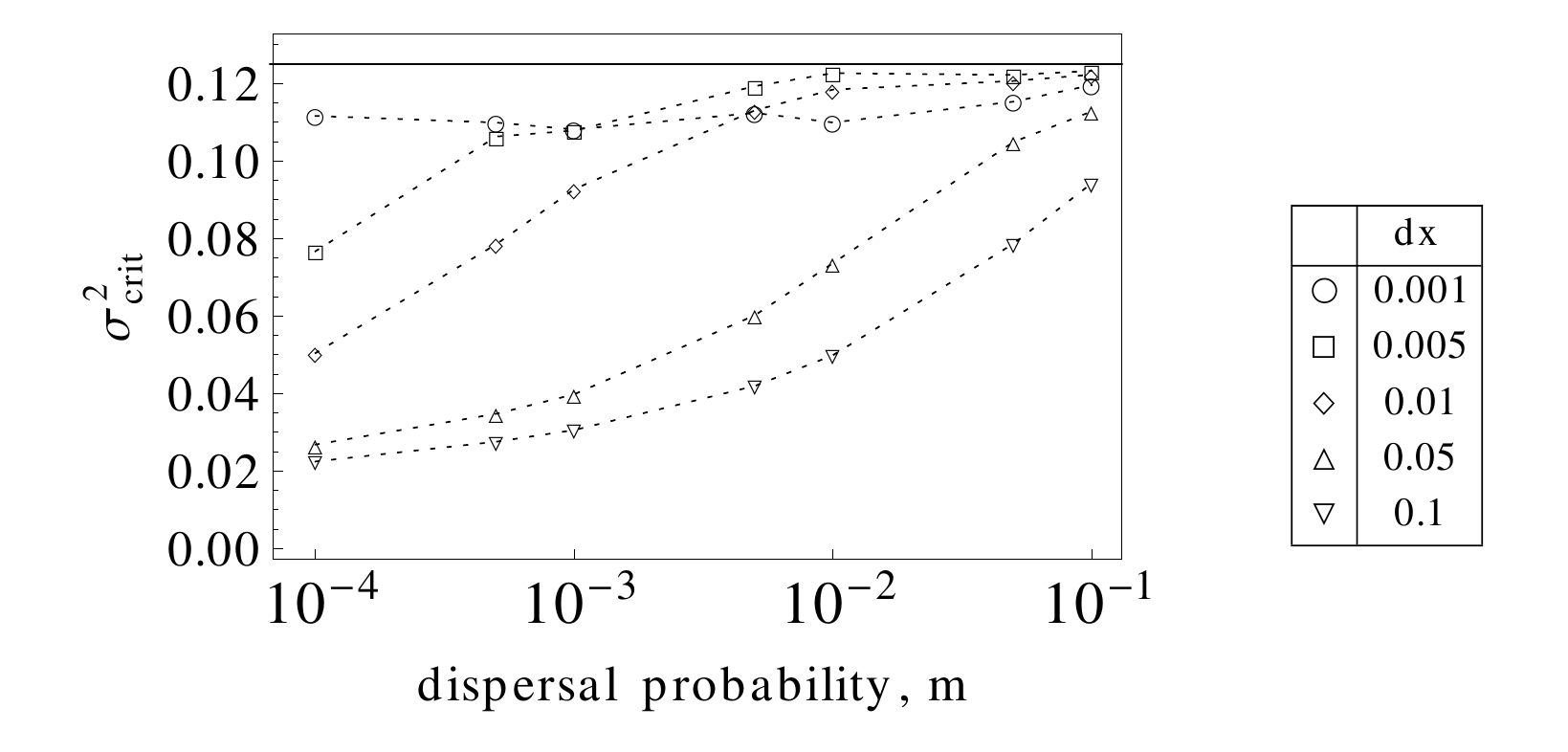}}
\put(0.03,0.31){\textbf B}
\end{picture}
\end{tabular}
\end{center}
\begin{center}

\end{center}
\caption{Critical strength of selection below which selection is disruptive  as a function of the dispersal probability, $m$. We assume  two independently identically Bernoulli distributed patches. (A) We compare the analytical formula for $\sigma^2_{\rm crit}$ (lines, equation \ref{eq_br_NF_TC}) to numerical calculations (plot markers). Circles, squares and diamonds correspond to a coefficient of temporal correlation of $\omega_{\rm T}=-0.3$, $0$ and $0.3$, respectively. Generally the results are in good agreement, except for very small $m$. (B) Here, we concentrate on the case without temporal correlations and zoom into the parameter region of very small $m$ (log-scale). The different plot markers vary the distance ${\rm d}x$ between resident at the singular point and mutant. The horizontal solid line gives the analytical expectation. The critical strength of selection decreases with $m$ for very small $m$. This effect becomes stronger as ${\rm d}x$ increases.
Parameters: Each data point averaged over 300 runs in (A) and 100 runs in (B). (A,B) $T=10^5$, $\gamma=0$. Other parameter values  as given in online appendix \ref{app_sc_num}. 
}
\label{fig_anavsnum}
\end{figure}

In Figure \ref{fig_anavsnum} we compare some of our analytical results from figure \ref{fig_TC}A to numerical calculations. We find that the results are in close agreement. This is also true for figure \ref{fig_TC}B (not shown). Discrepancy only occurs for very strong temporal correlation (not shown) and for very small dispersal probabilities (figure \ref{fig_anavsnum}B). The former represents the fact that for the numerical calculation we approximate the infinite time average by a long time series and that this approximation becomes increasingly unreliable as temporal correlations increase. The latter 
stems from a discrepancy between the assumptions  in the analytical and numerical approaches: In the analytical treatment we derive a condition for disruptive selection locally around the singular point, but for the numerical calculations we need to assume a discrete distance between mutant and resident. In the case of identically distributed patches, the range of mutants that can coexist with the singular strategy decreases  with decreasing dispersal probability $m$. As a consequence, the numerical value of $\sigma^2_{\text{crit}}$ is smaller than predicted by the analytical results for very small $m$. This effect becomes stronger the larger the distance between mutant and resident  in the numerical calculations (figure \ref{fig_anavsnum}B).

\subsection{Individual based simulations}\label{app_sim}

Extensive individual based computer simulations are performed for two main reasons: (1) The robustness of all analytical results is tested against violation of the adaptive dynamics assumptions. In particular, we investigate the effect of population size, mutation rate, mutational effect size, mode of reproduction and number of loci. (2) The structure of the genetic polymorphism after branching is investigated. Simulations are performed on the Vienna Scientific Cluster (VSC) using Matlab R2011a (Mathworks \citeyear{Mathworks:11}). The commented source code of the simulation function is given in section \ref{app_sc_ind}.

The implementation of the model follows closely the model description given in the main text. It can be summarized by the following steps. For each adult individual, a random draw determines whether it survives to the next time step or dies. The number of offspring gametes of each adult is determined according to equation \eqref{eq:selection}, where the trait of an individual is the sum of the values at each locus. We assume that each locus can take a set of discrete values on the real axis. The distance between two values is a parameter -- the minimal mutational step size. We assume that the loci are equally spaced on linear chromosomes. From the total number of offspring gametes in each patch, we draw the gametes that replace the dying adult individuals. All other offspring die. With a certain probability, the surviving gametes are the product of one or more recombination events between the parental chromosomes. The surviving gametes of each patch fuse to form diploids. With a certain probability, the values at a locus of an individual have undergone mutation. Mutations are drawn from a Gaussian distribution with a certain variance and are rounded to the minimal mutational step size. We also investigate the case of fixed mutational step-sizes.  With probability $m$ the newly established individuals disperse globally between patches (note that we assume that $k_i=c_i$). 
Alternatively, we also investigate the case that gametes disperse instead of zygotes; we do  not detect any difference in the results. 
We also treat the case of haploid clonal reproduction where the steps of gamete fusion and recombination are simply omitted. 

We investigate many different parameter combinations. Table \ref{tbl_sim_params} gives an overview of the most important parameters. The total set of parameters is commented in the source code below. The most important results of the simulation study are discussed in the second half of the results section.

\subsection{Supplementary figures} To restrict the parameter space, all simualtion figures in the main text assumed full dispersal ($m=1$). Figure \ref{fig_restricted_Fixed} investigates the accuracy of the analytical branching conditions for different values of the dispersal parameter. The observed increase in genetic variance in the simulation study generally matches the branching condition. For identically distributed patch optima and very low dispersal (figure \ref{fig_restricted_Fixed}A, $m<0.1$) the increase in genetic variance is rather slow and the accuracy is thus difficult to judge. A possible reason for this is that the maximum phenotypic distance between genotypes that can coexist decreases with decreasing $m$ (cf. figure \ref{fig_anavsnum}B) and that environmental stochasticity is strong in such a scenario.

\setlength{\unitlength}{\textwidth}
\begin{figure}[H]
\begin{center}
\begin{tabular}{>{\centering\arraybackslash} m{.43\textwidth}>{\centering\arraybackslash} m{.58\textwidth}}
\begin{picture}(0.49,0.32)
\put(0,0){\includegraphics[width=0.45\textwidth]{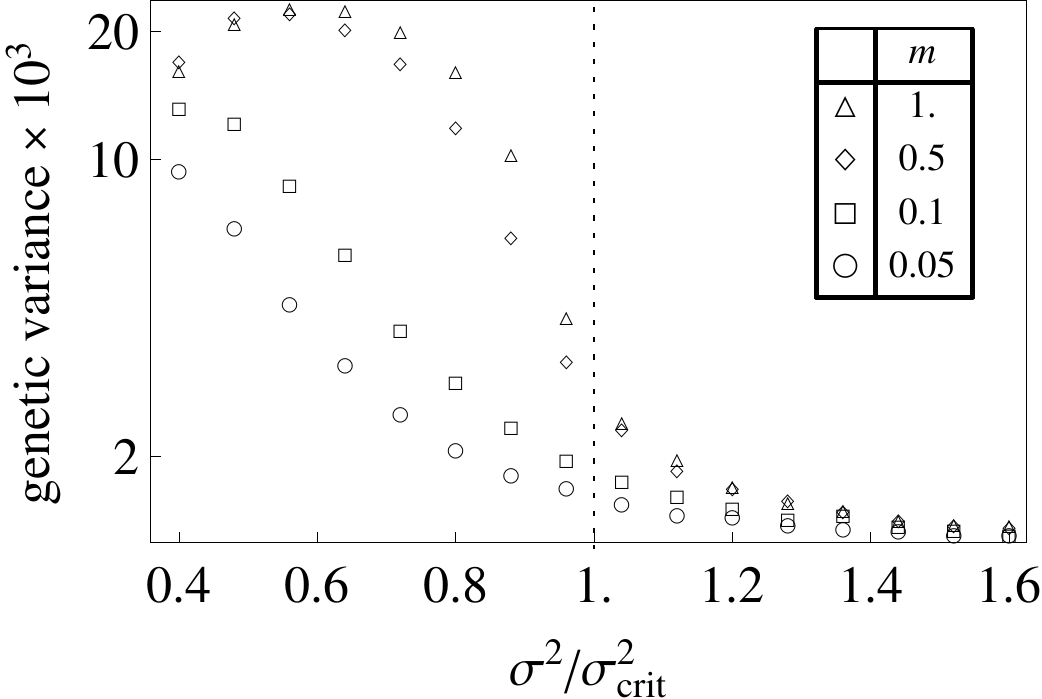}}
\put(0.03,0.31){\textbf A}
\end{picture}
&
\begin{picture}(0.495,0.32)
\put(0,0){\includegraphics[width=0.45\textwidth]{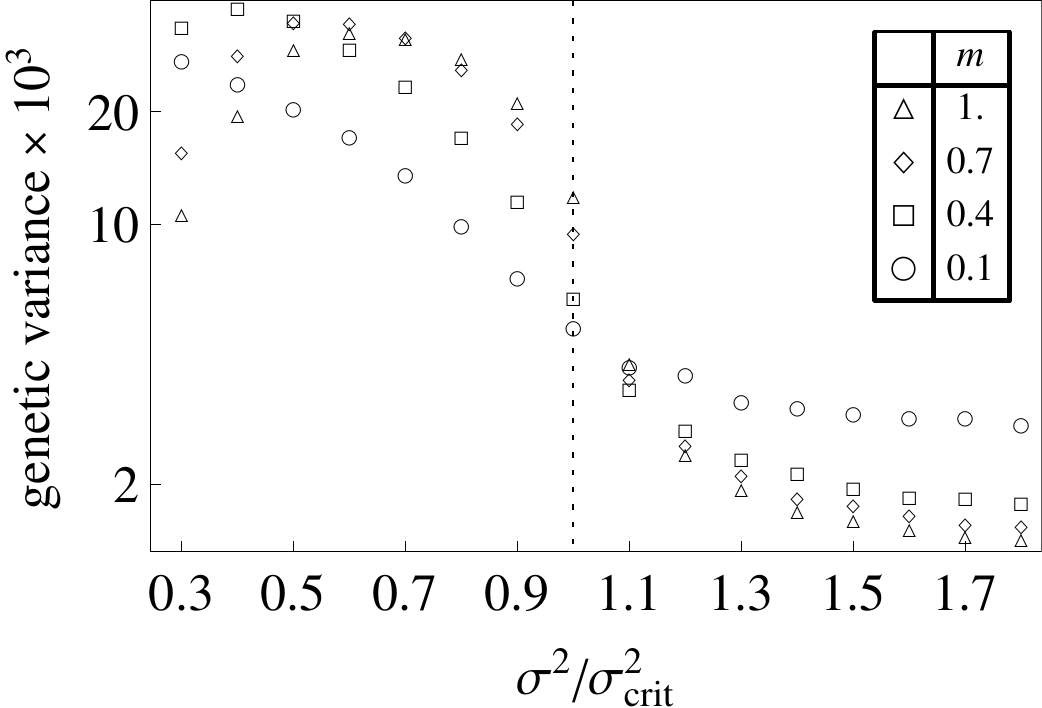}}
\put(0.03,0.31){\textbf B}
\end{picture}
\end{tabular}
\end{center}
\caption{
Long-term average genetic variance (on log-scale) measured from individual based simulations as a function of the strength of Gaussian stabilizing selection for (A) identically and (B) differently  distributed selective optima across two patches (cf. figure \ref{fig_TC}A vs. B). 
Spatial and temporal correlations are absent. 
Recall that for different environmental distributions (panel B) $\sigma^2_{\rm{crit}}$ depends on $m$, while for identically distributed patch optima (panel A) it is independent of $m$.
The dotted vertical line indicates the analytically expected branching point  $\sigma^2/\sigma^2_{\rm{crit}}=1$ (cf. equation \ref{eq_br_gauss_noTC}). 
The analytical condition for evolutionary branching matches with the observed increase in genetic variance in individual based simulations. 
As an exception, for identically distributed patches and very low dispersal (panel A, $m \le0.1$, squares \& circles) the precise onset of the increase in genetic variance is difficult to judge and  the amount of genetic variation is significantly reduced.
Parameter values: $\gamma=0$, $k=2$, $rec=0.05$, $\mu_{\rm trait}=0.001$; other parameters as given in table \ref{tbl_sim_params}.
}
\label{fig_restricted_Fixed}
\end{figure}

Figure \ref{fig_popsize_mP0p001} and \ref{fig_popsize_appendix} show additional simulation runs investigating the influence of population size on branching. The plots are analogous to figure \ref{fig_popsize}A and figure \ref{fig_popsize}B, respectively, except for some parameter variations. See plot legends for details. Overall, the results are very similar to figure \ref{fig_popsize}. As an exception, we observe that in the case of  lower mutation rates and smaller mutational effect sizes, deviations from our branching conditions appear already for smaller total population sizes (figure \ref{fig_popsize_mP0p001}A).

\setlength{\unitlength}{\textwidth}
\begin{figure}[H]
\begin{center}
\begin{tabular}{>{\centering\arraybackslash} m{.43\textwidth}>{\centering\arraybackslash} m{.58\textwidth}}
\begin{picture}(0.49,0.31)
\put(0,0){\includegraphics[width=0.45\textwidth]{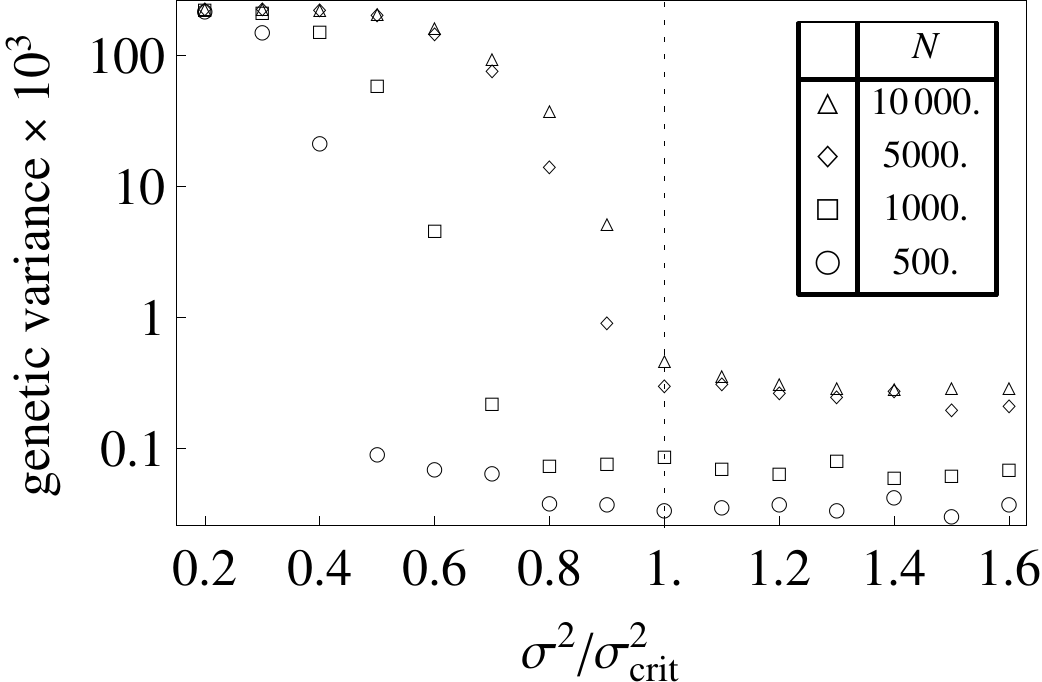}}
\put(0.03,0.3){\textbf A}
\end{picture}
&
\begin{picture}(0.49,0.3)
\put(0,0){\includegraphics[width=0.44\textwidth]{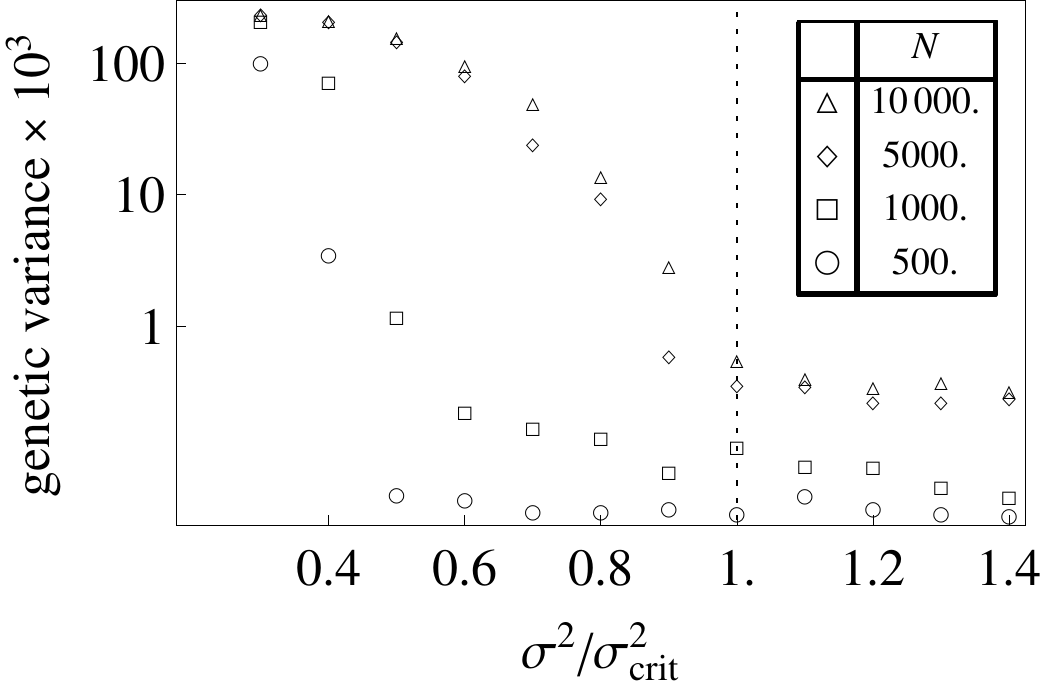}}
\put(0.03,0.3){\textbf B}
\end{picture}
\end{tabular}
\end{center}
\caption{ 
Long-term average genetic variance in a population (on log-scale) from individual-based simulations as a function of the strength of Gaussian stabilizing selection for different population sizes.  The plots are equivalent to figure  \ref{fig_popsize}A except for the following parameter variations. Panel (A) assumes lower mutation rate ($\mu_{\rm trait}=0.001$) and smaller mutational effect-size ($\sigma_\mu=0.01$). Panel (B) assumes differently distributed selective optima across patches (cf. figure \ref{fig_TC}B), a lower dispersal probability ($m=0.5$) and a higher number of patches ($n=10$). (A) For smaller mutational parameters, deviations from our branching conditions appear already for $N= 1000$. (B) Results are similar to \ref{fig_popsize}A. Parameter values: (A) $\gamma=0.5$, (B) $\gamma=0$; other parameters as in figure  \ref{fig_popsize}A.}
\label{fig_popsize_mP0p001}
\end{figure}

\setlength{\unitlength}{\textwidth}
\begin{figure}[H]
\begin{center}
\begin{tabular}{>{\centering\arraybackslash} m{.49\textwidth}>{\centering\arraybackslash} m{.49\textwidth}}
\begin{picture}(0.49,0.31)
\put(0,0){\includegraphics[width=0.45\textwidth]{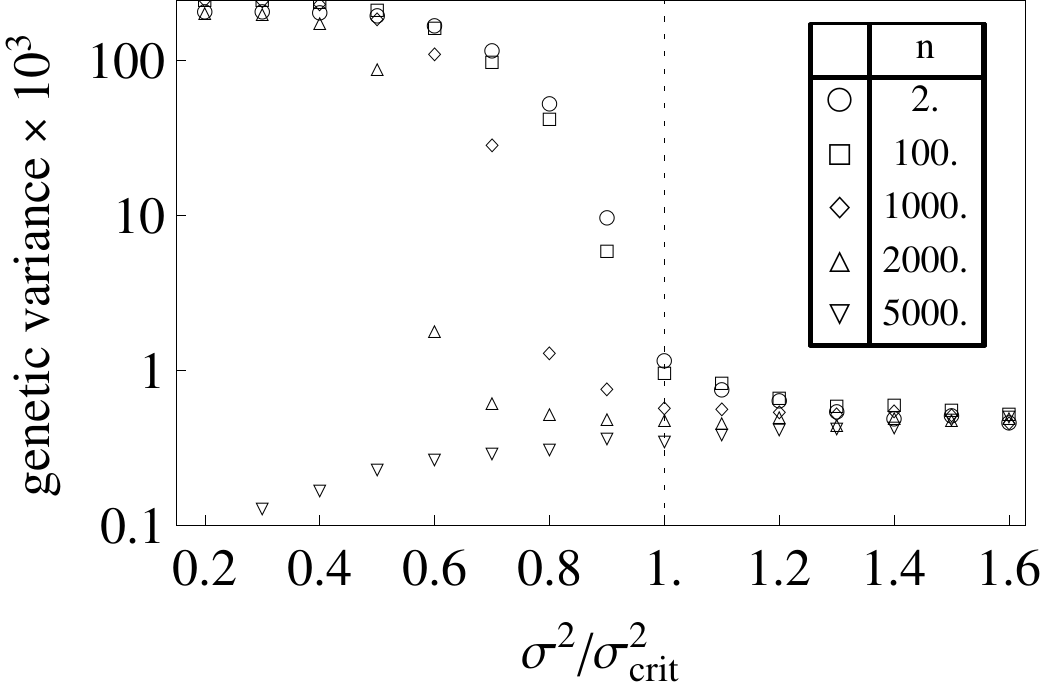}}
\put(0.03,0.3){\textbf A}
\end{picture}
&
\begin{picture}(0.49,0.3)
\put(0,0){\includegraphics[width=0.44\textwidth]{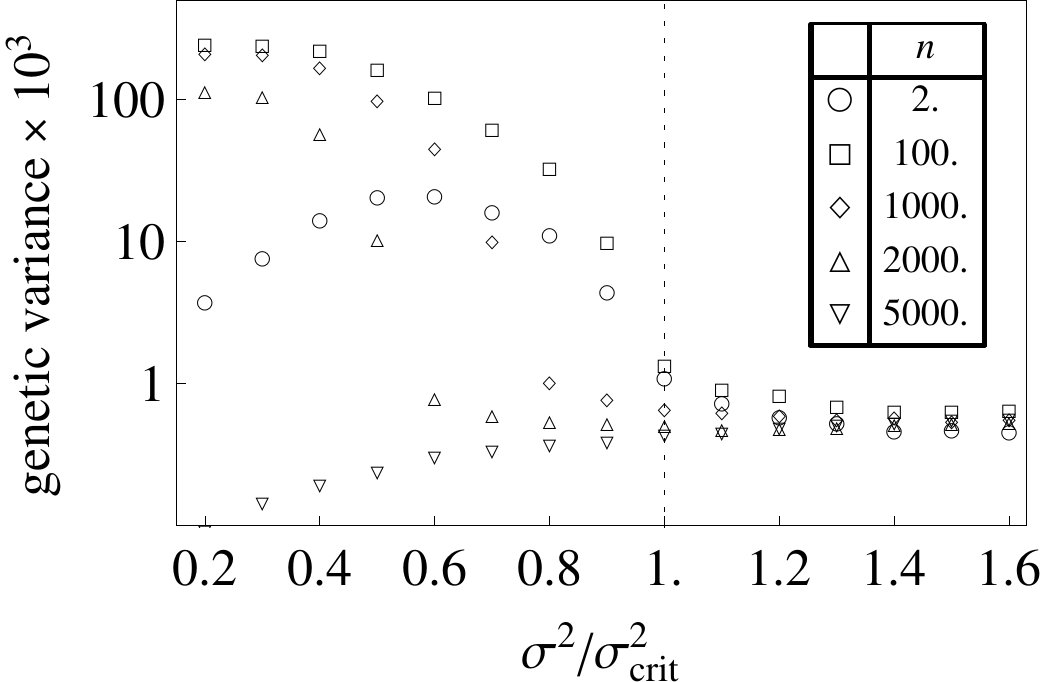}}
\put(0.03,0.3){\textbf B}
\end{picture}
\end{tabular}
\end{center}
\caption{ 
Long-term average genetic variance in a population (on log-scale) as a function of the strength of Gaussian stabilizing selection for different patch numbers at a constant total population size. The plots are equivalent to figure  \ref{fig_popsize}B except for the following parameter variations. Panel (A) assumes lower mutation rate ($\mu_{\rm trait}=0.001$) and smaller mutational effect-size ($\sigma_\mu=0.01$). In panel (B) generation overlap is absent ($\gamma=0$). Results are qualitatively similar to figure \ref{fig_popsize}B. Deviations from the branching condition can only be detected for very small local population sizes ($n>1000$, less than 10 individuals per patch). We added a data point for two individuals per patch ($n=5000$). Under these conditions there is no negative frequency dependence and thus no adaptive genetic polymorphism.
Other parameter values as in figure \ref{fig_popsize}B.}
\label{fig_popsize_appendix}
\end{figure}

\subsection{Source code for numerical calculations} \label{app_sc_num}

Below we give the source code used for numerical calculations of $\sigma^2_{\text{crit}}$ using Mathematica \citep{Mathematica:11}. Lines 4-5 give the definition of $r(x, \theta_{it})$ (equation \ref{eq:selection}) and lines 11-16 give the definition of ${\sf L}(x,\phi_{1t},..,\phi_{nt},\theta_{1t},..,\theta_{nt})$ (equations \ref{lji} and \ref{lii}). Note that we use $\Theta=(\theta_1,..,\theta_n)$. Lines 20-32 define a function that computes $w(y,x)$ (the variables are called $xm$ and $xr$) with the method described in \citet{Metz:08c} and lines 38-47 give a function using a nested intervals method to calculate $\sigma^2_{\text{crit}}$ for given parameter values. Explanation of the parameters and variables are given as comments in the code above each function.
Note that the function to calculate $\sigma^2_{\text{crit}}$ presented here assumes that there are no temporal and spatial correlations. Adapting the function to a given type of correlations is straight forward. Lines 51-54 give a typical calculation of the critical strength of selection as a function of the dispersal probability $m$. If not otherwise stated the given parameters are used for all calculations.

[Please refer to the online supplementary of the pubslished version for the source code, or contact the authors directly.]

\subsection{Source code for individual based simulations}\label{app_sc_ind}

In the following, we give the source code of the individual based simulation function, written in Matlab (Mathworks \citeyear{Mathworks:11}). Note that the source code presented here is optimized for readability rather that computational efficiency. For instance, clonal haploid and sexual diploid reproduction are given in a single function.

[Please refer to the online supplementary of the pubslished version for the source code, or contact the authors directly.]

\end{appendix}

\bibliography{literature}

\begin{thebibliography}{100}
\providecommand{\natexlab}[1]{#1}

\bibitem[{Abrams et~al.(1993)Abrams, Matsuda, and Harada}]{Abrams:93}
Abrams, P.~A., H.~Matsuda, and Y.~Harada. 1993.
\newblock Evolutionary unstable fitness maxima and stable fitness minima of
  continuous traits.
\newblock Evolutionary Ecology 7:465--487.

\bibitem[{Abrams et~al.(2013)Abrams, Tucker, and Gilbert}]{Abrams:13}
Abrams, P.~A., C.~M. Tucker, and B.~Gilbert. 2013.
\newblock Evolution of the storage effect.
\newblock Evolution 67:315--327.

\bibitem[{Ajar(2003)}]{Ajar:03}
Ajar, E. 2003.
\newblock Analysis of disruptive selection in subdivided populations.
\newblock Bmc Evolutionary Biology 3:22.

\bibitem[{Bergelson et~al.(2001)Bergelson, Kreitman, Stahl, and
  Tian}]{Bergelson:01}
Bergelson, J., M.~Kreitman, E.~A. Stahl, and D.~C. Tian. 2001.
\newblock Evolutionary dynamics of plant r-genes.
\newblock Science 292:2281--2285.

\bibitem[{Bolnick et~al.(2003)Bolnick, Svanb\"ack, Fordyce, Yang, Davis,
  Hulsey, and Forister}]{Bolnick:03}
Bolnick, D.~I., R.~Svanb\"ack, J.~A. Fordyce, L.~H. Yang, J.~M. Davis, C.~D.
  Hulsey, and M.~L. Forister. 2003.
\newblock The ecology of individuals: Incidence and implications of individual
  specialization.
\newblock The American Naturalist 161:1--28.

\bibitem[{Byers(2005)}]{Byers:05}
Byers, D.~L. 2005.
\newblock Evolution in heterogeneous environments and the potential of
  maintenance of genetic variation in traits of adaptive significance.
\newblock Genetica 123:107--124.

\bibitem[{Chesson(1984)}]{Chesson:84}
Chesson, P.~L. 1984.
\newblock The storage effect in stochastic population models.
\newblock Lecture Notes in Biomathematics 54:76--89.

\bibitem[{Chesson(1985)}]{Chesson:85}
---{}---{}---. 1985.
\newblock Coexistence of competitors in spatially and temporally varying
  environments - a look at the combined effects of different sorts of
  variability.
\newblock Theoretical Population Biology 28:263--287.

\bibitem[{Chesson(1994)}]{Chesson:94}
---{}---{}---. 1994.
\newblock Multispecies competition in variable environments.
\newblock Theoretical Population Biology 45:227--276.

\bibitem[{Chesson(2000{\natexlab{\emph{a}}})}]{Chesson:00b}
---{}---{}---. 2000{\natexlab{\emph{a}}}.
\newblock General theory of competitive coexistence in spatially-varying
  environments.
\newblock Theoretical Population Biology 58:211--237.

\bibitem[{Chesson(2000{\natexlab{\emph{b}}})}]{Chesson:00a}
---{}---{}---. 2000{\natexlab{\emph{b}}}.
\newblock Mechanisms of maintenance of species diversity.
\newblock Annual Review of Ecology and Systematics 31:343--368.

\bibitem[{Chesson and Warner(1981)}]{Chesson:81}
Chesson, P.~L., and R.~R. Warner. 1981.
\newblock Environmental variability promotes coexistence in lottery competitive
  systems.
\newblock The American Naturalist 117:923--943.

\bibitem[{Christiansen(1974)}]{Christiansen:74}
Christiansen, F.~B. 1974.
\newblock Sufficient conditions for protected polymorphism in a subdivided
  population.
\newblock The American Naturalist 108:157--166.

\bibitem[{Christiansen(1975)}]{Christiansen:75}
---{}---{}---. 1975.
\newblock Hard and soft selection in a subdivided population.
\newblock American Naturalist 109:11--16.

\bibitem[{Claessen et~al.(2007)Claessen, Andersson, Persson, and
  de~Roos}]{Claessen:07}
Claessen, D., J.~Andersson, L.~Persson, and A.~de~Roos. 2007.
\newblock Delayed evolutionary branching in small populations.
\newblock Evolutionary Ecology Research 9:51--69.

\bibitem[{Comins and Noble(1985)}]{Comins:85}
Comins, H.~N., and I.~R. Noble. 1985.
\newblock Dispersal, variability, and transient niches - species coexistence in
  a uniformly variable environment.
\newblock The American Naturalist 126:706--723.

\bibitem[{Cook(2003)}]{Cook:03}
Cook, L.~M. 2003.
\newblock The rise and fall of the carbonaria form of the peppered moth.
\newblock Quarterly Review of Biology 78:399--417.

\bibitem[{Cook and Hartl(1974)}]{Cook:74}
Cook, R.~D., and D.~J. Hartl. 1974.
\newblock Uncorrelated random environments and their effects on gene frequency.
\newblock Evolution 28:265--274.

\bibitem[{Day(2000)}]{Day:00}
Day, T. 2000.
\newblock Competition and the effect of spatial resource heterogeneity on
  evolutionary diversification.
\newblock The American Naturalist 155:790--803.

\bibitem[{Day(2001)}]{Day:01}
---{}---{}---. 2001.
\newblock Population structure inhibits evolutionary diversification under
  competition for resources.
\newblock Genetica 112-113:71--86.

\bibitem[{Deakin(1966)}]{Deakin:66}
Deakin, M. A.~B. 1966.
\newblock Sufficient conditions for genetic polymorphism.
\newblock The American Naturalist 100:690--692.

\bibitem[{Deakin(1968)}]{Deakin:68}
---{}---{}---. 1968.
\newblock Genetic polymorphism in a subdivided population.
\newblock Australian Journal of Biological Sciences 21:165--168.

\bibitem[{D\'ebarre and Gandon(2010)}]{Debarre:10}
D\'ebarre, F., and S.~Gandon. 2010.
\newblock Evolution of specialization in a spatially continuous environment.
\newblock J Evol Biol 23:1090--1099.

\bibitem[{D\'ebarre and Gandon(2011)}]{Debarre:11}
---{}---{}---. 2011.
\newblock Evolution in heterogeneous environments: between soft and hard
  selection.
\newblock The American Naturalist 177:E84--E97.

\bibitem[{Dempster(1955)}]{Dempster:55}
Dempster, E.~R. 1955.
\newblock Maintenance of genetic heterogeneity.
\newblock Cold Spring Harb Symp Quant Biol 20:25--31.

\bibitem[{Dieckmann and Doebeli(1999)}]{Dieckmann:99}
Dieckmann, U., and M.~Doebeli. 1999.
\newblock On the origin of species by sympatric speciation.
\newblock Nature 400:354--357.

\bibitem[{Dieckmann and Law(1996)}]{Dieckmann:96}
Dieckmann, U., and R.~Law. 1996.
\newblock The dynamical theory of coevolution: A derivation from stochastic
  ecological processes.
\newblock Journal of Mathematical Biology 34:579--612.

\bibitem[{Diekmann(2004)}]{Diekmann:04}
Diekmann, O. 2004.
\newblock A beginners guide to adaptive dynamics.
\newblock Pages 47--86 \emph{in} R.~Rudnicki, ed. Mathematical Modelling of
  Population Dynamics, vol.~63 of \emph{Banach Center Publications}. Polish
  Academy of Sciences, Warszawa.

\bibitem[{Ellner and Hairston(1994)}]{Ellner:94}
Ellner, S., and N.~G. Hairston. 1994.
\newblock Role of overlapping generations in maintaining genetic variation in a
  fluctuating environment.
\newblock The American Naturalist 143:403--417.

\bibitem[{Eshel(1983)}]{Eshel:83}
Eshel, I. 1983.
\newblock Evolutionary and continuous stability.
\newblock Journal of Theoretical Biology 103:99--111.

\bibitem[{Felsenstein(1976)}]{Felsenstein:76}
Felsenstein, J. 1976.
\newblock The theoretical population genetics of variable selection and
  migration.
\newblock Annual Review in Genetics 10:253--280.

\bibitem[{Geritz et~al.(1998)Geritz, Kisdi, Mesz\'ena, and Metz}]{Geritz:98}
Geritz, S., {\'E}.~Kisdi, G.~Mesz\'ena, and J.~Metz. 1998.
\newblock Evolutionarily singular strategies and the adaptive growth and
  branching of the evolutionary tree.
\newblock Evolutionary Ecology 12:35--57.

\bibitem[{Geritz(2005)}]{Geritz:05}
Geritz, S. A.~H. 2005.
\newblock Resident-invader dynamics and the coexistence of similar strategies.
\newblock Journal of Mathematical Biology 50:67--82.

\bibitem[{Gillespie(1973{\natexlab{\emph{a}}})}]{Gillespie:73b}
Gillespie, J. 1973{\natexlab{\emph{a}}}.
\newblock Natural-selection with varying selection coefficients - haploid
  model.
\newblock Genetical Research 21:115--120.

\bibitem[{Gillespie and Langley(1976)}]{Gillespie:76}
Gillespie, J., and C.~Langley. 1976.
\newblock Multilocus behavior in random environments. 1. {R}andom levene
  models.
\newblock Genetics 82:123--137.

\bibitem[{Gillespie and Turelli(1989)}]{Gillespie:89}
Gillespie, J., and M.~Turelli. 1989.
\newblock Genotype-environment interactions and the maintenance of polygenic
  variation.
\newblock Genetics 121:129--138.

\bibitem[{Gillespie(1973{\natexlab{\emph{b}}})}]{Gillespie:73}
Gillespie, J.~H. 1973{\natexlab{\emph{b}}}.
\newblock Polymorphism in random environments.
\newblock Theoretical Population Biology 4:193--195.

\bibitem[{Gillespie(1974)}]{Gillespie:74}
---{}---{}---. 1974.
\newblock Polymorphism in patchy environments.
\newblock The American Naturalist 108:145--151.

\bibitem[{Gillespie(1975)}]{Gillespie:75}
---{}---{}---. 1975.
\newblock Role of migration in genetic structure of populations in temporarily
  and spatially varying environments. 1. {C}onditions for polymorphism.
\newblock The American Naturalist 109:127--135.

\bibitem[{Gillespie(1976)}]{Gillespie:76b}
---{}---{}---. 1976.
\newblock Role of migration in genetic structure of populations in temporally
  and spatially varying environments. 2. {I}sland models.
\newblock Theoretical Population Biology 10:227--238.

\bibitem[{Gliddon and Strobeck(1975)}]{Gliddon:75}
Gliddon, C., and C.~Strobeck. 1975.
\newblock Necessary and sufficient conditions for multiple-niche polymorphism
  in haploids.
\newblock The American Naturalist 109:233--235.

\bibitem[{Hedrick(1978)}]{Hedrick:78}
Hedrick, P.~W. 1978.
\newblock Genetic-variation in a heterogeneous environment. 5. {S}patial
  heterogeneity in finite populations.
\newblock Genetics 89:389--401.

\bibitem[{Hedrick(2006)}]{Hedrick:06}
---{}---{}---. 2006.
\newblock Genetic polymorphism in heterogeneous environments: The age of
  genomics.
\newblock Annual Review of Ecology, Evolution and Systematics 37:67--93.

\bibitem[{Hedrick et~al.(1976)Hedrick, Ginewan, and Ewing}]{Hedrick:76}
Hedrick, P.~W., M.~E. Ginewan, and E.~P. Ewing. 1976.
\newblock Genetic-polymorphism in heterogeneous environments.
\newblock Annual Review of Ecology and Systematics 7:1--32.

\bibitem[{Houle(1992)}]{Houle:92}
Houle, D. 1992.
\newblock Comparing evolvability and variability of quantitative traits.
\newblock Genetics 130:195--204.

\bibitem[{Johansson and Ripa(2006)}]{Johansson:06}
Johansson, J., and J.~Ripa. 2006.
\newblock Will sympatric speciation fail due to stochastic competitive
  exclusion?
\newblock American Naturalist 168:572--578.

\bibitem[{Johansson et~al.(2010)Johansson, Ripa, and Kucklander}]{Johansson:10}
Johansson, J., J.~Ripa, and N.~Kucklander. 2010.
\newblock The risk of competitive exclusion during evolutionary branching:
  Effects of resource variability, correlation and autocorrelation.
\newblock Theoretical Population Biology 77:95--104.

\bibitem[{Karlin(1982)}]{Karlin:82}
Karlin, S. 1982.
\newblock Classifications of selection migration structures and conditions for
  a protected polymorphism.
\newblock Evolutionary Biology 14:61--204.

\bibitem[{Karlin and Levikson(1974)}]{Karlin:74}
Karlin, S., and B.~Levikson. 1974.
\newblock Temporal fluctuations in selection intensities - case of small
  population-size.
\newblock Theoretical Population Biology 6:383--412.

\bibitem[{Kassen(2002)}]{Kassen:02}
Kassen, R. 2002.
\newblock The experimental evolution of specialists, generalists, and the
  maintenance of diversity.
\newblock Journal of Evolutionary Biology 15:173--190.

\bibitem[{Kimura(1954)}]{Kimura:54}
Kimura, M. 1954.
\newblock Process leading to quasi-fixation of genes in natural populations due
  to random fluctuation of selection intensities.
\newblock Genetics 39:280--295.

\bibitem[{Kimura(1955)}]{Kimura:55}
---{}---{}---. 1955.
\newblock Stochastic processes and distribution of gene frequencies under
  natural selection.
\newblock Cold Spring Harbor Symposia On Quantitative Biology 20:33--53.

\bibitem[{Kisdi(2002)}]{Kisdi:02}
Kisdi, {\'E}. 2002.
\newblock Dispersal: Risk spreading versus local adaptation.
\newblock The American Naturalist 159:579--596.

\bibitem[{Kisdi and Geritz(1999)}]{Kisdi:99}
Kisdi, {\'E}., and S.~A.~H. Geritz. 1999.
\newblock Adaptive dynamics in allele space: Evolution of genetic polymorphism
  by small mutations in a heterogeneous environment.
\newblock Evolution 53:993--1008.

\bibitem[{Kopp and Hermisson(2006)}]{Kopp:06}
Kopp, M., and J.~Hermisson. 2006.
\newblock The evolution of genetic architecture under frequency-dependent
  disruptive selection.
\newblock Evolution 60:1537--1550.

\bibitem[{Leimar(2005)}]{Leimar:05}
Leimar, O. 2005.
\newblock The evolution of phenotypic polymorphism: Randomized strategies
  versus evolutionary branching.
\newblock The American Naturalist 165:669--681.

\bibitem[{Leimar(2008)}]{Leimar:08}
---{}---{}---. 2008.
\newblock Environmental and genetic cues in the evolution of phenotypic
  polymorphism.
\newblock Evolutionary Ecology .

\bibitem[{Leimar et~al.(2006)Leimar, Van~Dooren, and Hammerstein}]{Leimar:06}
Leimar, O., T.~J.~M. Van~Dooren, and P.~Hammerstein. 2006.
\newblock A new perspective on developmental plasticity and the principles of
  adaptive morph determination.
\newblock The American Naturalist 167:367--376.

\bibitem[{Levene(1953)}]{Levene:53}
Levene, H. 1953.
\newblock Genetic equilibrium when more than one ecological niche is available.
\newblock The American Naturalist 87:331--333.

\bibitem[{Levins(1962)}]{Levins:62}
Levins, R. 1962.
\newblock Theory of fitness in a heterogeneous environment. {I}. {T}he fitness
  set and the adaptive function.
\newblock The American Naturalist 96:361--373.

\bibitem[{Mackay(1981)}]{Mackay:81}
Mackay, T. F.~C. 1981.
\newblock Genetic-variation in varying environments.
\newblock Genetical Research 37:79--93.

\bibitem[{{MathWorks, Inc.}(2011)}]{Mathworks:11}
{MathWorks, Inc.} 2011.
\newblock MATLAB: the language of technical computing. Desktop tools and
  development environment, version R2011a, vol. 7.12.0.635.
\newblock MathWorks.

\bibitem[{McKenzie(1996)}]{McKenzie:96}
McKenzie, J.~A. 1996.
\newblock Ecological and evolutionary aspects of insecticide resistance.
\newblock R. G. Landes Co., Austin, Texas, USA.

\bibitem[{Mesz\'ena et~al.(1997)Mesz\'ena, Czibula, and Geritz}]{Meszena:97}
Mesz\'ena, G., I.~Czibula, and S.~A.~H. Geritz. 1997.
\newblock Adaptive dynamics in a 2-patch environment: a toy model for
  allopatric and parapatric speciation.
\newblock Journal of Biological Systems 5:265--284.

\bibitem[{Metz(2008)}]{Metz:08c}
Metz, J. 2008.
\newblock Fitness.
\newblock Pages 1599--1612 \emph{in} S.~J{\o}rgensen and B.~Fath, eds.
  Evolutionary Ecology. Vol. [2] of Encyclopedia of Ecology. Elsevier.

\bibitem[{Metz et~al.(1996)Metz, Geritz, Mesz\'ena, Jacobs, and
  Van~Heerwaarden}]{Metz:96a}
Metz, J., S.~Geritz, G.~Mesz\'ena, F.~Jacobs, and J.~Van~Heerwaarden. 1996.
\newblock Adaptive dynamics: A geometrical study of the consequences of nearly
  faithful reproduction.
\newblock Pages 183--231 \emph{in} S.~van Strien and S.~Verduyn~Lunel, eds.
  Stochastic and spatial structures of dynamical systems, Proceedings of the
  Royal Dutch Academy of Science. North Holland, Dordrecht, Netherlands;
  available at http://www.iiasa.ac.at/Research/ADN/Series.html.

\bibitem[{Metz et~al.(1992)Metz, Nisbet, and Geritz}]{Metz:92}
Metz, J., R.~Nisbet, and S.~Geritz. 1992.
\newblock How should we define `fitness' for general ecological scenarios?
\newblock Trends in Ecology and Evolution 7:198--202.

\bibitem[{Nachman et~al.(2003)Nachman, Hoekstra, and D'Agostino}]{Nachman:03}
Nachman, M.~W., H.~E. Hoekstra, and S.~L. D'Agostino. 2003.
\newblock The genetic basis of adaptive melanism in pocket mice.
\newblock Proceedings of the National Academy of Sciences of the United States
  of America 100:5268--5273.

\bibitem[{Nevo(1978)}]{Nevo:78}
Nevo, E. 1978.
\newblock Genetic-variation in natural-populations - patterns and theory.
\newblock Theoretical Population Biology 13:121--177.

\bibitem[{Nilsson and Ripa(2010{\natexlab{\emph{a}}})}]{Nilsson:10a}
Nilsson, J., and J.~Ripa. 2010{\natexlab{\emph{a}}}.
\newblock Adaptive branching in source-sink habitats.
\newblock Evolutionary Ecology 24:479--489.

\bibitem[{Nilsson and Ripa(2010{\natexlab{\emph{b}}})}]{Nilsson:10b}
---{}---{}---. 2010{\natexlab{\emph{b}}}.
\newblock The origin of polymorphic crypsis in a heterogeneous environment.
\newblock Evolution 64:1386--1394.

\bibitem[{Nurmi and Parvinen(2008)}]{Nurmi:08}
Nurmi, T., and K.~Parvinen. 2008.
\newblock On the evolution of specialization with a mechanistic underpinning in
  structured metapopulations.
\newblock Theoretical Population Biology 73:222--243.

\bibitem[{Nurmi and Parvinen(2011)}]{Nurmi:11}
---{}---{}---. 2011.
\newblock Joint evolution of specialization and dispersal in structured
  metapopulations.
\newblock Journal of Theoretical Biology 275:78--92.

\bibitem[{Parvinen and Egas(2004)}]{Parvinen:04}
Parvinen, K., and M.~Egas. 2004.
\newblock Dispersal and the evolution of specialization in a two-habitat type
  metapopulation.
\newblock Theoretical Population Biology 66:233--248.

\bibitem[{Peischl and Schneider(2010)}]{Peischl:10}
Peischl, S., and K.~A. Schneider. 2010.
\newblock Evolution of dominance under frequency-dependent intraspecific
  competition in an assortatively mating population.
\newblock Evolution 64:561--582.

\bibitem[{Pelz et~al.(2005)Pelz, Rost, Hunerberg, Fregin, Heiberg, Baert,
  MacNicoll, Prescott, Walker, Oldenburg, and Muller}]{Pelz:05}
Pelz, H.~J., S.~Rost, M.~Hunerberg, A.~Fregin, A.~C. Heiberg, K.~Baert, A.~D.
  MacNicoll, C.~V. Prescott, A.~S. Walker, J.~Oldenburg, and C.~R. Muller.
  2005.
\newblock The genetic basis of resistance to anticoagulants in rodents.
\newblock Genetics 170:1839--1847.

\bibitem[{Pennings et~al.(2008)Pennings, Kopp, Mesz\'ena, Dieckmann, and
  Hermisson}]{Pennings:08}
Pennings, P.~S., M.~Kopp, G.~Mesz\'ena, U.~Dieckmann, and J.~Hermisson. 2008.
\newblock An analytically tractable model for competitive speciation.
\newblock The American Naturalist 171:E44--E71.

\bibitem[{Ravign\'e et~al.(2009)Ravign\'e, Dieckmann, and
  Olivieri}]{Ravigne:09}
Ravign\'e, V., U.~Dieckmann, and I.~Olivieri. 2009.
\newblock Live where you thrive: Joint evolution of habitat choice and local
  adaptation facilitates specialization and promotes diversity.
\newblock The American Naturalist 174:E141--E169.

\bibitem[{Rueffler et~al.(2006)Rueffler, Van~Dooren, Leimar, and
  Abrams}]{Rueffler:06b}
Rueffler, C., T.~J.~M. Van~Dooren, O.~Leimar, and P.~A. Abrams. 2006.
\newblock Disruptive selection and then what?
\newblock Trends in Ecology and Evolution 21:238--245.

\bibitem[{Schreiber(2010)}]{Schreiber:10}
Schreiber, S.~J. 2010.
\newblock Interactive effects of temporal correlations, spatial heterogeneity
  and dispersal on population persistence.
\newblock Proceedings of the Royal Society B 277:1907--1914.

\bibitem[{Seger and Brockmann(1987)}]{Seger:87}
Seger, J., and H.~J. Brockmann. 1987.
\newblock What is bet-hedging?
\newblock Oxford Surveys in Evolutionary Biology 4:182--211.

\bibitem[{Snyder and Chesson(2003)}]{Snyder:03}
Snyder, R.~E., and P.~Chesson. 2003.
\newblock Local dispersal can facilitate coexistence in the presence of
  permanent spatial heterogeneity.
\newblock Ecology Letters 6:301--309.

\bibitem[{Spichtig and Kawecki(2004)}]{Spichtig:04}
Spichtig, M., and T.~J. Kawecki. 2004.
\newblock The maintenance (or not) of polygenic variation by soft selection in
  heterogenous environments.
\newblock The American Naturalist 164:70--84.

\bibitem[{Spurgin and Richardson(2010)}]{Spurgin:10}
Spurgin, L.~G., and D.~S. Richardson. 2010.
\newblock How pathogens drive genetic diversity: Mhc, mechanisms and
  misunderstandings.
\newblock Proceedings of the Royal Society B 277:979--988.

\bibitem[{Svardal et~al.(2011)Svardal, Rueffler, and Hermisson}]{Svardal:11}
Svardal, H., C.~Rueffler, and J.~Hermisson. 2011.
\newblock Comparing environmental and genetic variance as adaptive response to
  fluctuating selection.
\newblock Evolution 65:2492--2513.

\bibitem[{Szilagyi and Meszena(2010)}]{Szilagyi:10}
Szilagyi, A., and G.~Meszena. 2010.
\newblock Coexistence in a fluctuating environment by the effect of relative
  nonlinearity a minimal model.
\newblock Journal of Theoretical Biology 267:502--512.

\bibitem[{Taylor(2008)}]{Taylor:08}
Taylor, J.~E. 2008.
\newblock Environmental variation, fluctuating selection and genetic drift in
  subdivided populations.
\newblock Theoretical Population Biology 74:233--250.

\bibitem[{Tuljapurkar(1990)}]{Tuljapurkar:90}
Tuljapurkar, S. 1990.
\newblock Population Dynamics in variable environments, vol.~85 of
  \emph{Lecture Notes in Biomathematics}.
\newblock Springer Verlag, Berlin, Germany.

\bibitem[{Van~Dooren(1999)}]{Dooren:99}
Van~Dooren, T. J.~M. 1999.
\newblock The evolutionary ecology of dominance.
\newblock Journal of Theoretical Biology 198:519--532.

\bibitem[{van Doorn and Dieckmann(2006)}]{VanDoorn:06}
van Doorn, G., and U.~Dieckmann. 2006.
\newblock The long-term evolution of multilocus traits under
  frequency-dependent disruptive selection.
\newblock Evolution 60:2226--2238.

\bibitem[{van Doorn et~al.(2004)van Doorn, Dieckmann, and
  Weissing}]{VanDoorn:04}
van Doorn, S.~G., U.~Dieckmann, and F.~J. Weissing. 2004.
\newblock Sympatric speciation by sexual selection: A critical re-evaluation.
\newblock The American Naturalist 163:709--725.

\bibitem[{Via and Lande(1987)}]{Via:87}
Via, S., and R.~Lande. 1987.
\newblock Evolution of genetic-variability in a spatially heterogeneous
  environment - effects of genotype-environment interaction.
\newblock Genetical Research 49:147--156.

\bibitem[{Vignieri et~al.(2010)Vignieri, Larson, and Hoekstra}]{Vignieri:10}
Vignieri, S.~N., J.~G. Larson, and H.~E. Hoekstra. 2010.
\newblock The selective advantage of crypsis in mice.
\newblock Evolution 64:2153--2158.

\bibitem[{Wakano and Iwasa(2013)}]{Wakano:13}
Wakano, J.~Y., and Y.~Iwasa. 2013.
\newblock Evolutionary branching in a finite population: Deterministic
  branching vs. stochastic branching.
\newblock Genetics 193:229--241.

\bibitem[{Wallace(1975)}]{Wallace:75}
Wallace, B. 1975.
\newblock Hard and soft selection revisited.
\newblock Evolution 29:465-- 473.

\bibitem[{West-Eberhard(2003)}]{West-Eberhard:03}
West-Eberhard, M.~J. 2003.
\newblock Developmental Plasticity and Evolution.
\newblock Oxford University Press.

\bibitem[{{Wolfram Research, Inc.}(2011)}]{Mathematica:11}
{Wolfram Research, Inc.} 2011.
\newblock Mathematica Edition: Version 8.0.
\newblock {Champaign, IL}.

\bibitem[{Wright(1943)}]{Wright:43}
Wright, S. 1943.
\newblock Isolation by distance.
\newblock Genetics 28:114--138.

\bibitem[{Wright(1948)}]{Wright:48}
---{}---{}---. 1948.
\newblock On the roles of directed and random changes in gene frequency in the
  genetics of populations.
\newblock Evolution 2:279--294.

\bibitem[{Yeaman and Whitlock(2011)}]{Yeaman:11}
Yeaman, S., and M.~C. Whitlock. 2011.
\newblock The genetic architecture of adaptation under migration-selection
  balance.
\newblock Evolution 65:1897--1911.

\end{thebibliography}
\bibliographystyle{amnatnat}

\end{document}